\documentclass[11pt]{article}
\usepackage{epsfig}
\usepackage{html}
\def\utfit{{\bf{U}}\kern-.24em{\bf{T}}\kern-.21em{\it{fit}}\@}

\newcommand{\lsim}{
\mathrel{\hbox{\rlap{\hbox{\lower4pt\hbox{$\sim$}}}\hbox{$<$}}}}

\newcommand{\gsim}{
\mathrel{\hbox{\rlap{\hbox{\lower4pt\hbox{$\sim$}}}\hbox{$>$}}}}

\newcommand{\be}{\begin{equation}}
\newcommand{\ee}{\end{equation}}
\newcommand{\bi}{\begin{itemize}}
\newcommand{\ei}{\end{itemize}}

\def \kpn{K^+\rightarrow\pi^+\nu\bar\nu}
\def \klpn{K_{\rm L}\rightarrow\pi^0\nu\bar\nu}

\def \BXsgamma{B \to X_s \gamma}          
\def \bsll{b \to s l^+l^-}                      
\def \BXsll{B \to X_s l^+l^-}             

\def \BXsnunu{B \to X_s \nu \bar{\nu}}    
\def \BXdnunu{B \to X_d \nu \bar{\nu}}    
\def \Bsmumu{B_s \to \mu^{+} \mu^{-}}    
\def \Bdmumu{B_d \to \mu^{+} \mu^{-}}    
\def \BXclnu{B \to X_c l \bar{\nu}}       

\newcommand{\kmm}{K_{\rm L} \to \mu^+ \mu^-}

\renewcommand{\baselinestretch}{1.2}

\begin{document}
\begin{titlepage}
\vspace*{-1.5truecm}

\begin{flushright}
UCSD-PTH 05-08\\
TUM-HEP-585/05\\
ROMA-1404/05\\
hep-ph/0505110
\end{flushright}

\vspace*{0.2truecm}

\begin{center}
\boldmath

{\Large {\bf Upper Bounds on Rare $K$ and $B$ Decays from Minimal Flavour
Violation }} 

\unboldmath
\end{center}

\vspace{0.3truecm}

\begin{center}
{\bf Christoph Bobeth,${}^a$   Marcella Bona,${}^b$
   Andrzej J. Buras,${}^c$ \\
Thorsten Ewerth,${}^d $ Maurizio Pierini,${}^e$  Luca Silvestrini,${}^{c,f}$\\ 
and Andreas Weiler${}^{c}$} 
\vspace{0.3truecm}

${}^a$ {\sl Physics Department, University of California at San Diego, La Jolla, CA 92093, USA}

\vspace{0.05truecm}

${}^b$ {\sl Dipartimento di Fisica, Universit\`a di Torino and INFN,  Sezione di Torino, Via P. Giuria 1, I-10125  Torino, Italy}

\vspace{0.05truecm}

${}^c$ {\sl Physik Department, Technische Universit\"at M\"unchen,
D-85748 Garching, Germany}

\vspace{0.05truecm}

 ${}^d$ {\sl Institute of Theoretical Physics, Sidlerstrasse 5, CH-3012 Bern, Switzerland}

\vspace{0.05truecm}

${}^e$ {\sl Laboratoire de l'Acc\'el\'erateur Lin\'eaire, IN2P3-CNRS et Universit\'e de Paris-Sud, BP 34, 
F-91898 Orsay Cedex}

\vspace{0.05truecm}

${}^f$ {\sl Dipartimento di Fisica, Universit\`a di Roma ``La Sapienza'' and INFN, Sezione di Roma,
Piazzale A. Moro 2, 00185 Roma, Italy}

\end{center}

\vspace{0.3cm}
\begin{abstract}
\vspace{0.1cm}\noindent
We study the branching ratios of rare $K$ and 
$B$ decays in models with minimal flavour violation, 
using the presently available information from the universal unitarity triangle analysis and
{}from the measurements of $Br(B\to X_s\gamma)$, 
$Br(B\to X_s l^+l^-)$ and $Br(\kpn)$. We find the following upper bounds: 
$Br(\kpn)< 11.9 \times 10^{-11}$, 
$Br(\klpn)< 4.6 \times 10^{-11}$, 
$Br(\kmm)_{\mathrm SD}< 1.4 \times 10^{-9}$,
$Br(\BXsnunu)< 5.2 \times 10^{-5}$,
$Br(\BXdnunu)< 2.2 \times 10^{-6}$,
$Br(\Bsmumu)< 7.4 \times 10^{-9}$,
$Br(\Bdmumu)< 2.2 \times 10^{-10}$ at $95\%$ probability.
We analyze in detail various possible scenarios with positive or negative interference of Standard
Model and New Physics contributions, and show how an improvement of experimental data corresponding
to the projected 2010 B factory integrated luminosities will allow to disentangle and test these different possibilities.
Finally, anticipating that subsequently the leading role in constraining this kind of new physics 
will be taken over by the rare decays $\kpn$, $\klpn$ and 
$B_{s,d}\to\mu^+\mu^-$, that are dominated by the $Z^0$-penguin function
$C$, we also present plots for several branching ratios as functions of $C$.
 We point out an interesting
triple correlation between 
$K^+\to\pi^+\nu\bar\nu$, $B\to X_s\gamma$ and $B\to X_s l^+l^-$
present in MFV models.

\end{abstract}

\end{titlepage}
%
%
%


\setcounter{page}{1}
\pagenumbering{arabic}

%
%
\section{Introduction}\label{sec:intro}

\setcounter{equation}{0}

Recently, great experimental progress has been made in the study of
Flavour Changing Neutral Current (FCNC) decays, leading not only to an
impressive accuracy in the extraction of CKM parameters from the
Unitarity Triangle (UT) analysis \cite{UTfitSM,CKMfitter}, but also to
stringent constraints on models with extra sources of flavour and CP
violation, although an accidental agreement of the UT analysis with
the Standard Model (SM) cannot yet be excluded
\cite{hep-ph/0307195,UTfitNP}.

One is then naturally led to consider models with
Minimal Flavour Violation (MFV) \cite{UUT}, in which flavour
and CP violation is governed entirely by the CKM
matrix \cite{CAB,KM} and the relevant operators in effective
Hamiltonians for weak decays are the same as in the SM.

As pointed out in \cite{UUT}, there exists a universal unitarity
triangle (UUT) valid in all these models, that can be constructed
independently of the parameters specific to a given model.  Moreover,
there exist several relations between various branching ratios that
allow straightforward tests of these models. A review has been given
in \cite{Zakopane}.

This formulation of MFV agrees with the one of
\cite{AMGIISST,BOEWKRUR} except for the case of models with two Higgs
doublets at large $\tan \beta$, where also additional operators,
strongly suppressed in the SM, can contribute significantly and the
relations in question are not necessarily satisfied. In the present
paper, MFV will be defined as in \cite{UUT,Zakopane}.

As reviewed in \cite{Zakopane}, this class of models can be formulated
to a very good approximation in terms of eleven parameters: four
parameters of the CKM matrix and seven values of the universal master
functions $F_i(v)$ that parametrize the short distance contributions
to rare decays with $v$ denoting symbolically the parameters of a
given MFV model. However, as argued in \cite{Zakopane}, the new
physics contributions to the functions
\be\label{SCD}
S(v),\qquad  C(v),\qquad D^\prime(v),
\ee
representing respectively $\Delta F=2$ box diagrams, $Z^0$-penguin
diagrams and the magnetic photon penguin diagrams, are the most
relevant ones for phenomenology, with the remaining functions
producing only minor deviations from the SM in low-energy processes.
Several explicit calculations within models with MFV confirm this
conjecture. We have checked the impact of these additional functions
on our analysis, and we will comment on it in
Section~\ref{sec:numerics}.

Now, the existence of a UUT implies that the four CKM parameters can
be determined independently of the values of the functions in
(\ref{SCD}). Moreover, only $C(v)$ and $D^\prime(v)$ enter the
branching ratios for radiative and rare decays so that constraining
their values by (at least) two specific branching ratios allows to
obtain straightforwardly the ranges for all branching ratios within
the class of MFV models. Analyses of that type can be found in
\cite{Zakopane,AMGIISST,hep-ph/0303060}.\footnote{An alternative
  approach is to extract from rare decays the relevant
  Wilson coefficients
  \cite{hep-ph/0112300,hep-ph/0310219,hep-ph/0410155}.  However, since
  in MFV models these coefficients have nontrivial correlations among
  themselves, we find it more transparent to express the physical
  quantities in terms of the functions in eq.~(\ref{SCD}).}

The unique decay to determine the function $D^\prime(v)$ is $B\to
X_s\gamma$, whereas a number of decays such as $\kpn$, $\klpn$,
$B_{s,d}\to\mu^+\mu^-$, $B\to X_{s,d}\nu\bar\nu$ and $K_L\to\pi^0
l^+l^-$ can be used to determine $C(v)$.  The decays $B\to X_{s,d}
l^+l^-$ depend on both $C(v)$ and $D^\prime(v)$ and can be used
together with $B\to X_s\gamma$ and $\kpn$ to determine $C(v)$.

Eventually the decays $\kpn$ and $\klpn$, being the theoretically
cleanest ones \cite{BSU,gino}, will be used to determine $C(v)$.
However, so far only three events of $\kpn$ have been observed
\cite{Adler970,Adler02,E949} and no event of $\klpn$, with the same
comment applying to $B_{s,d}\to\mu^+\mu^-$, $B\to X_{s,d}\nu\bar\nu$
and $K_L\to\pi^0 l^+l^-$.  On the other hand the branching ratio for
$B\to X_s\gamma$ has been known for some time and the branching ratio
for $B\to X_{s} l^+l^-$ has been recently measured by Belle
\cite{hep-ex/0408119} and BaBar \cite{hep-ex/0404006} collaborations.
The latter, combined with $\kpn$, provide presently the best estimate
of the range for $C(v)$ within MFV models.

The main goals of the present paper are
\begin{itemize}
\item to calculate various branching ratios as functions of $C(v)$
  within MFV models,
\item
  to determine the allowed range for $C(v)$ from presently available data,
\item
  to find the upper bounds for the branching ratios of $\kpn$,
  $\klpn$, $B_{s,d}\to\mu^+\mu^-$, $B\to X_{s,d}\nu\bar\nu$ and
  $K_L\to\pi^0 l^+l^-$ within MFV models as defined here,
\item
  to assess the impact of future measurements on MFV models.
\end{itemize}

Our paper is organized as follows. Section 2 can be considered as a
guide to the literature, where the formulae for the branching
ratios in question can be found. In this Section we also give the list
of the input parameters. In Section 3 we present our numerical
analysis of various branching ratios as functions of $C(v)$ and their
expectation values and upper bounds. A brief summary of our
results is given in Section 4.

%
%
\section{Basic Formulae}

\setcounter{equation}{0}

In the MFV models considered here there are no new complex phases and
flavour changing transitions are governed by the CKM matrix.
Moreover, the only relevant operators are those already present in
the SM.  Consequently, new physics enters only through the Wilson
coefficients of the SM operators that can receive additional contributions
due to the exchange of new virtual particles beyond the SM ones.

Any weak decay amplitude can be then cast in the simple form
\be
{\mathrm{A(Decay)}}= \sum_i B_i \eta^i_{\mathrm{QCD}}V^i_{\mathrm{CKM}} 
 F_i(v),
\qquad
 F_i(v)=F^i_{\rm SM}+F^i_{\mathrm{New}}~~\mathrm{(real)},
\label{master}
\ee 
where $F_i(v)$ are the {\it  master functions} of MFV models \cite{Zakopane}
\be\label{masterf}
S(v),~X(v),~Y(v),~Z(v),~E(v),~ D'(v),~ E'(v)
\ee
with $v$ denoting collectively the parameters of a given MFV model.
Examples of models in this class are the Two Higgs Doublet Model II
and the Minimal Supersymmetric Standard Model (MSSM) 
without new sources of flavour violation and for small or moderate
$\tan\beta$. Also models with one universal extra
dimension \cite{BSW02,BPSW} and the simplest little Higgs models are
of MFV type \cite{LittleHiggs}.

In order to find the functions $F_i(v)$ in (\ref{masterf}), one first
looks at various functions resulting from penguin diagrams: $C$ ($Z^0$
penguin), $D$ ($\gamma$ penguin), $E$ (gluon penguin), $D'$
($\gamma$-magnetic penguin) and $E'$ (chromomagnetic
penguin). Subsequently box diagrams have to be considered.  Here we
have the box function $S$ ($\Delta F=2$ transitions), as well as
the $\Delta F=1$ box functions $B^{\nu\bar\nu}$ and $B^{l \bar l}$
relevant for decays with ${\nu\bar\nu}$ and ${l\bar l}$ in the
final state, respectively.

While the $\Delta F=2$ box function $S$ and the penguin functions $E$,
$D'$ and $E'$ are gauge independent, this is not the case for $C$, $D$
and the $\Delta F=1$ box diagram functions $B^{\nu\bar\nu}$ and
$B^{l \bar l}$.  In phenomenological applications it is more
convenient to work with gauge independent functions \cite{PBE}
\begin{equation}\label{XYZ} 
X(v)=C(v)+B^{\nu\bar\nu}(v),\qquad  
Y(v)  =C(v)+B^{l\bar l}(v), \qquad
Z(v)  =C(v)+\frac{1}{4}D(v).
\end{equation}

We have the following correspondence between the most interesting FCNC
processes and the master functions in the MFV models \cite{Zakopane,BH92}:
\begin{center}
\begin{tabular}{lcl}
$K^0-\bar K^0$-mixing ($\varepsilon_K$) 
&\qquad\qquad& $S(v)$ \\
$B_{d,s}^0-\bar B_{d,s}^0$-mixing ($\Delta M_{s,d}$) 
&\qquad\qquad& $S(v)$ \\
$K \to \pi \nu \bar\nu$, $B \to X_{d,s} \nu \bar\nu$ 
&\qquad\qquad& $X(v)$ \\
$K_{\rm L}\to \mu \bar\mu$, $B_{d,s} \to l\bar l$ &\qquad\qquad& $Y(v)$ \\
$K_{\rm L} \to \pi^0 l^+ l^-$ &\qquad\qquad& $Y(v)$, $Z(v)$, 
$E(v)$ \\
$\varepsilon'$, $\Delta S=1$ &\qquad\qquad& $X(v)$,
$Y(v)$, $Z(v)$,
$E(v)$\\
Nonleptonic $\Delta B=1$&\qquad\qquad& $X(v)$,
$Y(v)$, $Z(v)$,
$E(v)$,  $E'(v)$\\
$B \to X_s \gamma$ &\qquad\qquad& $D'(v)$, $E'(v)$ \\
$B \to X_s~{\rm gluon}$ &\qquad\qquad& $E'(v)$ \\ 
$B \to X_s l^+ l^-$ &\qquad\qquad&
$Y(v)$, $Z(v)$, $E(v)$, $D'(v)$, $E'(v)$
\end{tabular}
\end{center}
This table means that the observables like branching ratios, mass
differences $\Delta M_{d,s}$ in $B_{d,s}^0-\bar B_{d,s}^0$-mixing and
the CP violation parameters $\varepsilon$ and $\varepsilon'$, can all
be to a very good approximation expressed in terms of the
corresponding master functions and the relevant CKM factors. The
remaining entries in the formulae for these observables are
low-energy quantities such as the parameters $B_i$, that can be
calculated within the SM and the QCD factors $\eta^i_{\rm QCD}$
describing the renormalization group evolution of operators for scales
$\mu\le M_W$. These factors being universal can be calculated,
similarly to $B_i$, in the SM. The remaining, model-specific QCD
corrections can be absorbed in the functions $F_i$.

The formulae for the processes listed above in the SM, given in terms
of the master functions and CKM factors can be found in many papers.
The full list using the same notation is given in \cite{BBL}. An
update of these formulae with additional references is given in two
papers on universal extra dimensions \cite{BSW02,BPSW}, where one has
to replace $F_i(v,1/R)$ by $F_i(v)$ to obtain the formulae in a
general MFV model. 
In what follows we will use the formulae of \cite{BSW02,BPSW} except that:
\begin{itemize}
\item We will set the functions \be B^{\nu\bar\nu}(v), \qquad B^{l
    \bar l}(v), \qquad E(v) \ee to their SM values and we will trade
  the functions $D^\prime(v)$ and $E^\prime(v)$ for the low-energy
  coefficient $C_7^\mathrm{eff} (\mu_b)$ which enters both $b \to s
  \gamma$ and $\bsll$.  In this manner the only free variables are the
  functions listed in (\ref{SCD}), plus the $D(v)$ function. As
  remarked below, this latter function has only a minor impact on our
  analysis. We have also explored the possible impact of NP
  contributions to $B^{\nu\bar\nu}(v)$ and $B^{l \bar l}(v)$, as will be
  discussed at the end of Section~\ref{sec:numerics}.
\item In obtaining $Br(\kpn)$ we have included the recently calculated 
   long distance contributions~\cite{Isidori:2005xm} that enhance the branching ratio 
   by roughly $6\%$. This amounts  effectively to a charm parameter of
  $P_c=0.43\pm 0.07$.
\item
We will use the formula for $(K_{\rm L} \to \pi^0 l^+ l^-)_{\rm CPV}$ 
from \cite{BI03,Isidori:2004rb}. 
\item We will use the complete NLO formulae for $B\to X_s\gamma$ from
  \cite{b2sgammanlo}.
\item We will use the complete NNLO formulae for $B\to X_s l^+l^-$
  from \cite{b2sllnnlo,hep-ph/0310219}.
\end{itemize}

\begin{table*}[hbt]
\vspace{0.4cm}
\begin{center}
\begin{tabular}{|c||c|c|c|}
\hline
{Branching Ratios} &  Formula &   Reference & Parameters
 \\ \hline
$Br(\kpn)$ &  $(4.24)$ & \cite{BSW02} &  $Br(K^+ \to \pi^0 e^+ \nu)$, $m_c$
\\ \hline
$Br(\klpn)$ &  $(4.27)$ & \cite{BSW02} & $Br(K^+ \to \pi^0 e^+ \nu)$
\\ \hline
$Br(\kmm)_{\rm SD} $ &  $(4.32)$ &  \cite{BSW02} & $m_c$
\\ \hline
$Br(K_{\rm L} \to \pi^0 l^+ l^-)_{\rm CPV}$ &  $(43)$ & 
 \cite{Isidori:2004rb} & see \cite{Isidori:2004rb}
\\ \hline
$Br(B\to X_s\nu\bar\nu)$ &  $(4.29)$ & \cite{BSW02} & $Br(\BXclnu)$
\\ \hline
$Br(B\to X_d\nu\bar\nu)$ &  $(4.29)$ &  \cite{BSW02} & $Br(\BXclnu)$
\\ \hline
$Br(B_s\to \mu^+\mu^-)$ &  $(4.30)$ &   \cite{BSW02} & $F_{B_{s}}$
\\ \hline
$Br(B_d\to \mu^+\mu^-)$ &  $(4.30)$ &  \cite{BSW02} & $F_{B_{d}}$
\\ \hline
\end{tabular}
\end{center}
\caption[]{\it  Guide to the formulae. See text for
explanations. The dependence of all branching ratios on CKM parameters and
the top quark mass is not explicitly reported.
}
\label{guide}
\end{table*}

In Table~\ref{guide} we indicate where the formulae in question can be
found and which additional input parameters are involved in them.  In
Table~\ref{tab:parameters} we give the numerical 
values of all the parameters involved in the analysis.

\begin{table*}[htbp!]
\begin{center}
\begin{tabular}{|c||c|c|}
\hline
         Parameter                          &  Value                            
     & Gaussian ($\sigma$)              \\
     \hline
$\lambda$                                & $0.2255$            
     & $0.0014$                \\
$\vert V_{cb} \vert$                                & $0.0415$   
     &$0.0007$   \\ \hline
$\bar \rho$                                & $0.191$            
     & $0.046$                \\
$\bar \eta$                                & $0.353$   
     &$0.028$   \\ \hline
$F_{B_{s}}$                                  & $230$ MeV                        
     & $30$ MeV              \\
$F_{B_{d}}$                                  & $189$ MeV                        
     & $27$ MeV                 \\
$Br(\BXclnu)$                                  & $0.1045$                              
     & $0.0021$                  \\
$Br(K^+ \to \pi^0 e^+ \nu)$                                 & $0.0487$            
     & $0.0006$           \\
$m_t^{\mathrm{pole}}$                                       & $178.0$ GeV
     & $4.3$ GeV                 \\
$\overline m_b$          & $4.21$ GeV
     & $0.08$ GeV                \\
$\overline m_c$                                       & $1.3$ GeV
     & $0.1$ GeV                  \\
$\alpha_s(M_Z)$                                  & 0.119                             
     & 0.003                         \\ \hline
\end{tabular} 
\end{center}
\caption {\it Values of the relevant parameters used in the analysis.}
\label{tab:parameters} 
\end{table*}

Finally, for the reader's convenience, and in order to show the
relative importance of NP contributions to the processes we consider,
we report below numerical formulae for the branching ratios in terms
of $F^i_{\mathrm{New}}$ in eq.~(\ref{master}).  These numerical
expressions have been obtained for central values of the parameters in
Table~\ref{tab:parameters}, as functions of $\Delta C \equiv
C(v)-C_{\mathrm{SM}}$, $\Delta C_7^\mathrm{eff} \equiv
C_7^\mathrm{eff}- C_{7\,\mathrm{SM}}^\mathrm{eff}$,  $\Delta D \equiv
D(v)-D_{\mathrm{SM}}$, $\Delta B^{l\bar
  l} \equiv B^{l\bar l}(v)-B^{l\bar l}_{\mathrm{SM}}$ and $\Delta
B^{\nu\bar \nu}\equiv B^{\nu\bar \nu}(v)-B^{\nu\bar
  \nu}_{\mathrm{SM}}$.  With the aid of eq.~(\ref{eq:numeric}), it is
possible to quickly check the impact of NP contributions in any given
MFV model. As a first insight, we see that the dependence of
  $Br(\BXsll)$ on $\Delta D$ is relatively weak, as can be read off
  from the small prefactors in the formulae below. From
eq.~(\ref{eq:numeric}) one can also check whether the NP contribution
to box diagrams in any given model is large enough as to modify
significantly our results obtained for $\Delta B^{l\bar l}=\Delta
B^{\nu\bar \nu}=0$ in the next Section.  Finally, these formulae allow
to understand the structure of our numerical results. We
have~\footnote{Notice that we have discarded terms with coefficients
  smaller than $0.1$ in $Br(\BXsll)$.}:
\begin{eqnarray}
Br(\BXsll, 0.04<q^2 (\mathrm{GeV})<1) &=& 1.16 \,\cdot 10^{-6}  \left( 
1+0.38 \,
(\Delta B^{l\bar l})^2+0.46 \,\Delta C_7^\mathrm{eff}
    \Delta B^{l\bar l} \right. \nonumber \\
 &&    +0.41 \,\Delta C 
\Delta B^{l\bar l}-3.47 \,\Delta C_7^\mathrm{eff} +0.56 \,
\Delta B^{l\bar l}+4.31
    (\Delta C_7^\mathrm{eff})^2 \nonumber \\
   && \left.+0.19 \,
(\Delta C)^2+0.38
    \Delta C-0.11 \,\Delta C_7^\mathrm{eff} \Delta D \right)\,, \nonumber \\
Br(\BXsll, 1<q^2 (\mathrm{GeV})<6) &=& 1.61 \,\cdot 10^{-6}  \left( 
1+1.33 \,
(\Delta B^{l\bar l})^2+1.26 \,\Delta C_7^\mathrm{eff}
    \Delta B^{l\bar l} \right. \nonumber \\
 &&    +1.43 \,\Delta C 
\Delta B^{l\bar l}-0.31 \,\Delta D
    \Delta B^{l\bar l}+2.08 \,
\Delta B^{l\bar l}+1.42
    (\Delta C_7^\mathrm{eff})^2 \nonumber \\
   && \left.+0.67 \,
(\Delta C)^2+1.36
    \Delta C-0.29 \,\Delta C_7^\mathrm{eff} 
\Delta D-0.18
    \Delta D \right)\,, \nonumber \\
Br(\BXsll, 14.4<q^2 (\mathrm{GeV})<25) &=& 3.70 \,\cdot 10^{-7}  \left( 
1+1.18 \,
(\Delta B^{l\bar l})^2+ 0.70 \,\Delta C_7^\mathrm{eff}
    \Delta B^{l\bar l} + 0.60 \,\Delta C_7^\mathrm{eff}\right. \nonumber \\
 &&    +1.27 \,\Delta C 
\Delta B^{l\bar l}-0.27 \,\Delta D
    \Delta B^{l\bar l}+2.18 \,
\Delta B^{l\bar l}+0.21
    (\Delta C_7^\mathrm{eff})^2 \nonumber \\
   && \left.+0.60 \,
(\Delta C)^2+1.24
    \Delta C-0.16 \,\Delta C_7^\mathrm{eff} 
\Delta D-0.24
    \Delta D \right)\,, \nonumber \\
Br( B_d \to  \mu^+ \mu^- ) &=& 1.08 \,\cdot 10^{-10} 
\left(1 \, +\Delta B^{l\bar l} + \Delta C \right)^2 \,,\nonumber \\
Br(B_s \to \mu^+ \mu^-)  &=& 3.76 \,\cdot 10^{-9} 
\left(1 \,+\Delta B^{l\bar l} + \Delta C  \right)^2\,,
\nonumber \\
Br(B \to X_d \nu \bar \nu)  &=& 1.50 \,\cdot 10^{-6} 
\left(1+ 0.65 \,(\Delta C+\Delta B^{\nu\bar \nu})\right)^2\,,
\nonumber \\
Br(B \to X_s \nu \bar \nu)  &=& 3.67 \,\cdot 10^{-5} 
\left(1+ 0.65 \,(\Delta C+\Delta B^{\nu\bar \nu}) \right)^2\,, \nonumber \\
Br(K^+ \to \pi^+ \nu \bar \nu) &=& 8.30 \cdot 10^{-11} 
\left(1+ 0.20 
(\Delta C+\Delta B^{\nu \bar \nu})^2+0.89 
(\Delta C+\Delta B^{\nu \bar \nu})\right)\,, \nonumber \\
Br(K_L \to \pi^0 \nu \bar \nu )&=& 3.10 \cdot 10^{-11}
(1+ 0.65 (\Delta C+\Delta B^{\nu \bar \nu}))^2\,, \nonumber \\
Br(K_L \to \mu^+\mu^-) &=& 8.58\cdot 10^{-10}
(1 + 0.82 ( \Delta C+\Delta B^{l \bar l}))^2\,.
\label{eq:numeric}
\end{eqnarray} 

%
%
\section{Numerical Analysis}
\label{sec:numerics}
\setcounter{equation}{0}

Our numerical analysis consists of three steps:
\begin{enumerate}
\item Extracting CKM parameters using the UUT analysis;
\item Determining the allowed range for $\Delta C$ and $\Delta
  C_7^\mathrm{eff}$ from presently available data;
\item Computing the expectation values of rare decays based on these
  allowed ranges.
\end{enumerate}
For the first step, we use the very recent results of the {\utfit}
collaboration on the UUT analysis \cite{UTfitNP}:
\begin{eqnarray}
\bar \rho &=& 0.191            
     \pm 0.046    \,,\qquad \bar \eta = 0.353 \pm 0.028.
\label{eq:uutpar}
\end{eqnarray} 
Since the UUT analysis is independent of loop functions, the above
results are in particular independent of the top quark mass.

In the second step, to minimize the theoretical input, we have traded
$D^\prime(v)$ and $E^\prime(v)$ for $C_7^\mathrm{eff}$, which
is the relevant low-energy quantity entering $Br(B\to X_s \gamma)$ and
$Br(B\to X_s l^+l^-)$. Concerning $Br(B\to X_s \gamma)$, we compare
the theoretical value with the experimental results of CLEO
\cite{hep-ex/0108032}, Belle \cite{hep-ex/0403004} and BaBar
\cite{babarbsg} in the corresponding kinematic ranges, adding a
conservative $10 \%$ flat theoretical error to the theoretical
prediction.  This error contains both the uncertainties due to the
cutoff in the photon spectrum \cite{b2sgammacut} and the ones related
to higher order effects, which are particularly large since we are
omitting here model-specific NLO terms for the NP contribution.  For
$Br(B\to X_s l^+ l^-)$, we use the experimental data in the $q^2$
regions $0.04 < q^2 (\mathrm{GeV}) < 1$, $1 < q^2 (\mathrm{GeV})< 6$
and $14.4 < q^2 (\mathrm{GeV})< 25$ to avoid the theoretical
uncertainty due to the presence of $ c \bar c$ resonances.

\begin{figure}[htb!]
\begin{center}
\includegraphics*[width=0.48\textwidth]{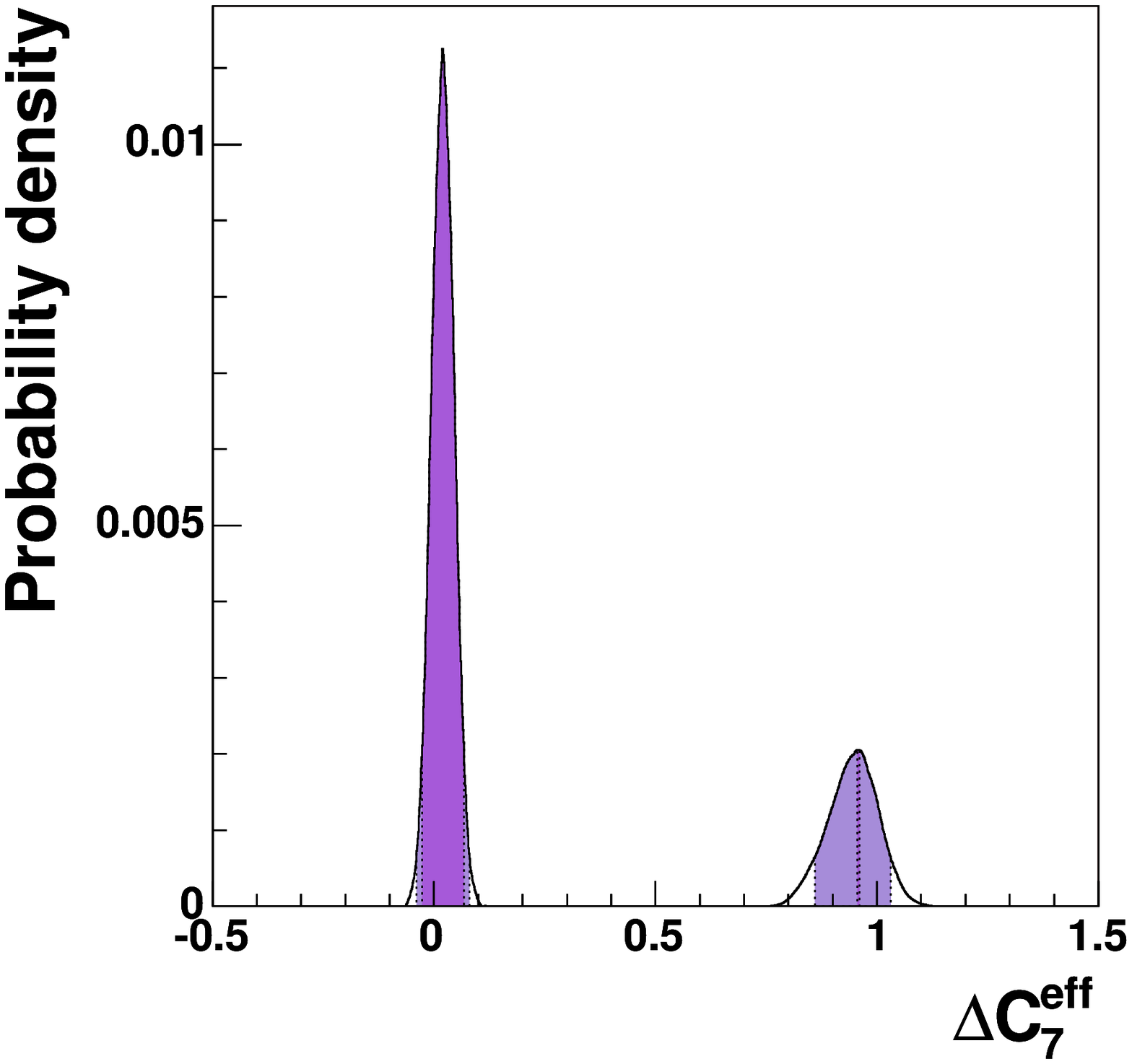}
\includegraphics*[width=0.48\textwidth]{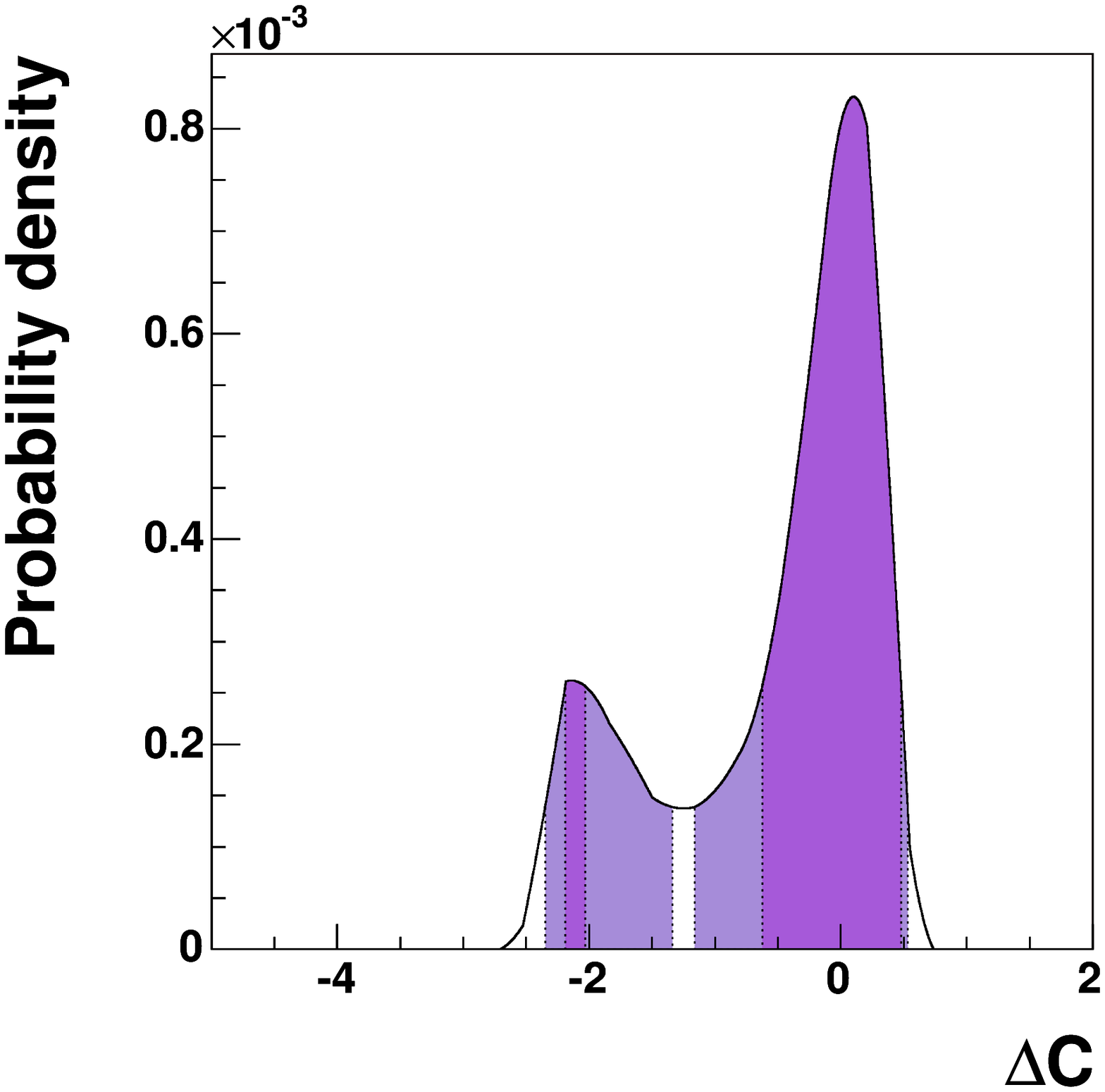}
\includegraphics*[width=0.70\textwidth]{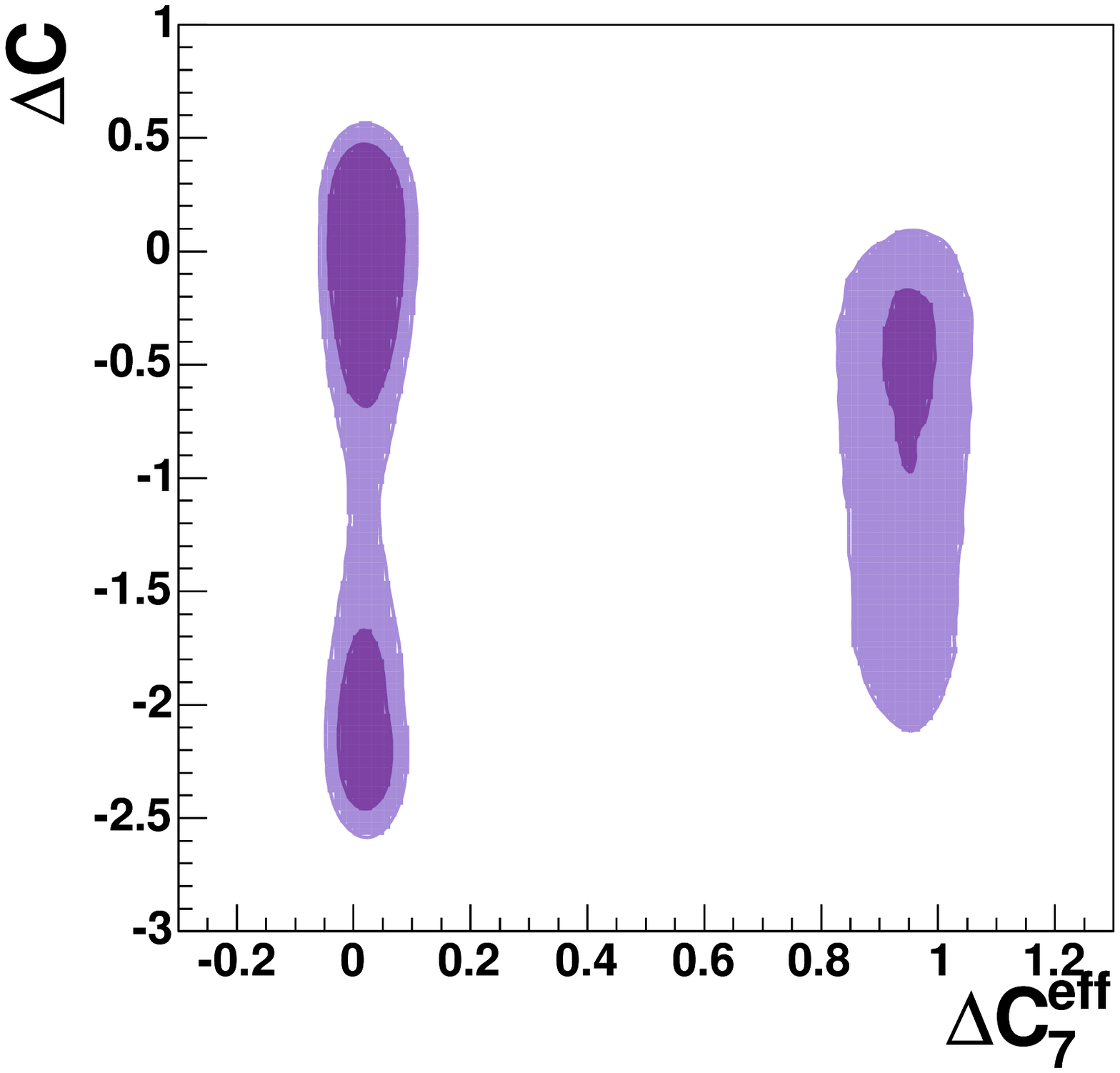}
\caption{%
  \it P.d.f.'s for $\Delta C_7^\mathrm{eff}$ (top-left), $\Delta C$
  (top-right) and $\Delta C$ vs.  $\Delta C_7^\mathrm{eff}$ (bottom).
  Dark (light) areas correspond to the $68\%$ ($95\%$) probability
  region.}
\label{fig:Cs}
\end{center}
\end{figure}

\begin{figure}[htb!]
\begin{center}
\includegraphics*[width=0.32\textwidth]{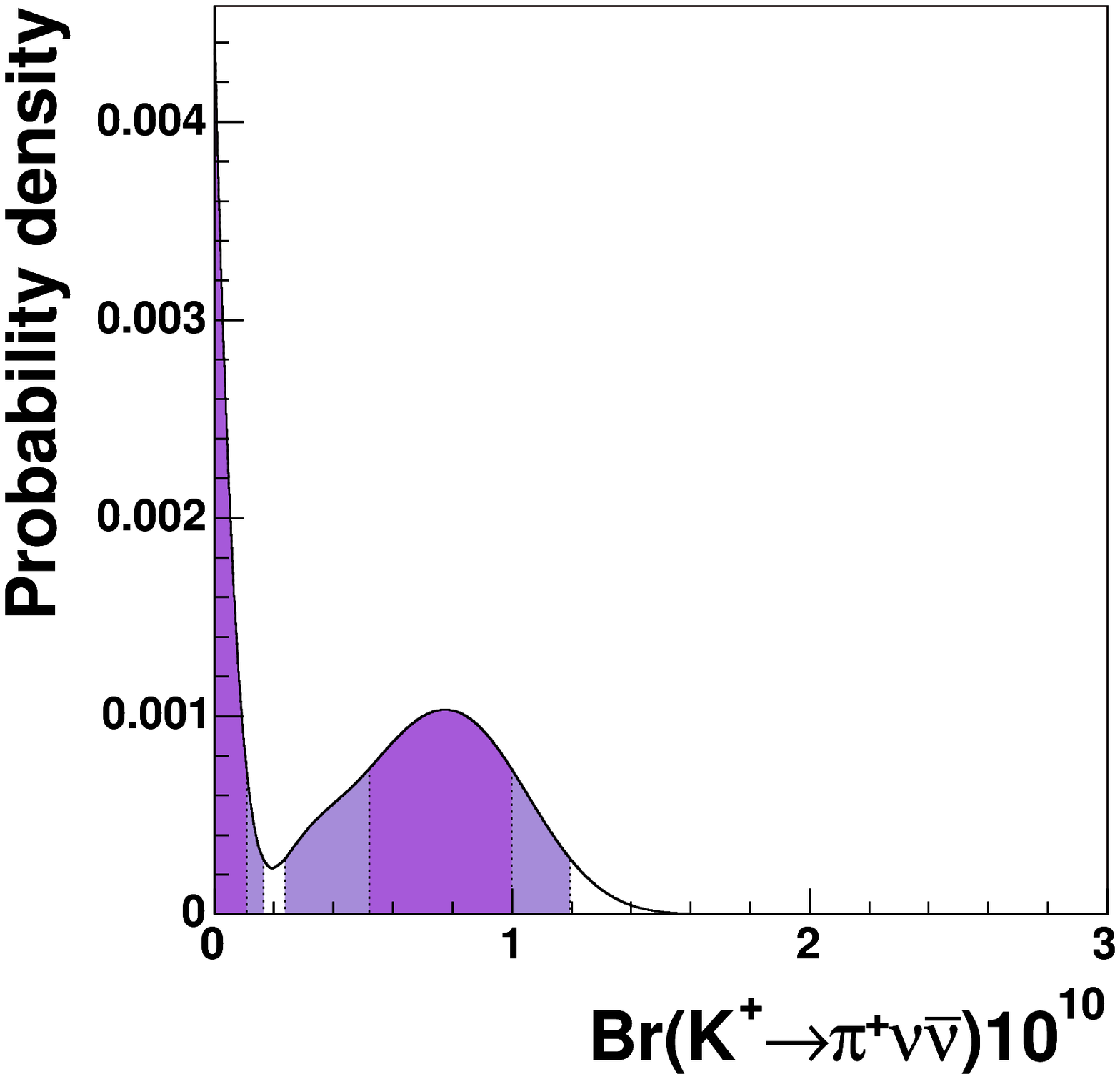}
\includegraphics*[width=0.32\textwidth]{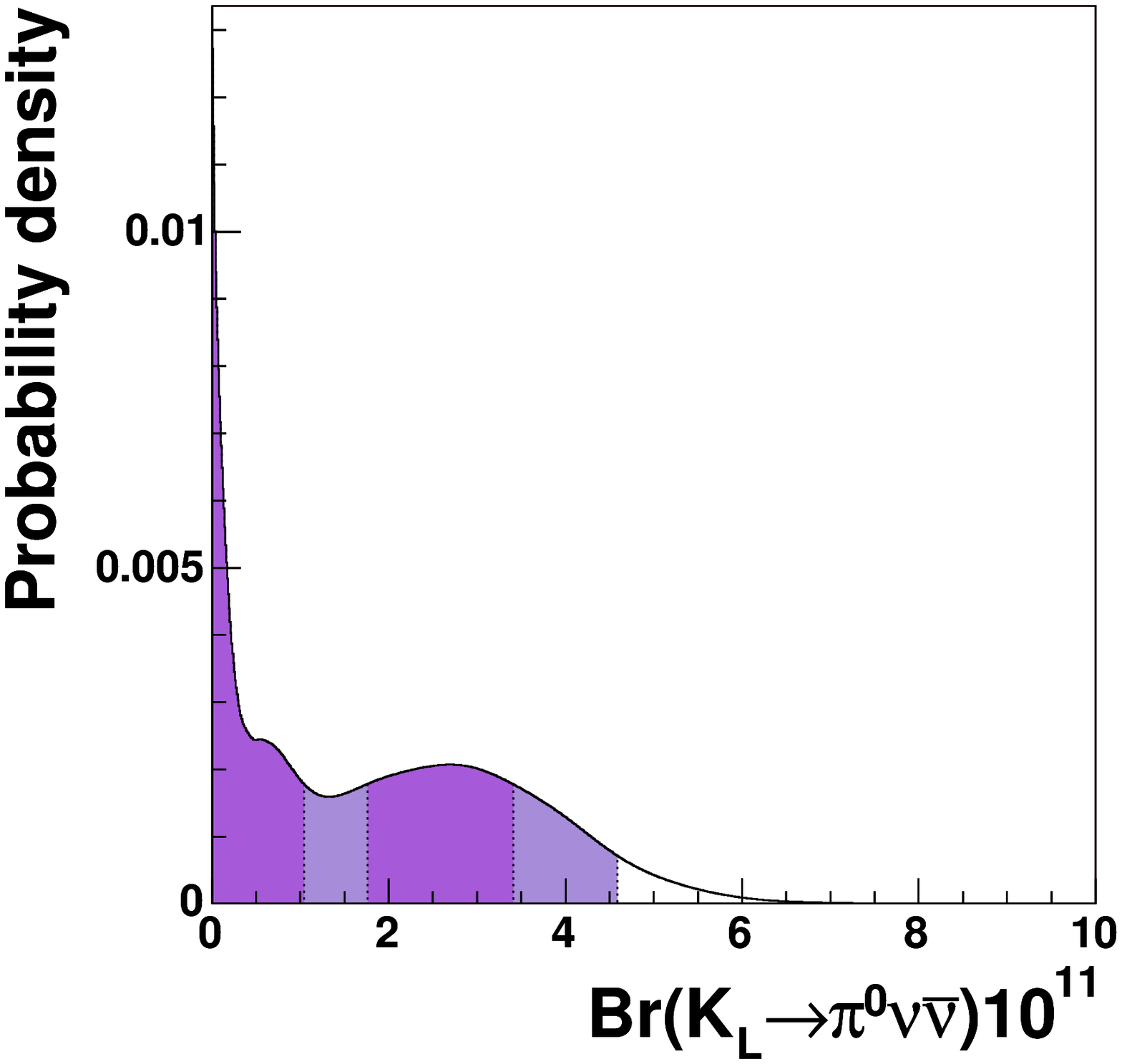}
\includegraphics*[width=0.32\textwidth]{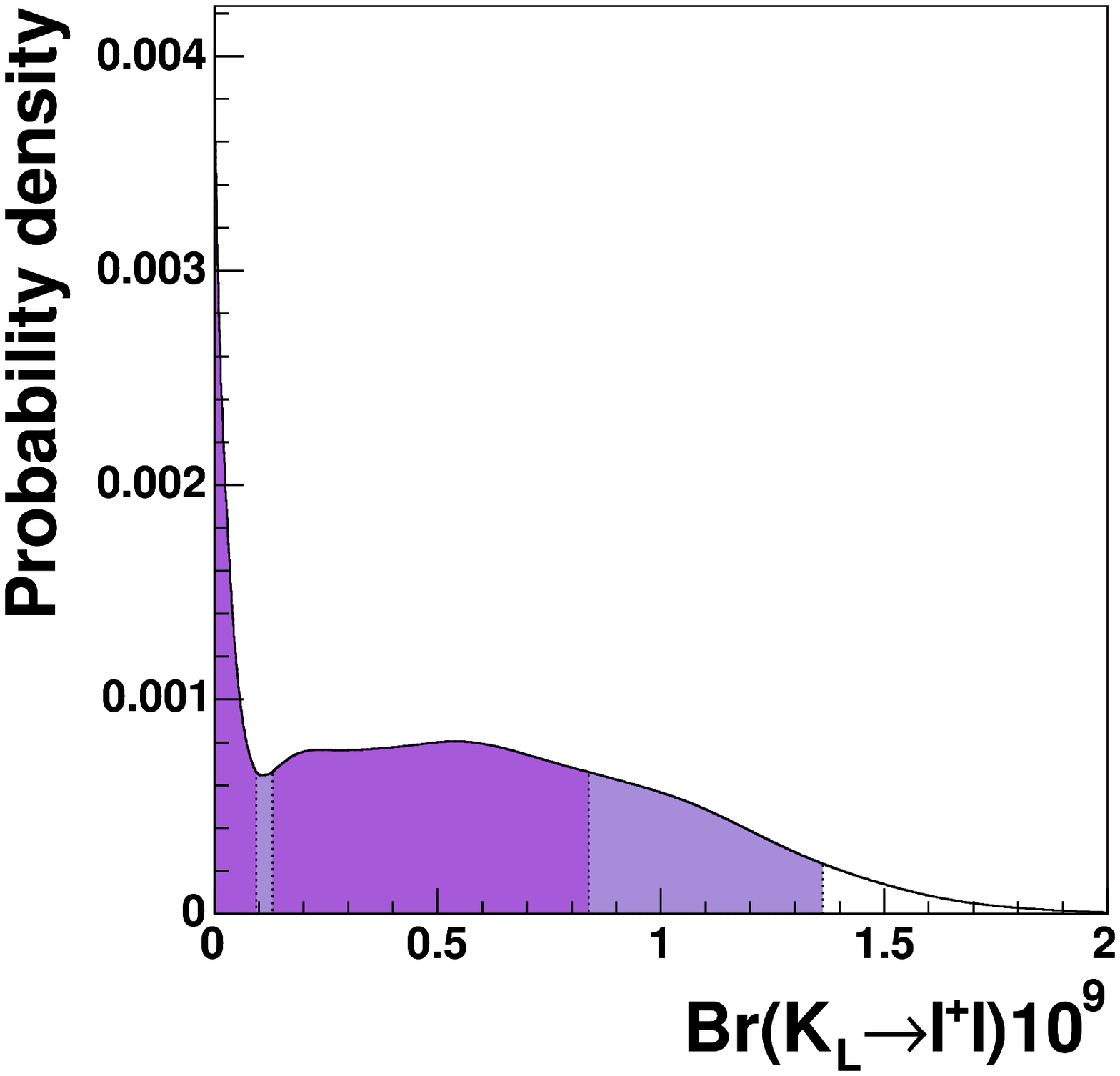}
\includegraphics*[width=0.32\textwidth]{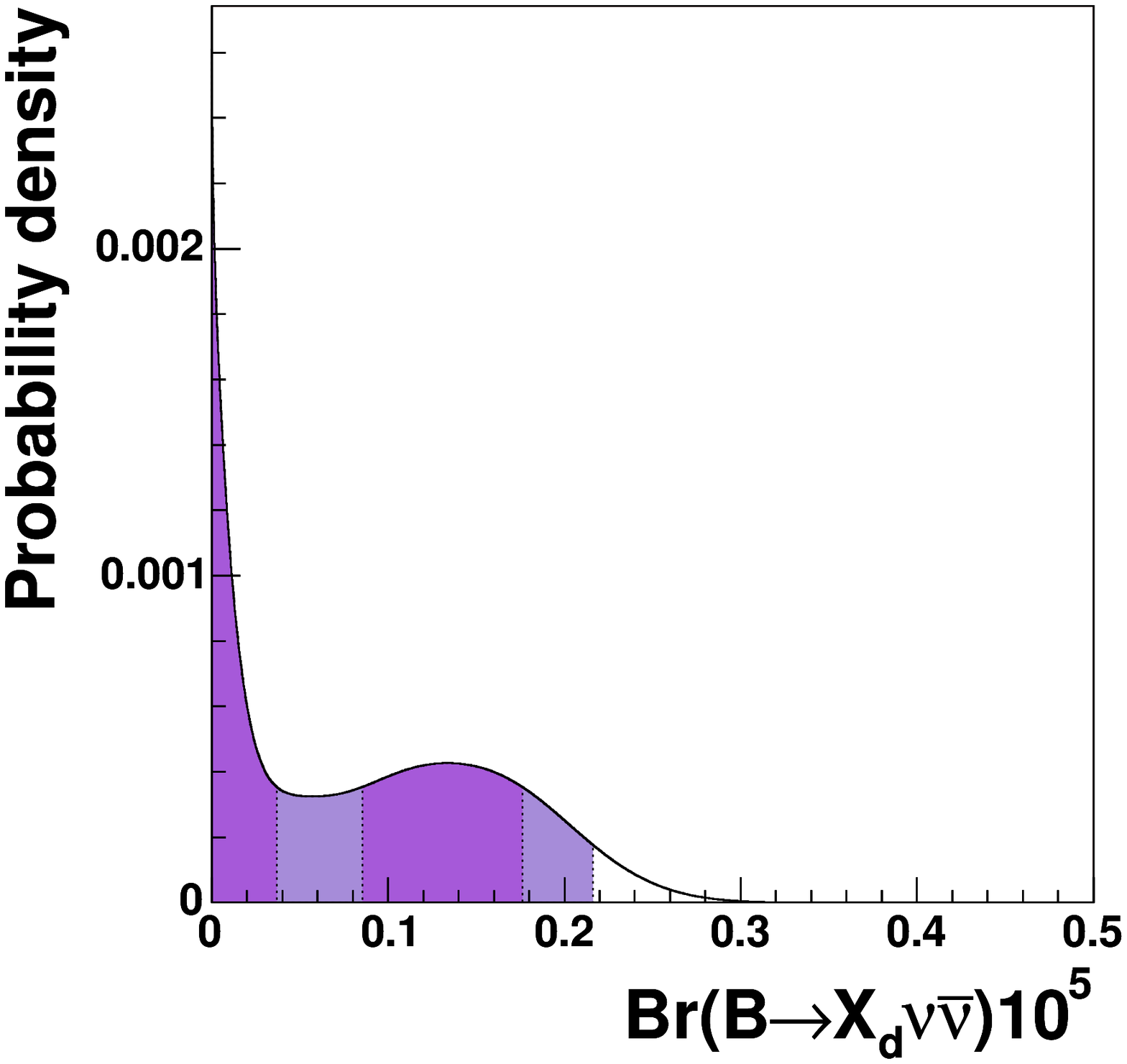}
\includegraphics*[width=0.32\textwidth]{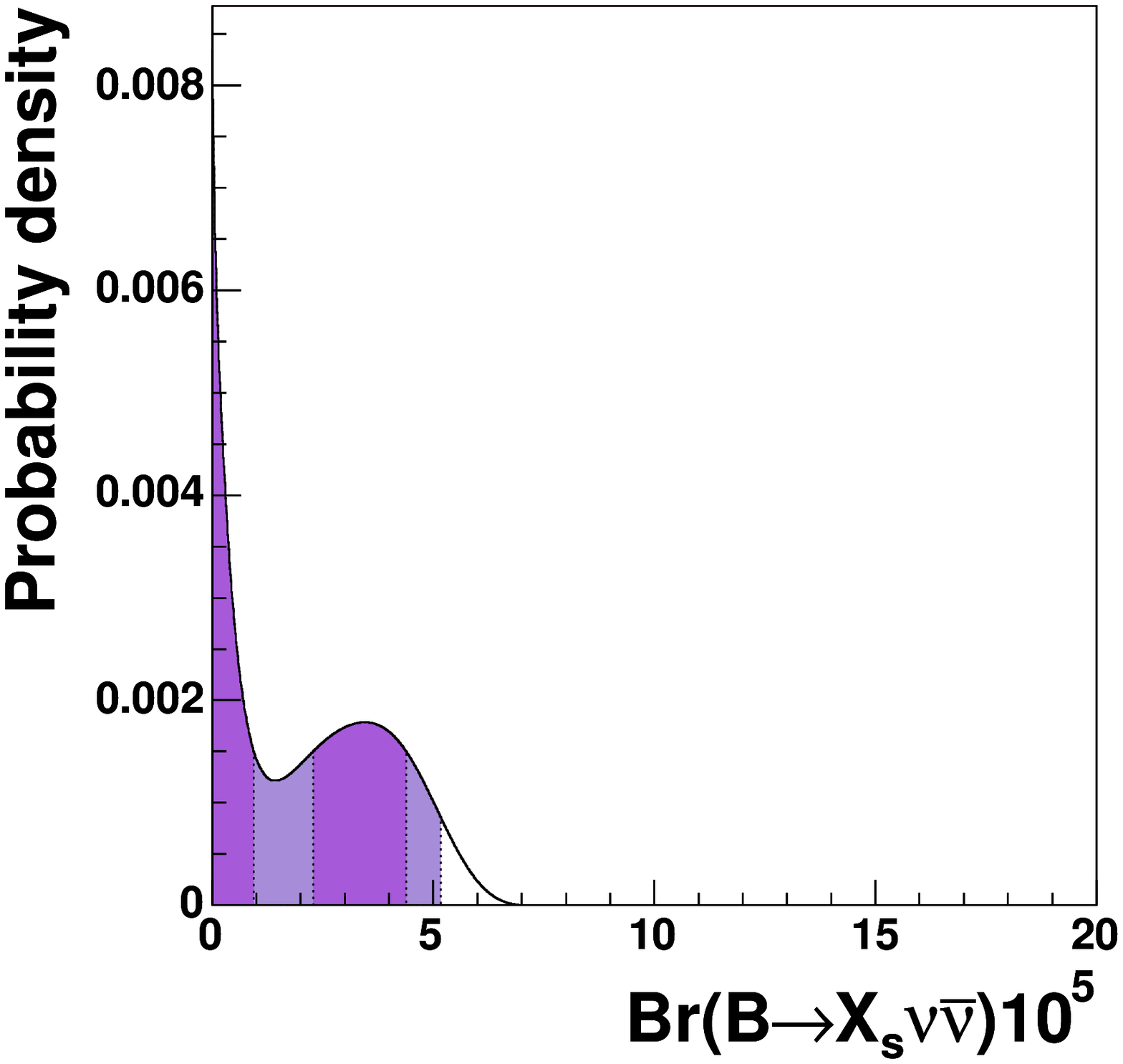}
\includegraphics*[width=0.32\textwidth]{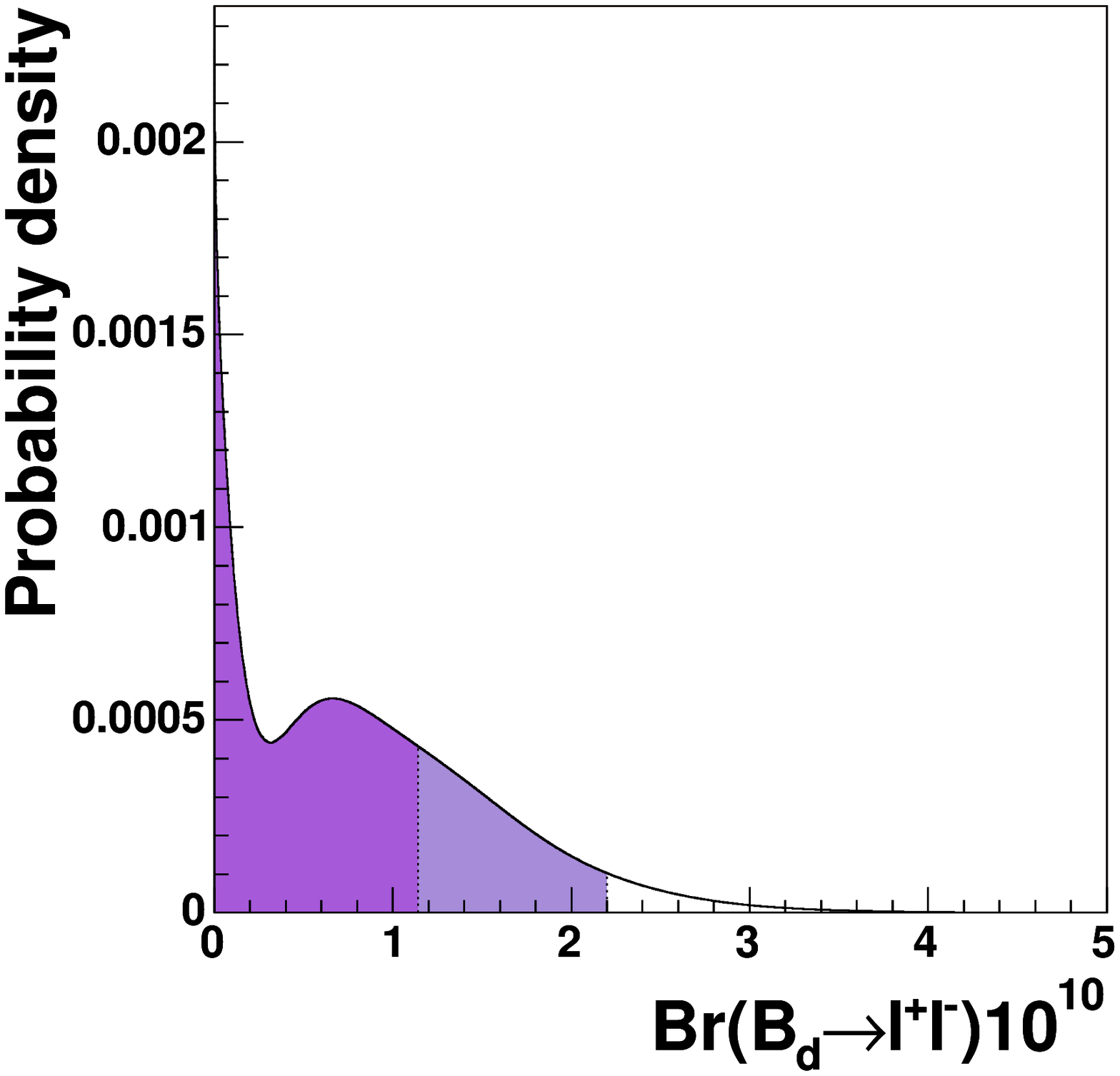}
\includegraphics*[width=0.32\textwidth]{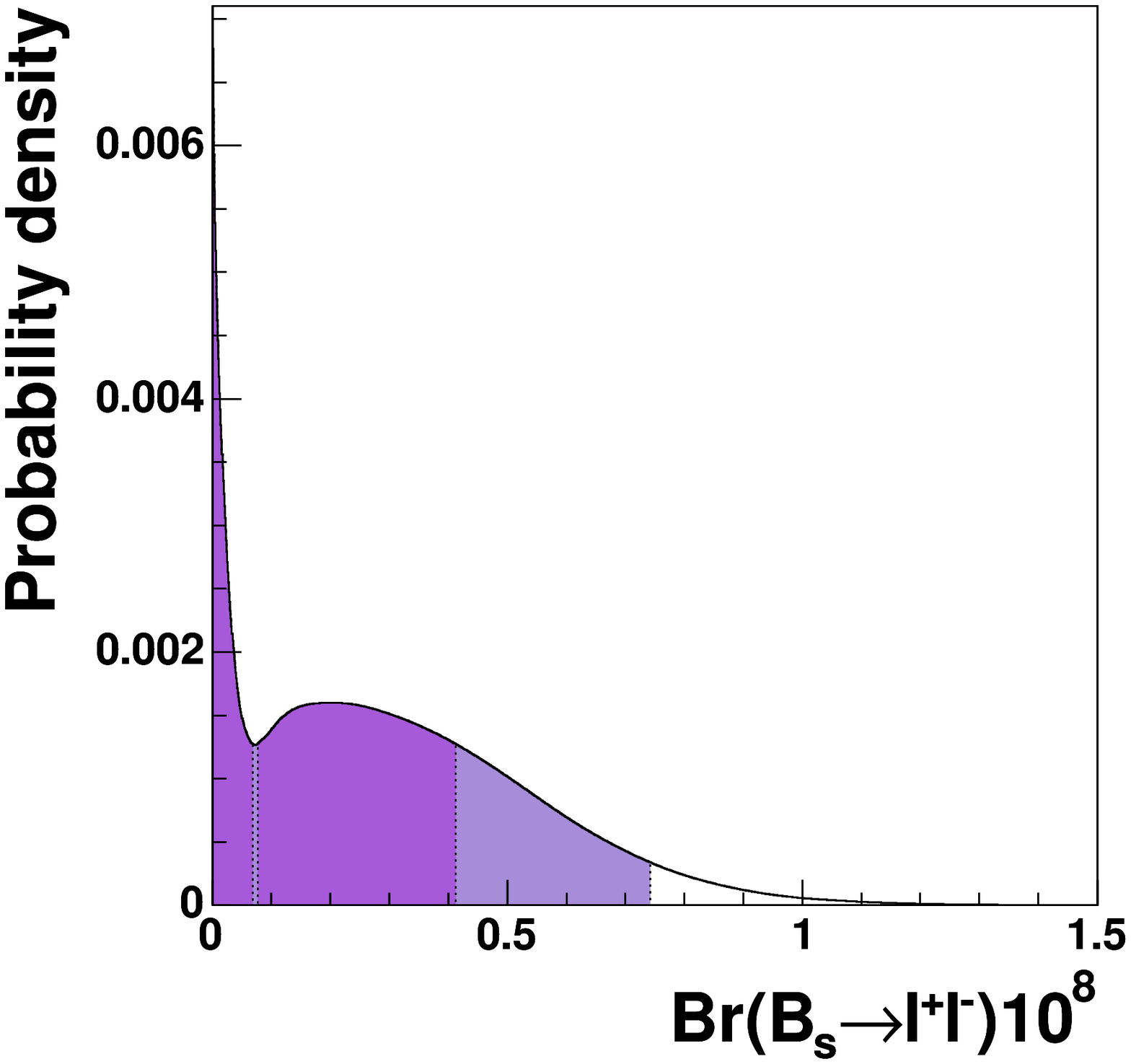}
\caption{%
\it P.d.f.'s for the branching ratios of the rare decays  
$Br(\kpn)$, $Br(\klpn)$, $Br(\kmm)_{\rm SD}$,   
$Br(B\to X_{d,s}\nu\bar\nu)$, and $Br(B_{d,s}\to \mu^+\mu^-)$. Dark (light) areas correspond to the
$68\%$ ($95\%$) probability region.}
\label{fig:BRs}
\end{center}
\end{figure}

\begin{figure}[htb!]
\begin{center}
\includegraphics*[width=0.48\textwidth]{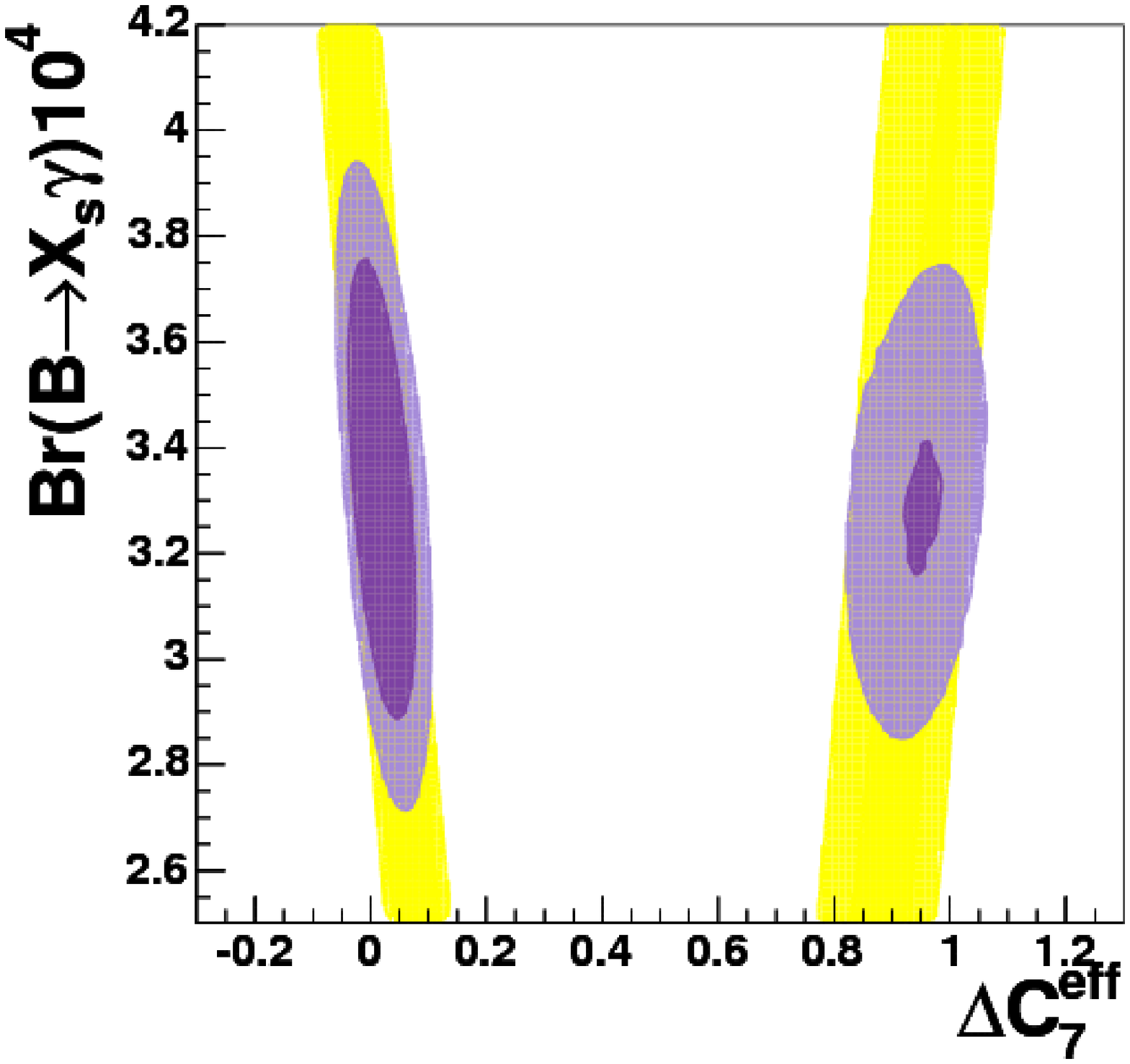}
\includegraphics*[width=0.48\textwidth]{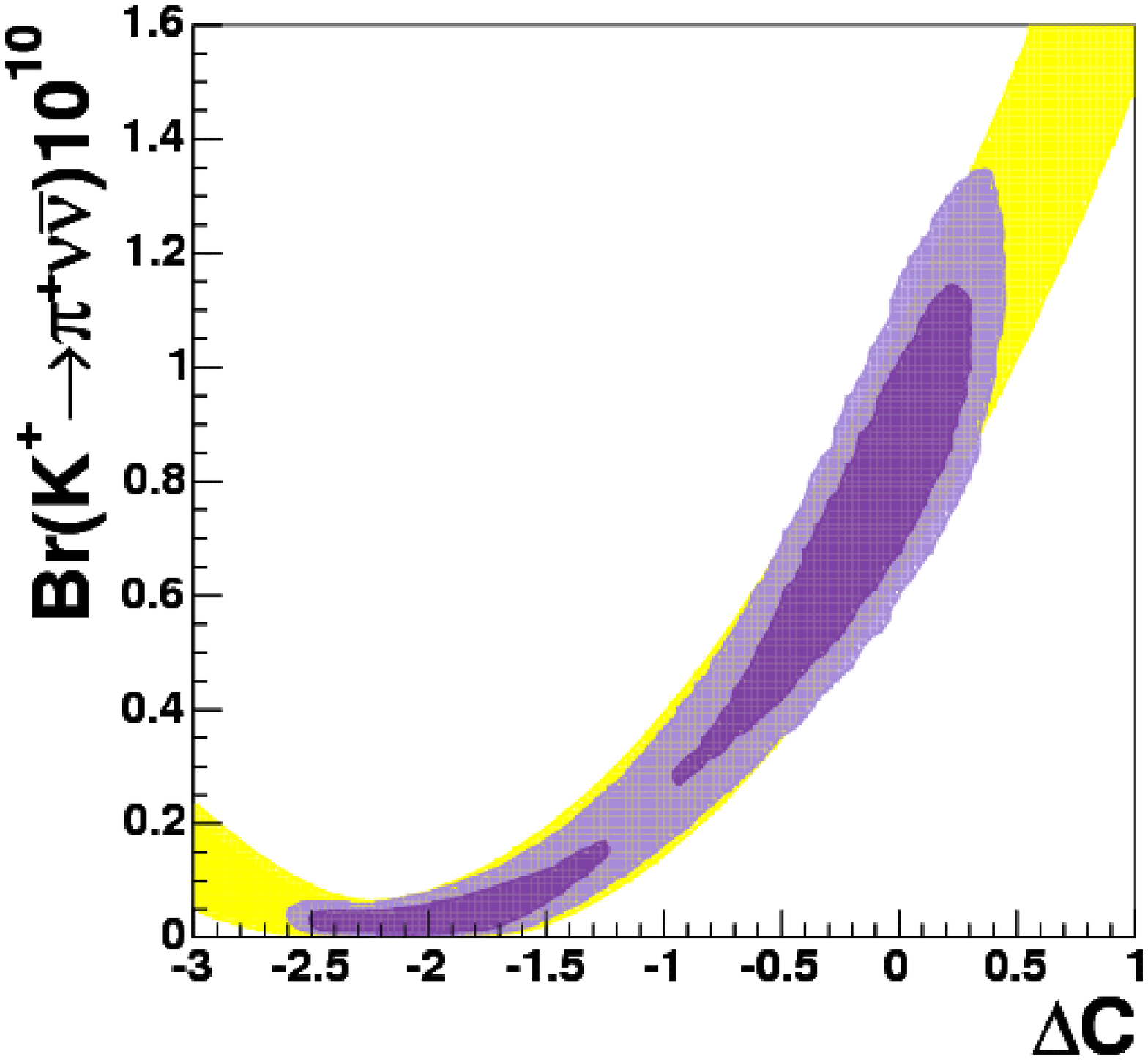}
\includegraphics*[width=0.32\textwidth]{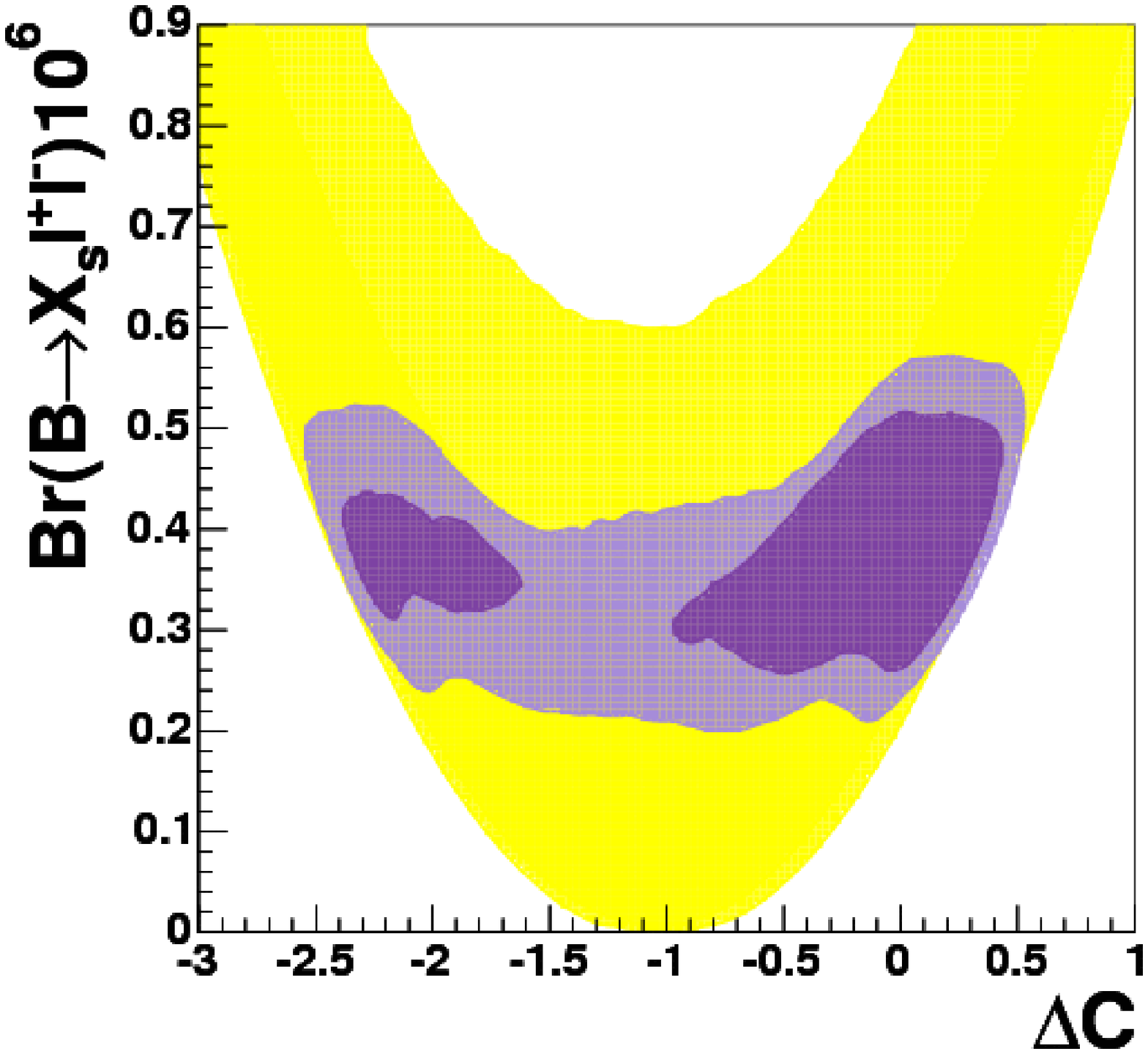}
\includegraphics*[width=0.32\textwidth]{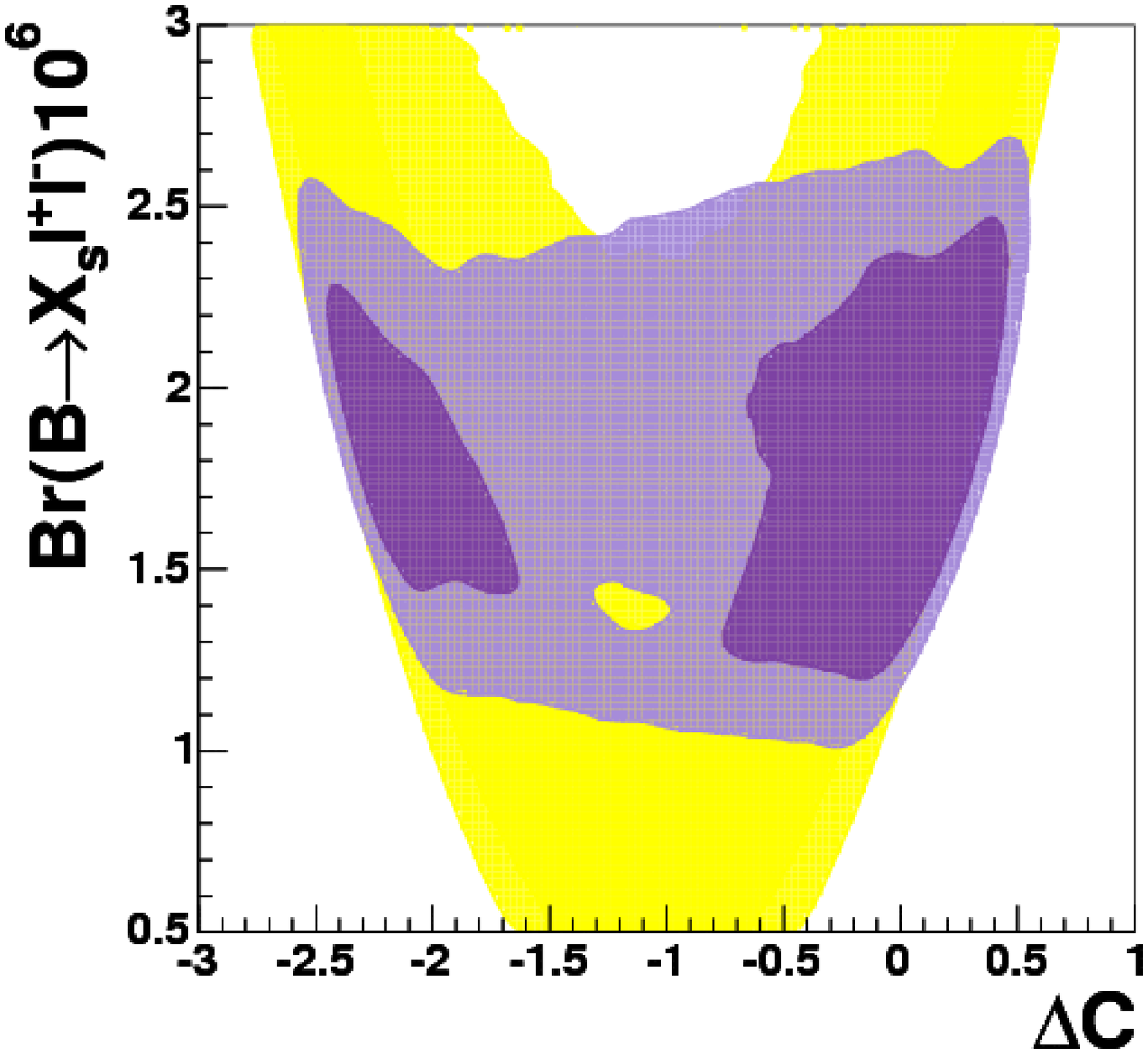}
\includegraphics*[width=0.32\textwidth]{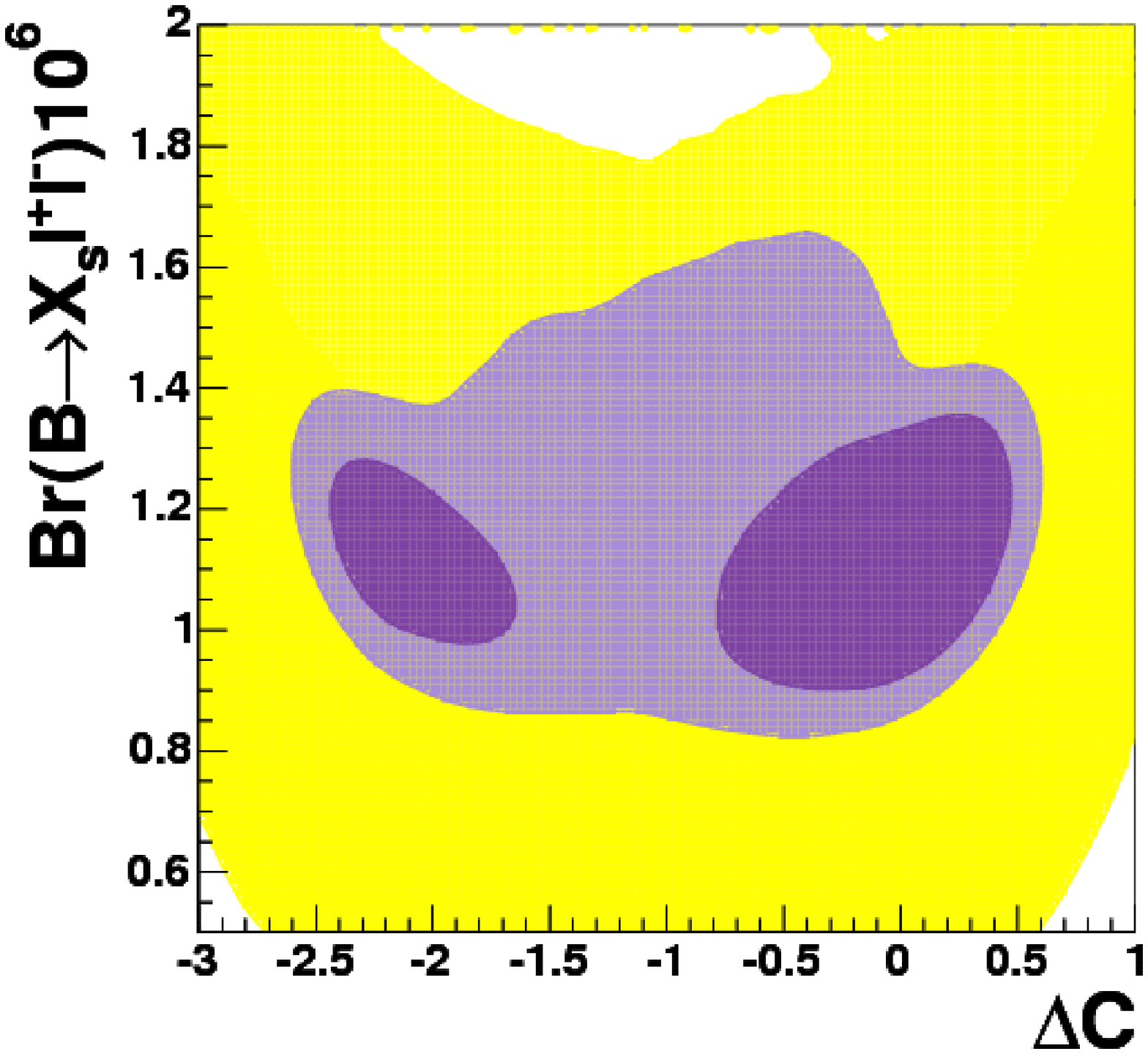}
\caption{%
  \it P.d.f.'s for the branching ratios of the rare decays used to
  constrain $\Delta C$ and $\Delta C_7^\mathrm{eff}$ as a function of
  these parameters: $Br(\BXsgamma)_{E_\gamma > 1.8 \mathrm{GeV}}$
  (top-left), $Br(\kpn)$ (top-right), $Br(\BXsll)_{14.4 < q^2
    \mathrm{(GeV)} < 25}$ (bottom-left), $Br(\BXsll)_{1 < q^2
    \mathrm{(GeV)} < 6}$ (bottom-center), $Br(\BXsll)_{0.04 < q^2
    \mathrm{(GeV)} < 1}$ (bottom-right). Dark (light) areas correspond
  to the $68\%$ ($95\%$) probability region. Very light areas
  correspond to the range obtained without using the experimental
  information.}
\label{fig:constraints}
\end{center}
\end{figure}

\begin{figure}[htb!]
\begin{center}
\includegraphics*[width=0.32\textwidth]{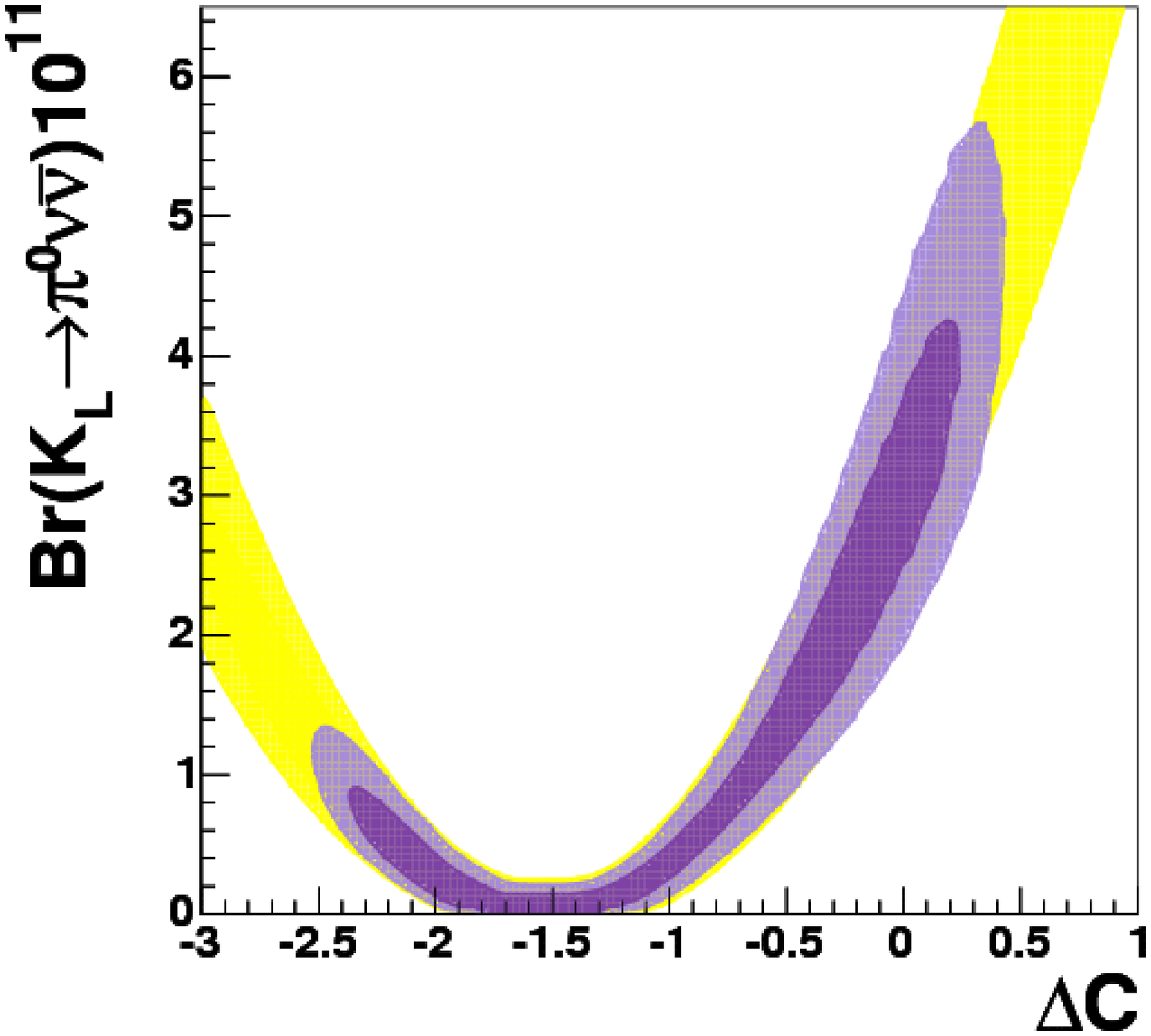}
\includegraphics*[width=0.32\textwidth]{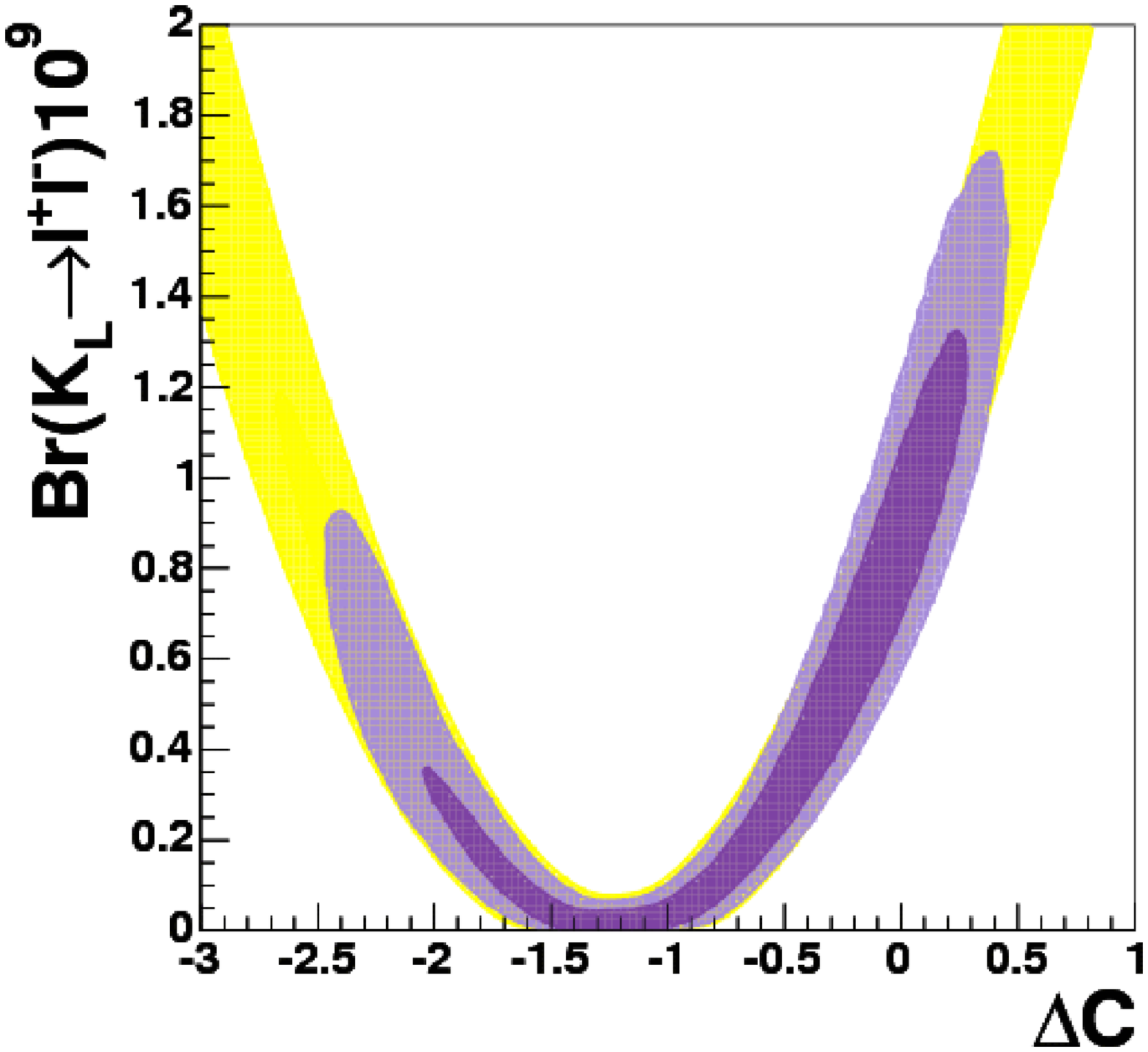}
\includegraphics*[width=0.32\textwidth]{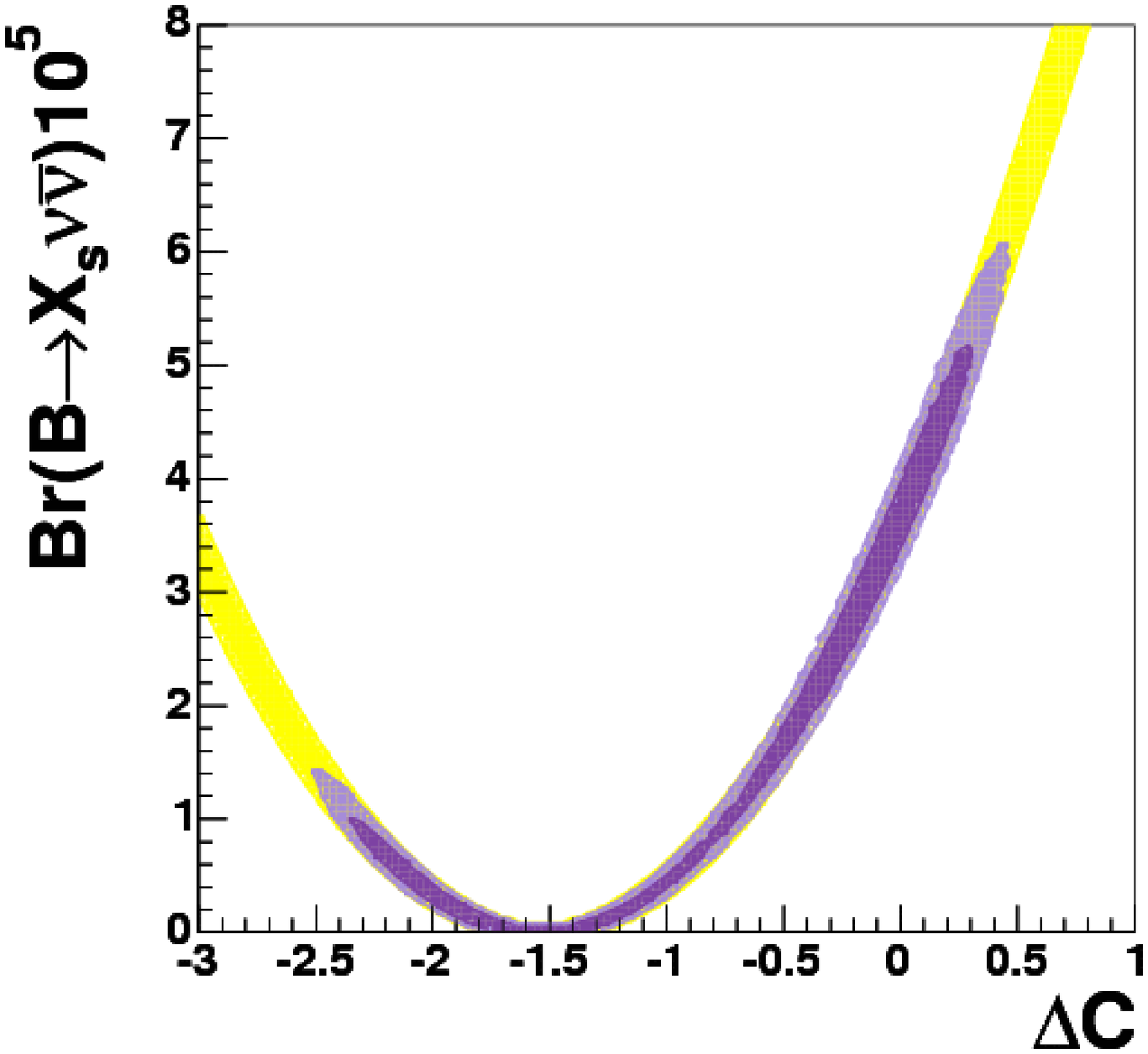}
\includegraphics*[width=0.32\textwidth]{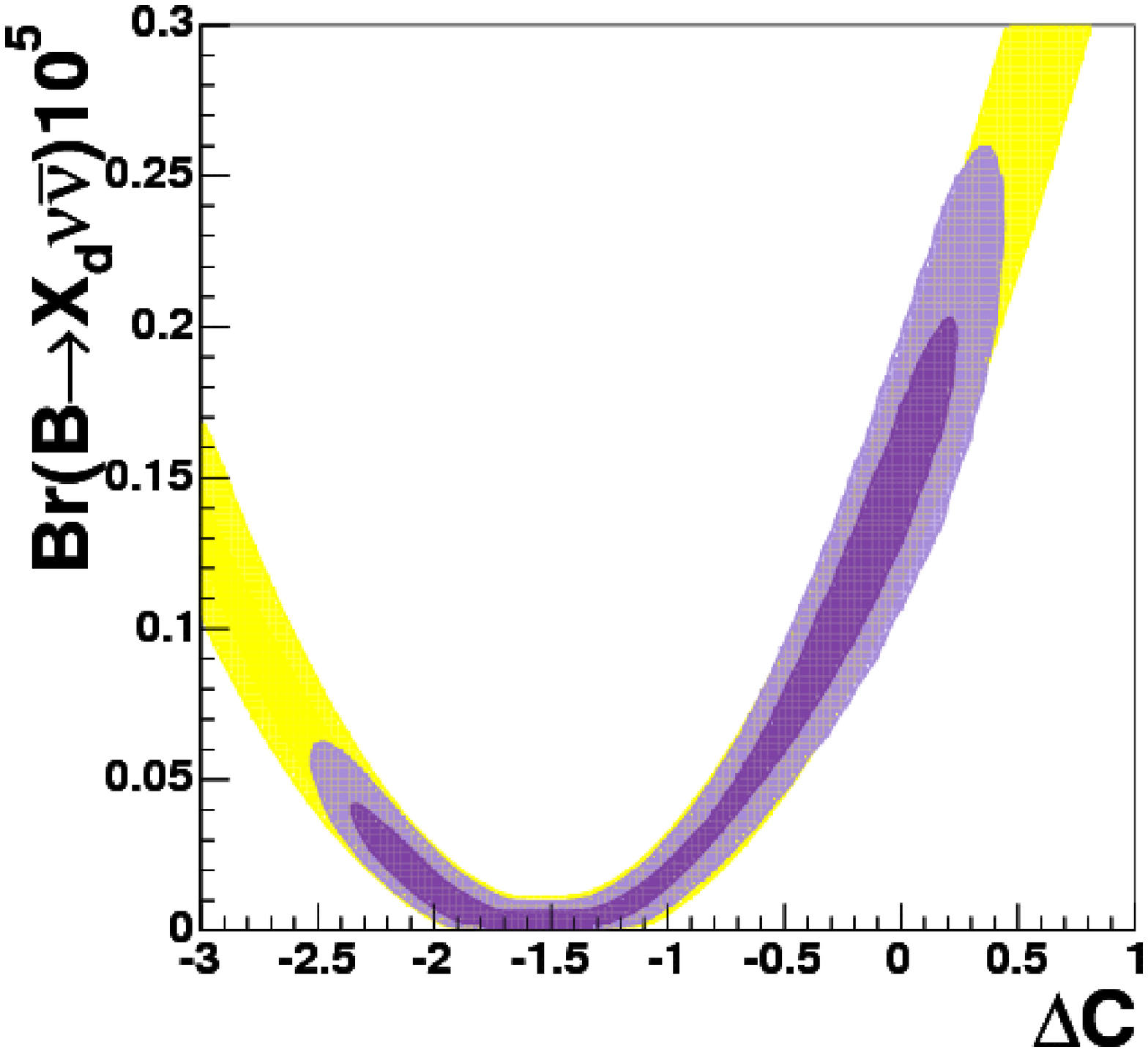}
\includegraphics*[width=0.32\textwidth]{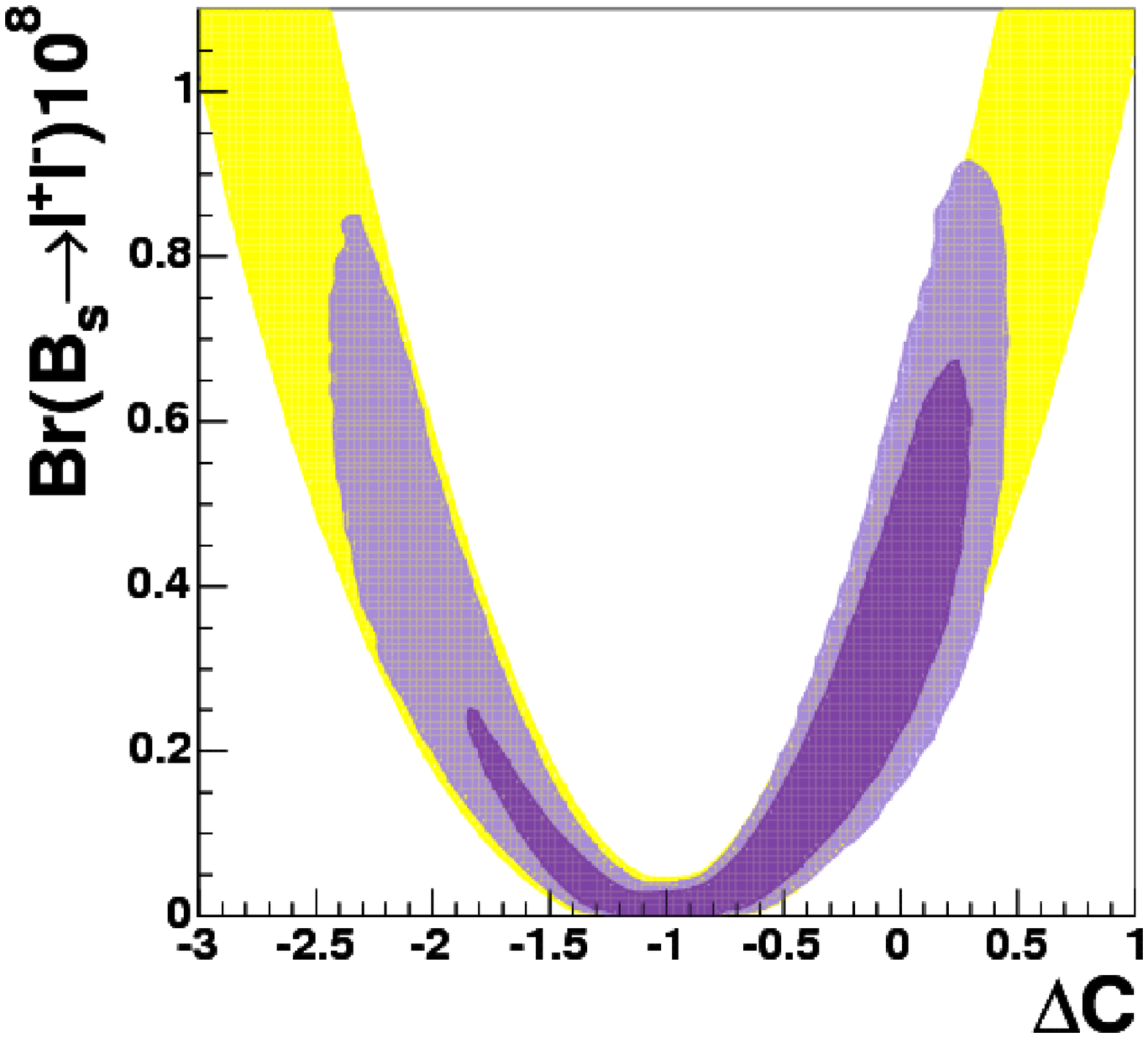}
\includegraphics*[width=0.32\textwidth]{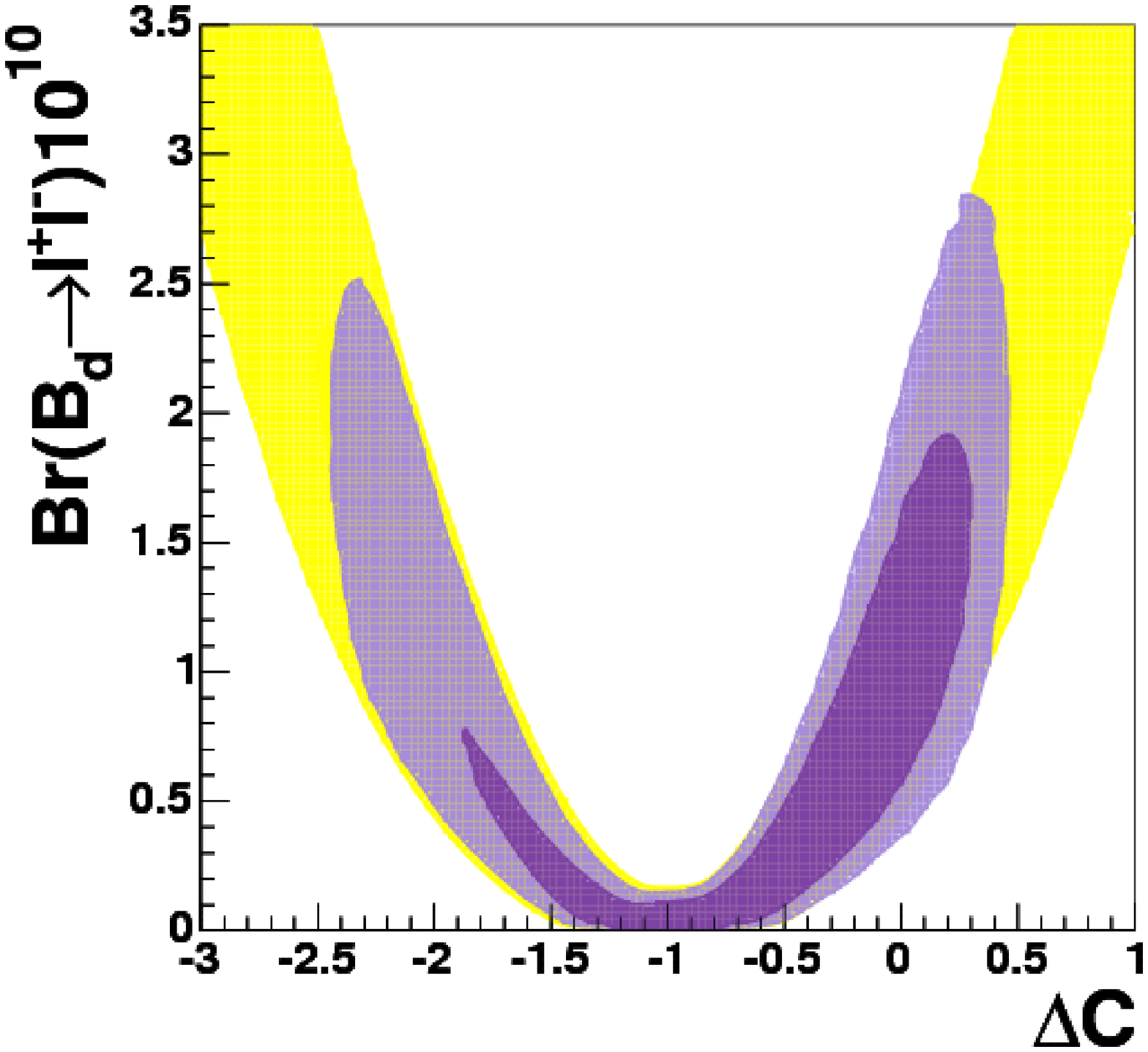}
\caption{%
  \it P.d.f.'s for the branching ratios of the rare decays
  $Br(\klpn)$, $Br(\kmm)_{\rm SD}$, $Br(B\to X_{d,s}\nu\bar\nu)$, and
  $Br(B_{d,s}\to \mu^+\mu^-)$ as a function of $\Delta C$.  Dark
  (light) areas correspond to the $68\%$ ($95\%$) probability region.
  Very light areas correspond to the range obtained without using the
  experimental information.}
\label{fig:raredec}
\end{center}
\end{figure}

\begin{table*}[t]
\small{
\begin{center}
\begin{tabular}{|c|c|c|c|c|}
\hline
{Branching Ratios} &  MFV (95\%) &  SM (68\%) &  SM (95\%) & exp
 \\ \hline
$Br(\kpn)\times 10^{11}$ & $< 11.9$ & $8.3 \pm 1.2$ &  $[6.1,10.9]$
& $(14.7^{+13.0}_{-8.9})$ \cite{E949}
\\ \hline
$Br(\klpn)\times 10^{11}$  & $< 4.59$ &  $3.08 \pm 0.56$ &  $[2.03,4.26]$ &
 $ < 5.9 \cdot10^{4}$  \cite{Harati:1999hd}
\\ \hline
$Br(\kmm)_{\rm SD}\times 10^{9} $ & $< 1.36$ & $0.87 \pm 0.13$ &  $[0.63,1.15]$ & -
\\ \hline
$Br(B\to X_s\nu\bar\nu)\times 10^{5}$ & $<5.17$ &  $3.66 \pm 0.21$ &  $[3.25,4.09]$
&  $<64 $ 
\cite{Barate:2000rc}
\\ \hline
$Br(B\to X_d\nu\bar\nu)\times 10^{6}$ &  $<2.17$ & $1.50 \pm 0.19$ &  $[1.12,1.91]$
& -
\\ \hline
$Br(B_s\to \mu^+\mu^-)\times 10^{9}$ &  $< 7.42$ & $3.67 \pm 1.01$ &  $[1.91,5.91]$
& $<2.7\cdot 10^{2}$  \cite{Herndon:2004tk}
\\ \hline
$Br(B_d\to \mu^+\mu^-)\times 10^{10}$ &  $< 2.20$ & $1.04 \pm 0.34$ &  $[0.47,1.81]$
& $<1.5 \cdot 10^3$ \cite{Herndon:2004tk}
\\ \hline
\end{tabular}
\caption[]{\it Upper bounds for rare decays in MFV models at $95 \%$
  probability, the corresponding values in the SM (using inputs from
  the UUT analysis) and the available experimental information. See
  the text for details.  }
\label{brMFV}
\end{center}
}
\end{table*}

The second and third steps are carried out using the approach of
ref.~\cite{utfitmethod}: taking $C(v)$, $C_7^\mathrm{eff}(\mu_b)$ and
$D(v)$ to have a flat \textit{a-priori} distribution and using the
available experimental data and theoretical inputs, we determine the
\textit{a-posteriori} probability density function (p.d.f.) for
$C(v)$, $C_7^\mathrm{eff}(\mu_b)$ and all the rare decays listed in
Table \ref{brMFV}.  Concerning $D(v)$, it plays only a marginal role
in these decays and therefore it is not well determined by the
analysis. We varied $\Delta D$ in the conservative range $\pm 4
  D_{\rm SM}$. Even this rather large variation has little impact on
  the extraction of the allowed range for $C (v)$.

In Figure \ref{fig:Cs} we plot the p.d.f. for $\Delta C(v)$ and
$\Delta C_7^\mathrm{eff}$, that represent $F^i_{\mathrm{New}}$
in (\ref{master}) and enter eq.~(\ref{eq:numeric}).  In Figure
\ref{fig:BRs} we plot the p.d.f. for the branching ratios.  The
corresponding upper bounds at $95 \%$ probability are reported in
Table \ref{brMFV}, where, for comparison, we also report the results
obtained within the SM, using the same CKM parameters obtained from
the UUT analysis.  Finally, in Figures \ref{fig:constraints} and
\ref{fig:raredec} we plot the branching ratios of the rare decays vs.
$C(v)$, to make the impact of future measurements on the determination
of $C(v)$ more transparent.

Let us now comment on our results. As can be seen from Figure
\ref{fig:Cs}, we have two possible solutions for $\Delta
C_7^\mathrm{eff}$, one very close to the SM, and the other
corresponding to reversing the sign of $C_7^\mathrm{eff}(\mu_b)$
(recall that $C_7^\mathrm{eff}(\mu_b)$ is negative in the SM and equal
to $C_7^\mathrm{eff}(\mu_b)\approx -0.33$).  The second solution is
disfavoured: it is barely accessible at $68 \%$ probability, in
accordance with the results of \cite{hep-ph/0410155}.  This result is
easy to understand. In the case of the second solution for $\Delta
C_7^{\rm eff}$, the branching ratio $Br(\BXsll)$ becomes larger than
the experimental value.  The full results are:
\begin{eqnarray}
\Delta C_7^\mathrm{eff} &=& (0.02 \pm 0.047) \cup (0.958 \pm 0.002) 
\mathrm{~at~68\%~probability,} 
\nonumber \\
\Delta C_7^\mathrm{eff} &=&  [-0.039,0.08] \cup [0.859,1.031] \mathrm{~at~95\%~probability.}
\label{eq:C7effrange}
\end{eqnarray} 

\begin{figure}[htb!]
\begin{center}
\includegraphics*[width=0.32\textwidth]{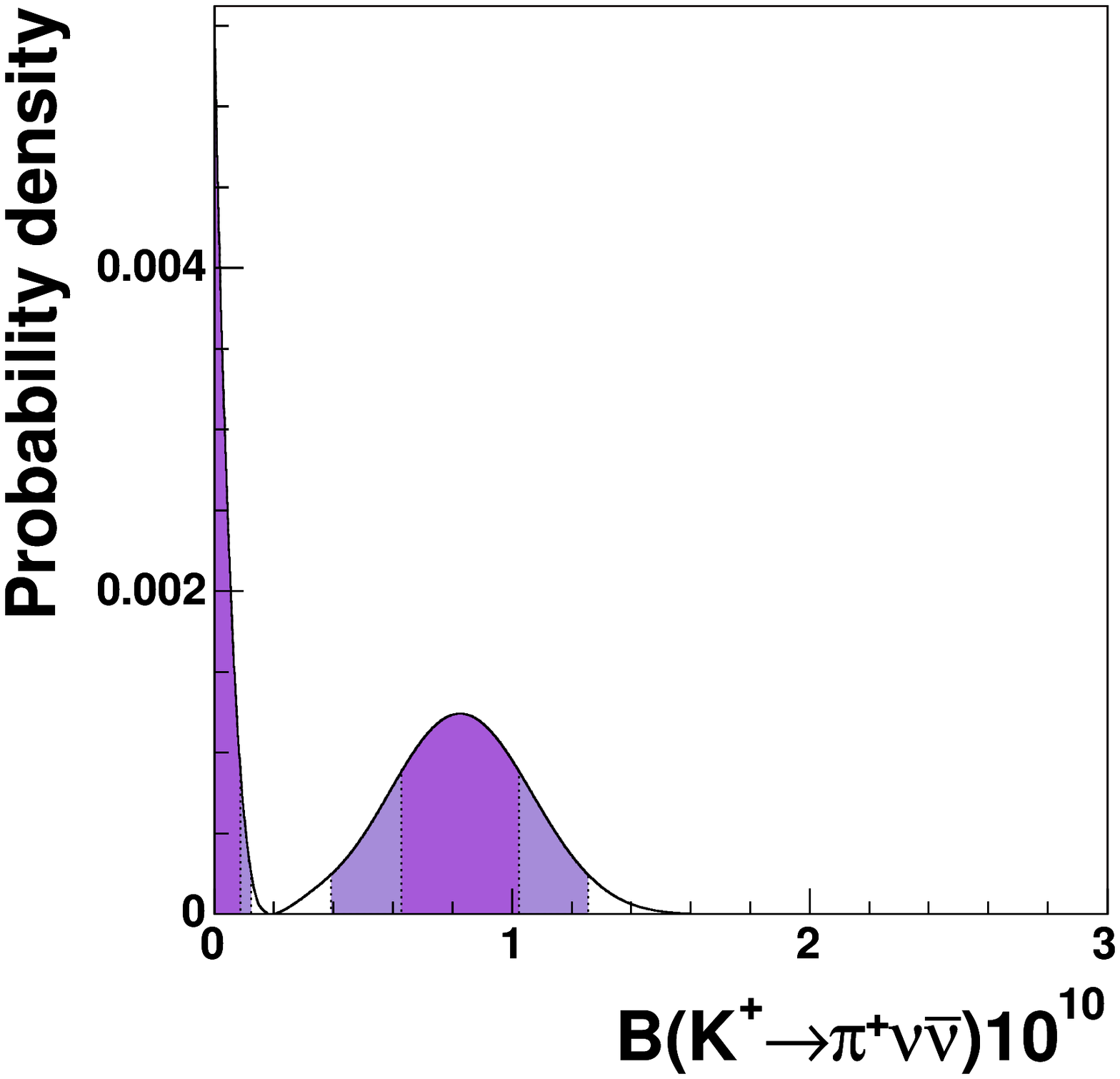}
\includegraphics*[width=0.32\textwidth]{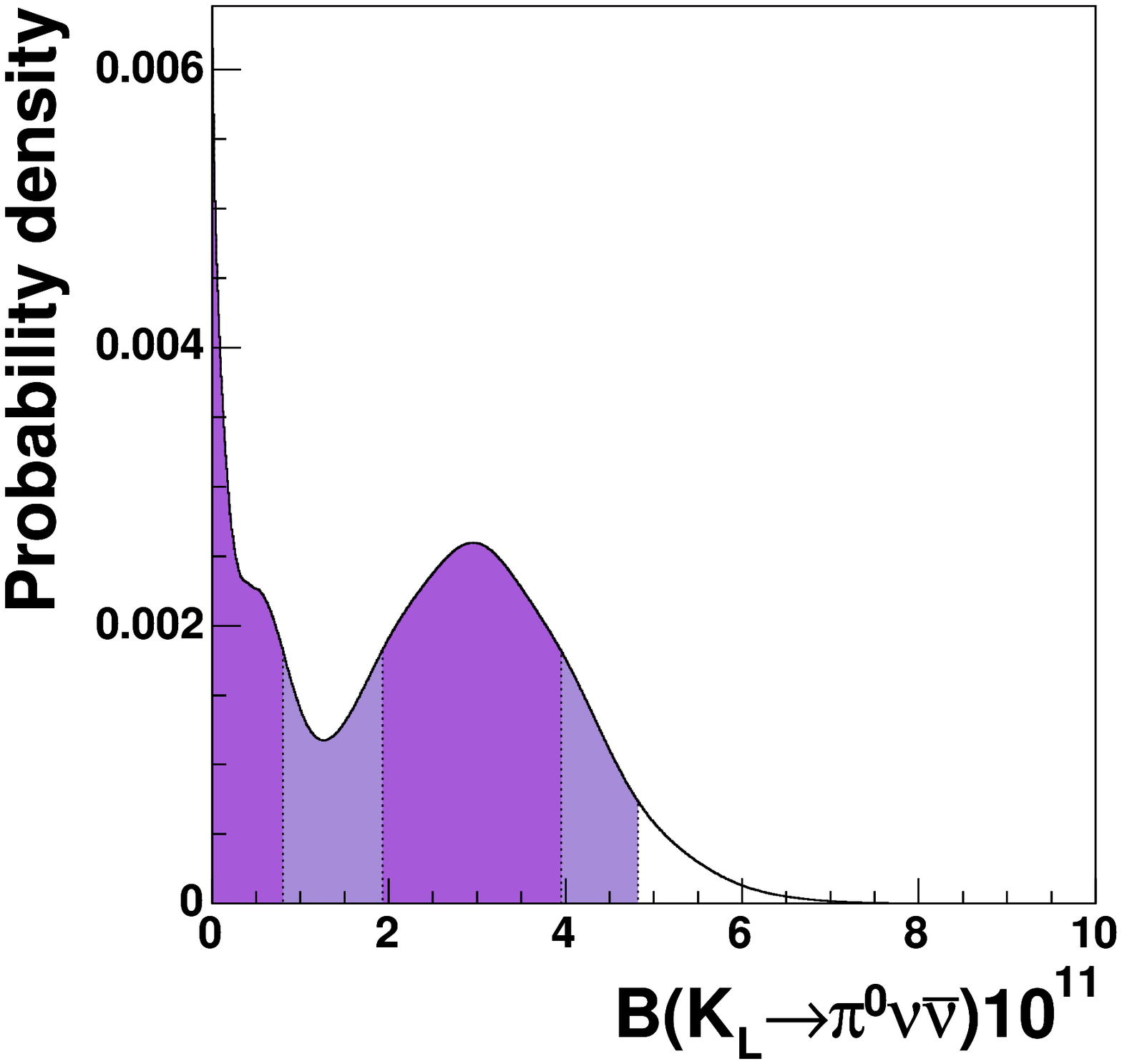}
\includegraphics*[width=0.32\textwidth]{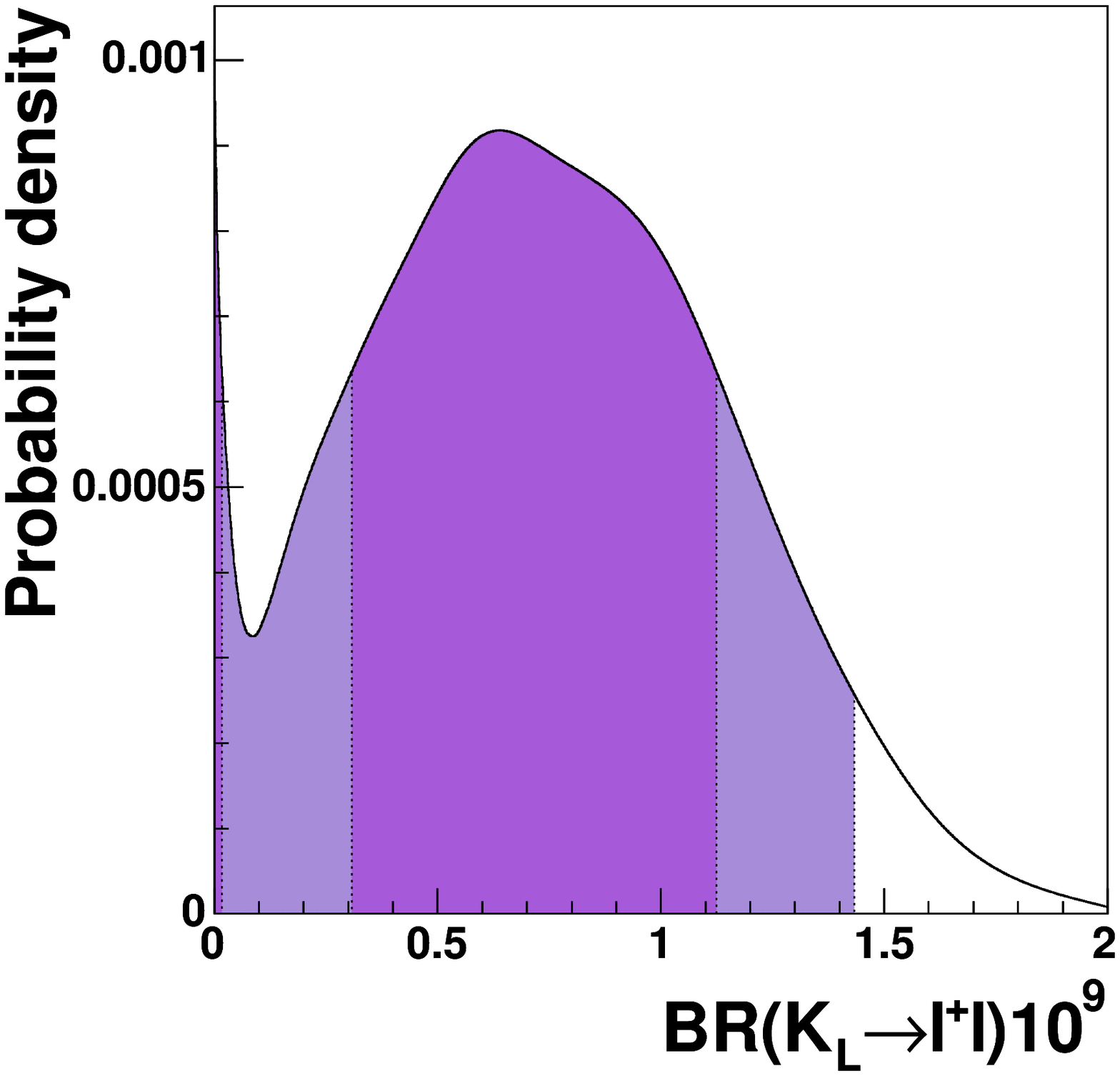}
\includegraphics*[width=0.32\textwidth]{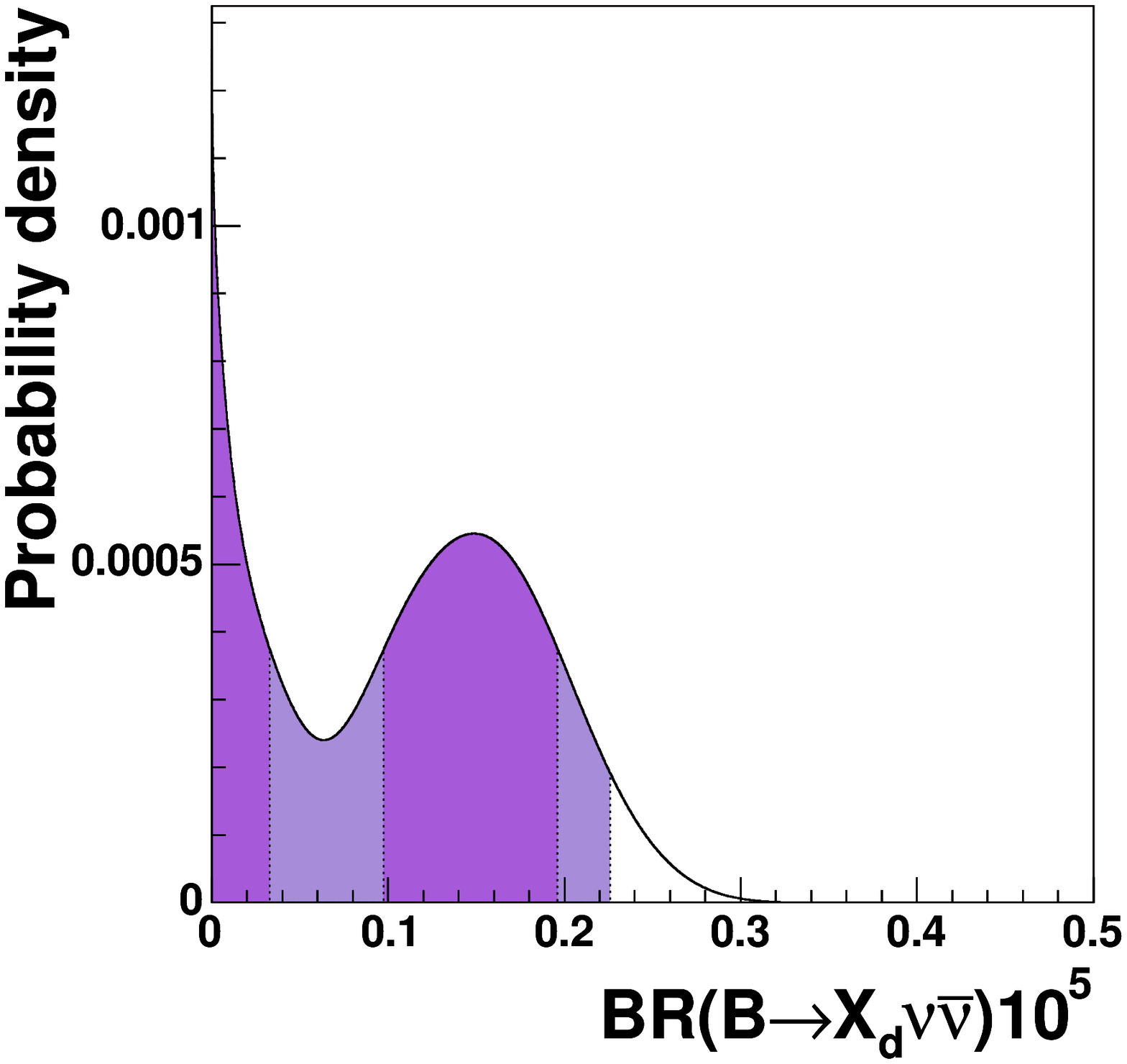}
\includegraphics*[width=0.32\textwidth]{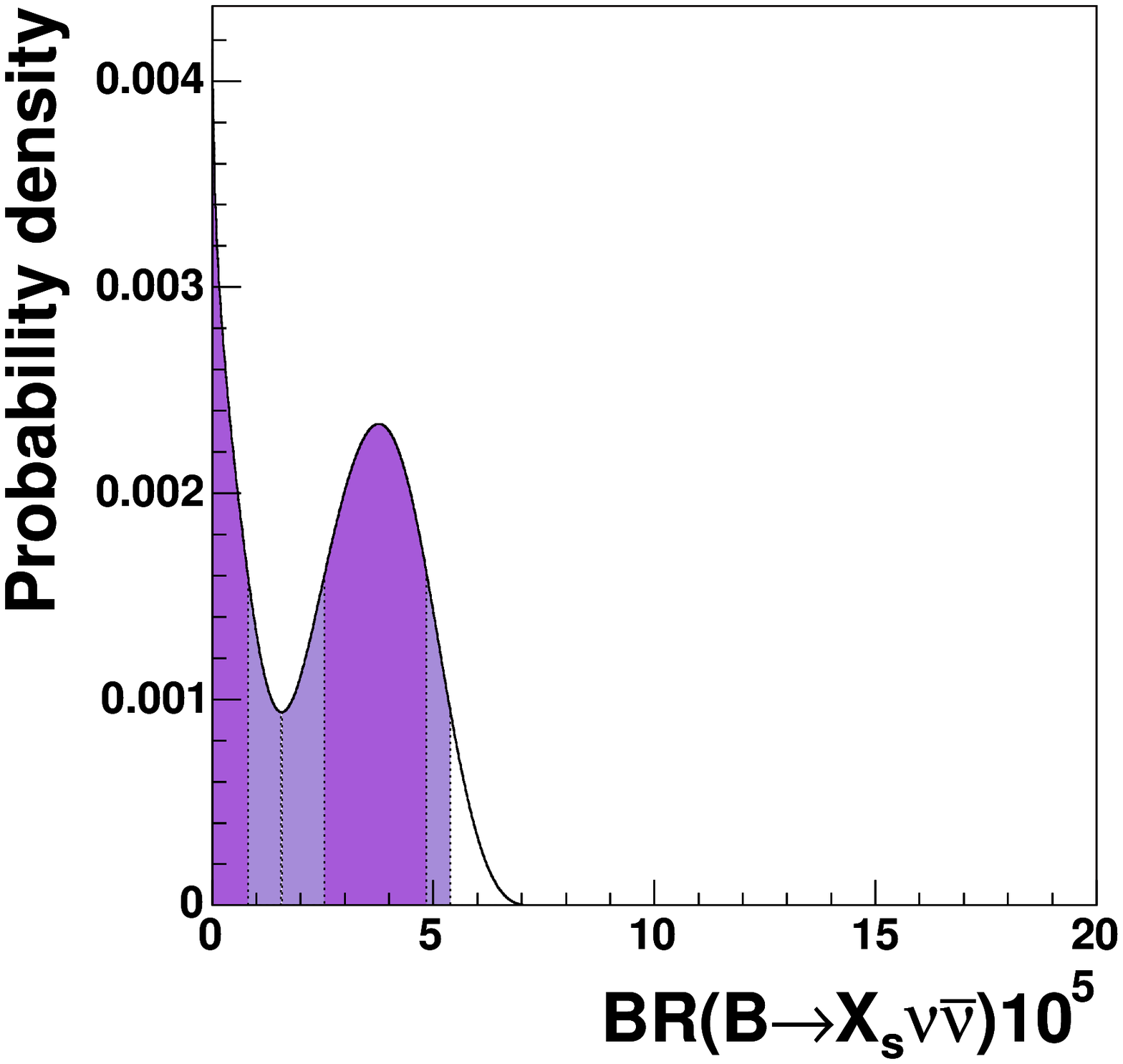}
\includegraphics*[width=0.32\textwidth]{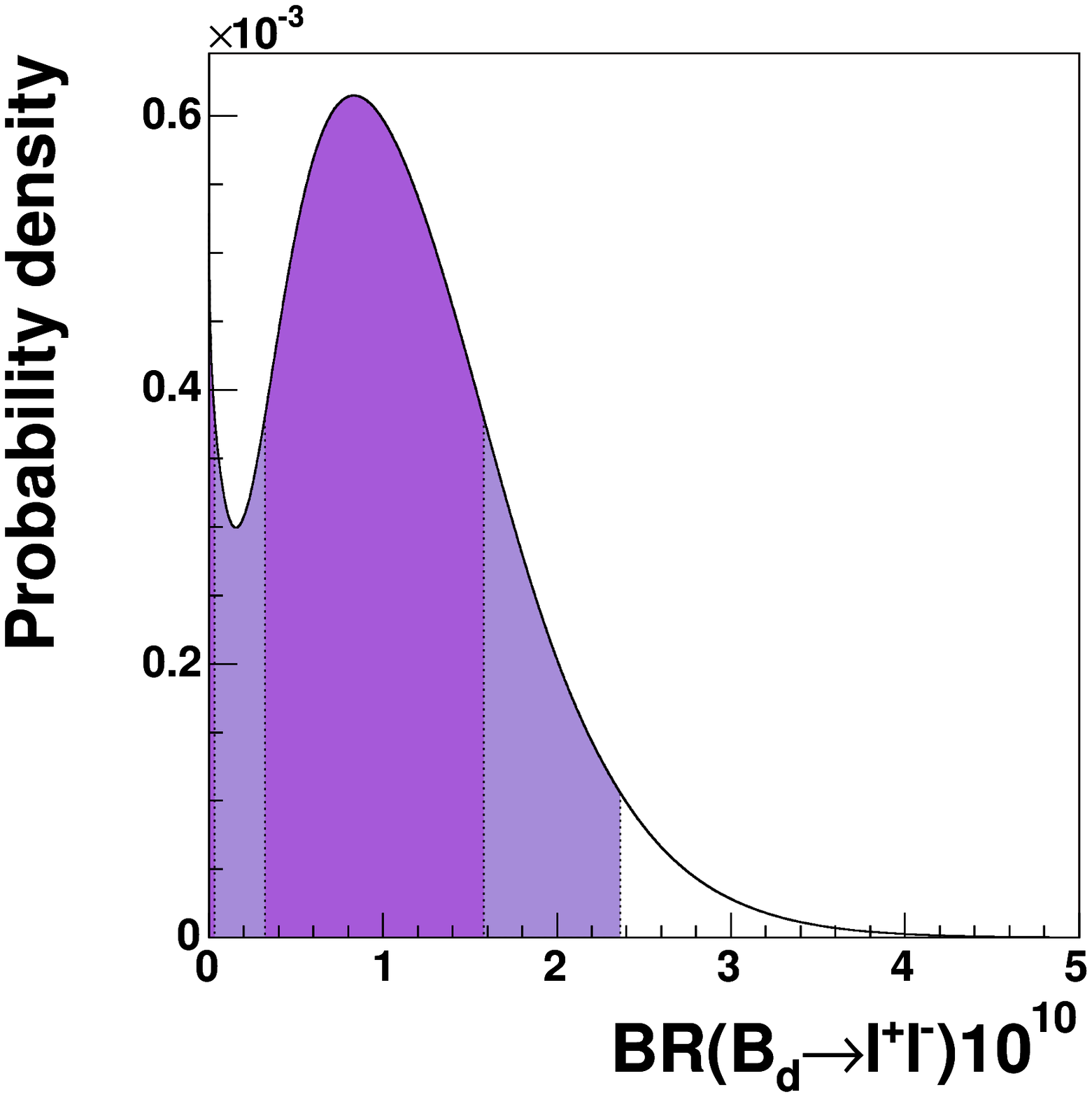}
\includegraphics*[width=0.32\textwidth]{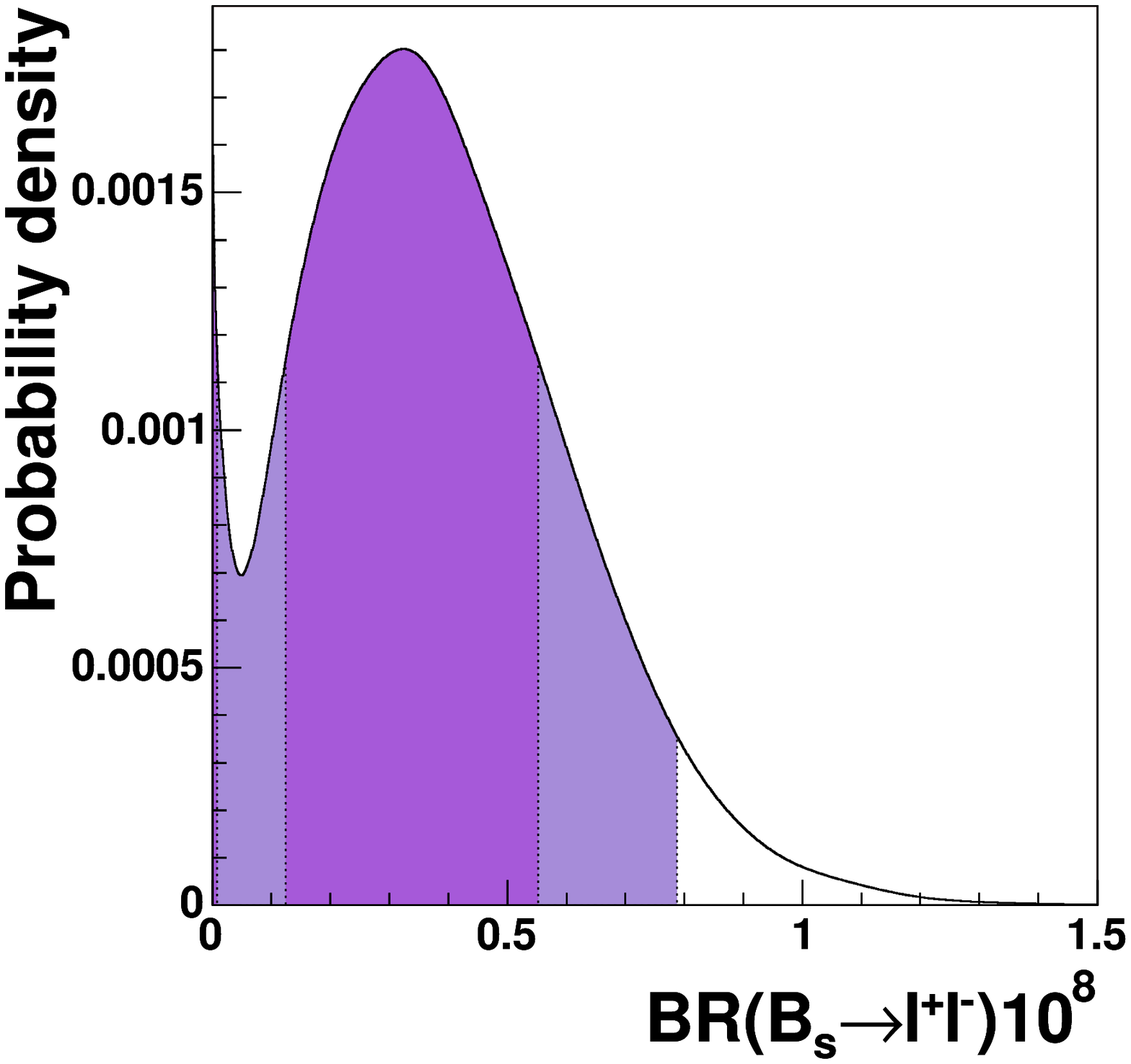}
\caption{%
\it P.d.f.'s for the branching ratios of the rare decays  
$Br(\kpn)$, $Br(\klpn)$, $Br(\kmm)_{\rm SD}$,  
$Br(B\to X_{d,s}\nu\bar\nu)$, and $Br(B_{d,s}\to \mu^+\mu^-)$  
considering only the LOW solution for $\Delta C_7^\mathrm{eff}$. 
Dark (light) areas correspond to the
$68\%$ ($95\%$) probability region.}
\label{fig:raredeclC7}
\end{center}
\end{figure}

\begin{figure}[htb!]
\begin{center}
\includegraphics*[width=0.32\textwidth]{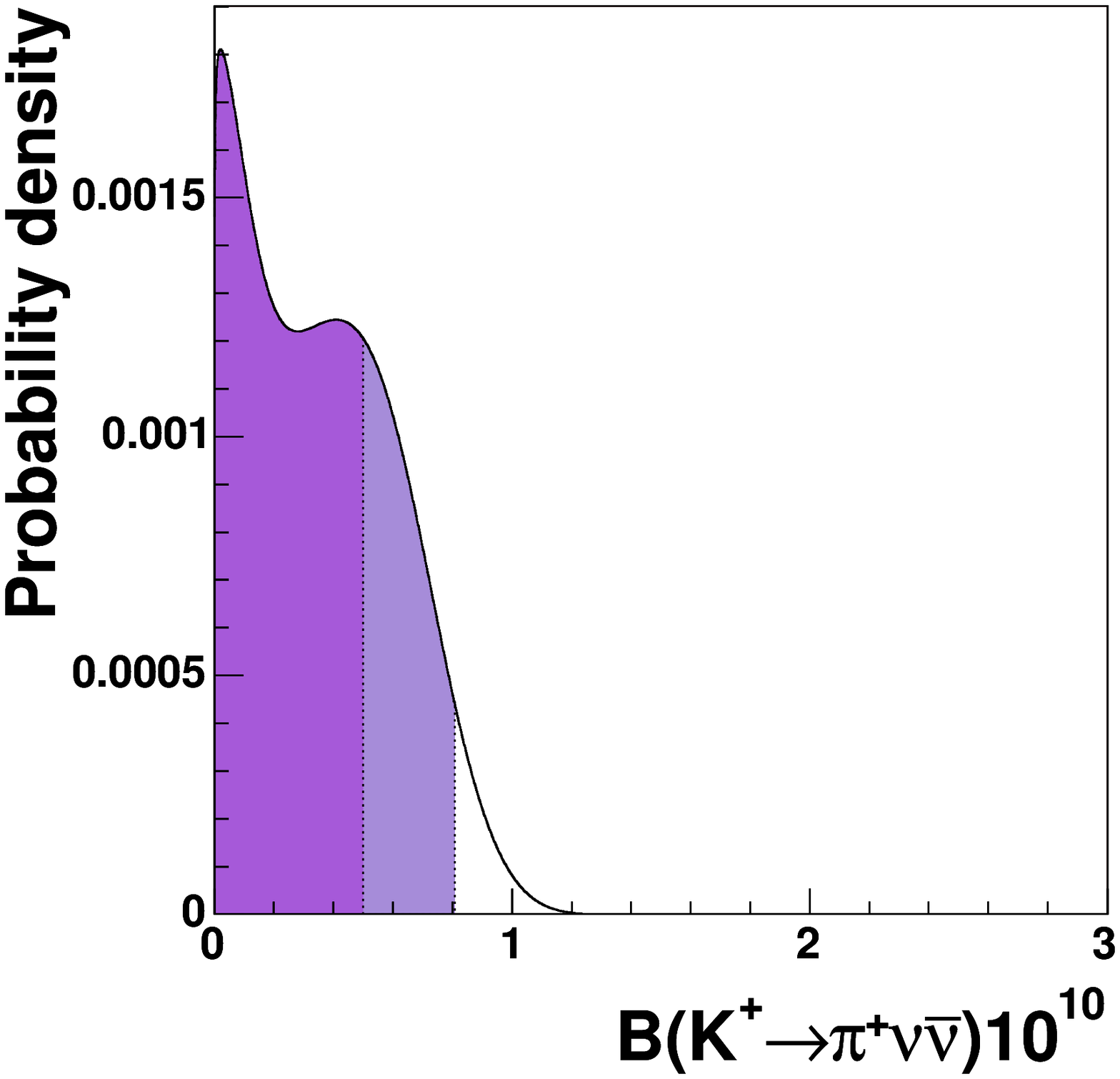}
\includegraphics*[width=0.32\textwidth]{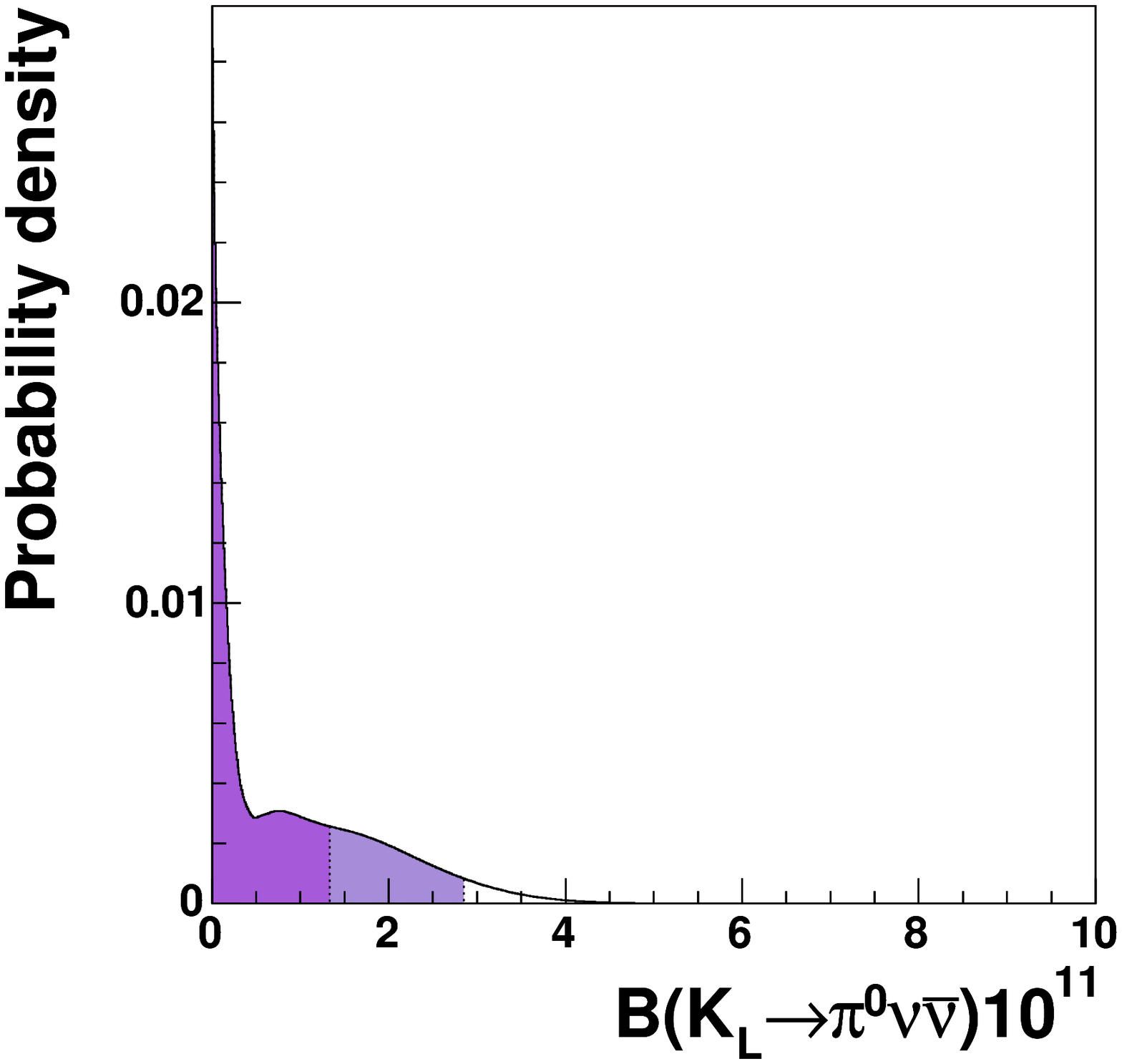}
\includegraphics*[width=0.32\textwidth]{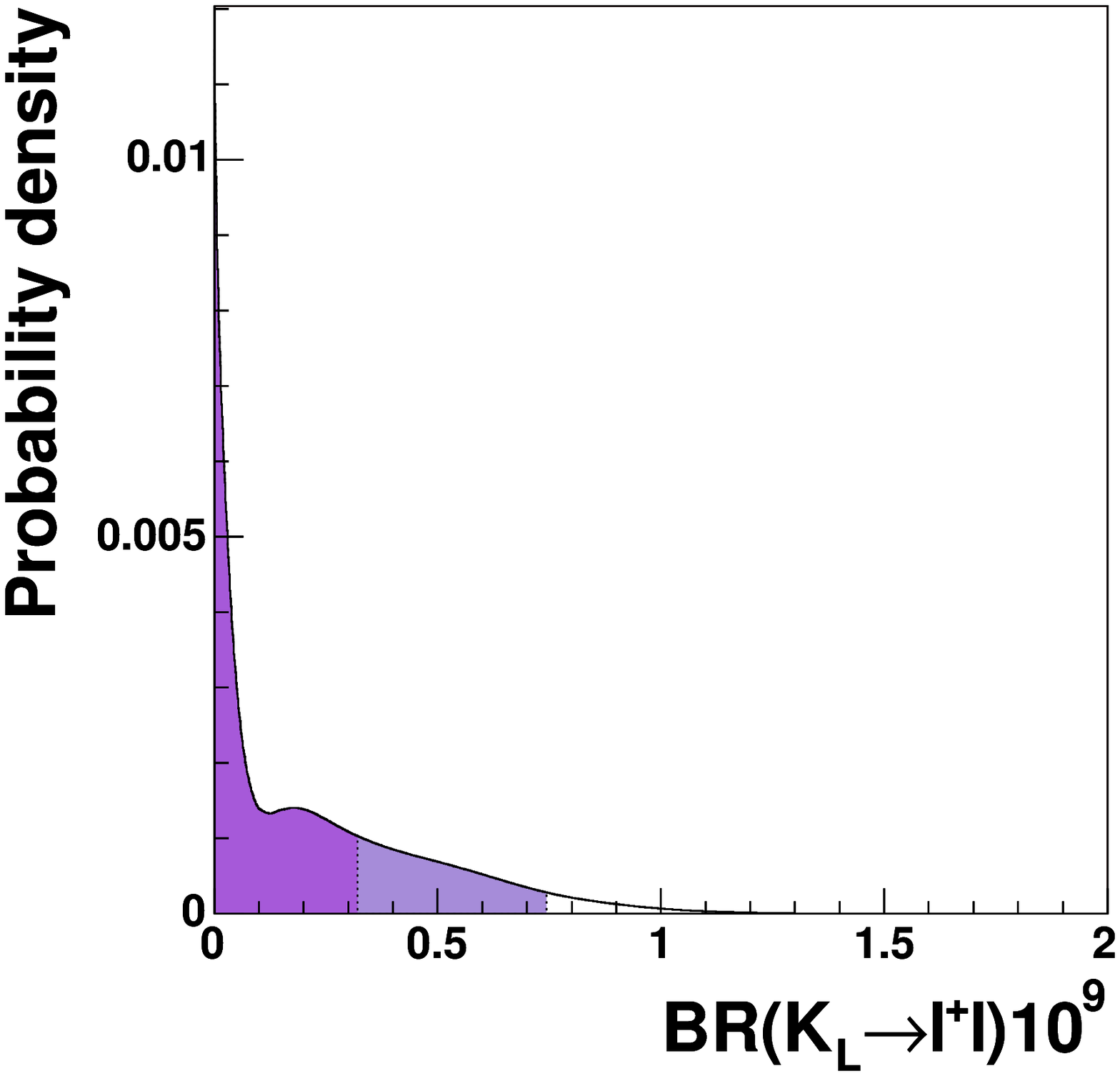}
\includegraphics*[width=0.32\textwidth]{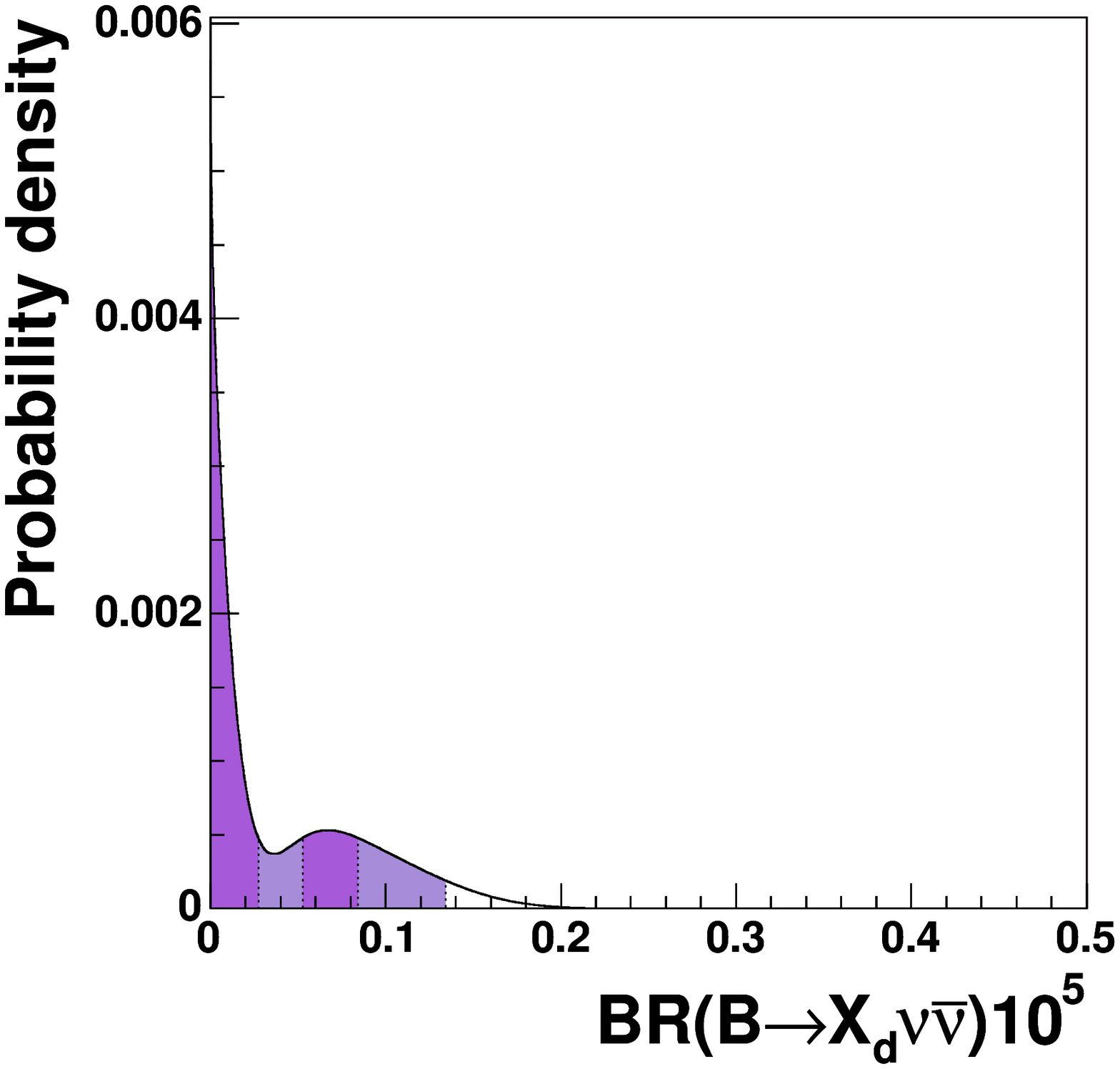}
\includegraphics*[width=0.32\textwidth]{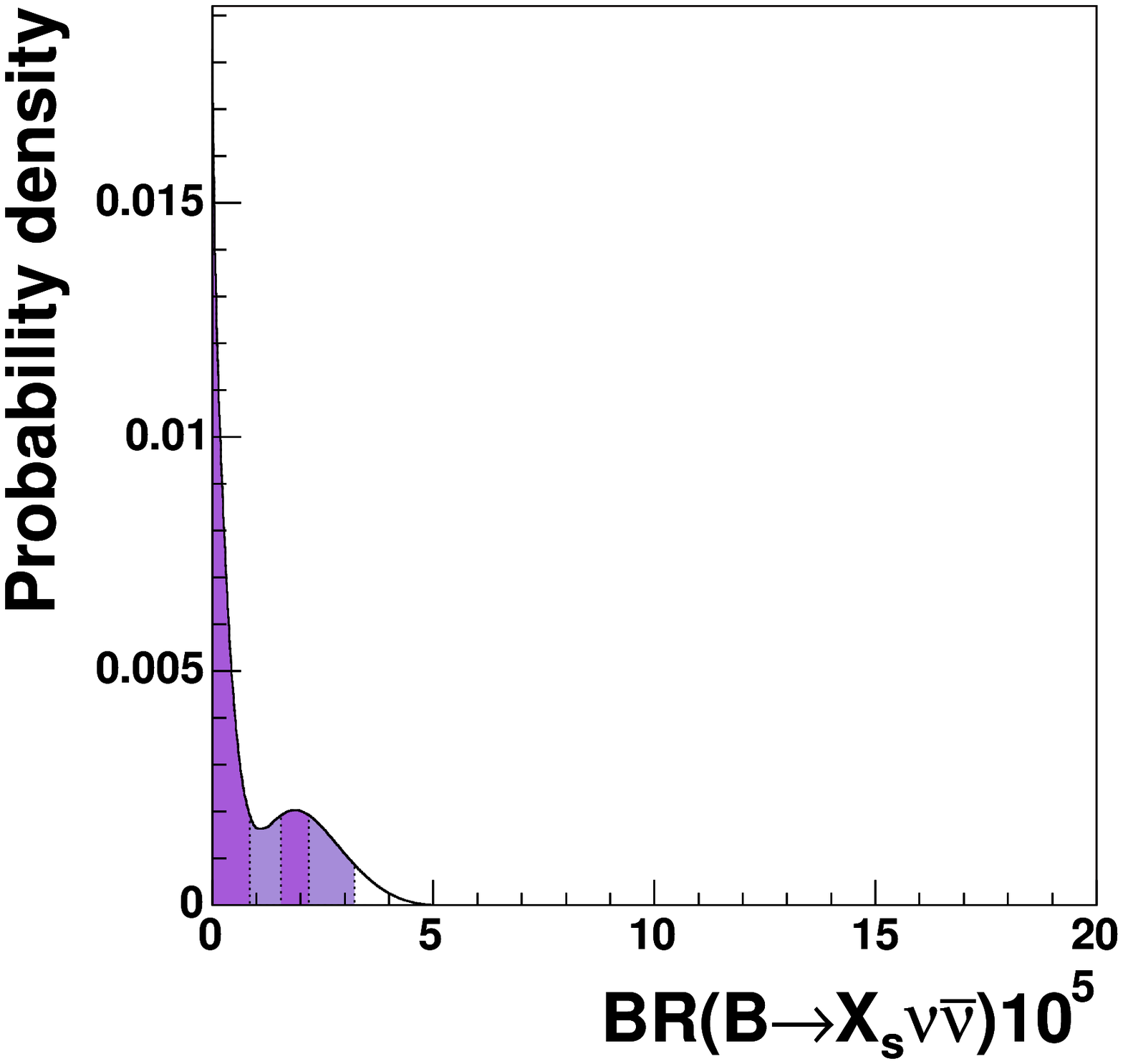}
\includegraphics*[width=0.32\textwidth]{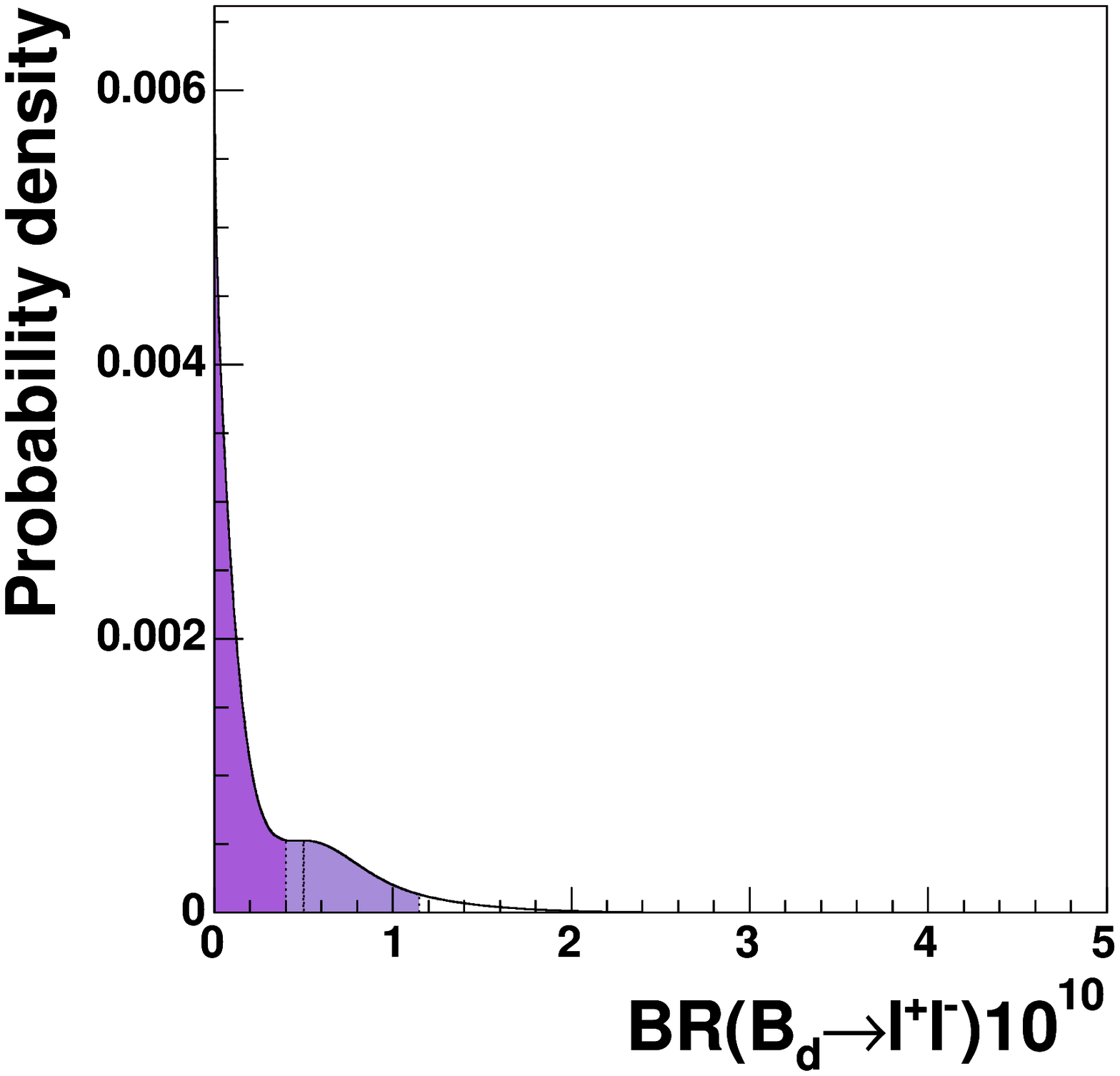}
\includegraphics*[width=0.32\textwidth]{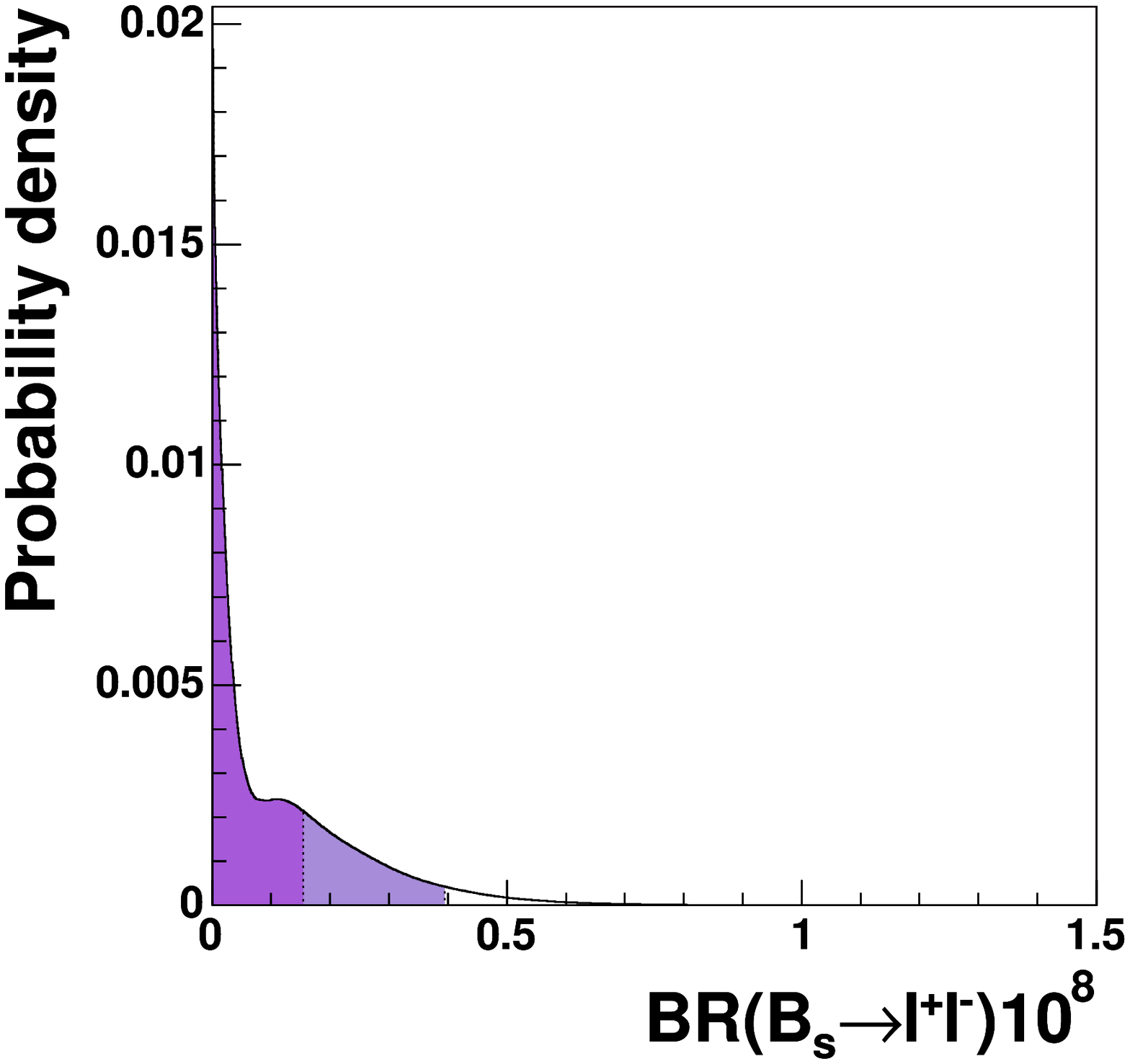}
\caption{%
\it P.d.f.'s for the branching ratios of the rare decays  
$Br(\kpn)$, $Br(\klpn)$, $Br(\kmm)_{\rm SD}$,  
$Br(B\to X_{d,s}\nu\bar\nu)$, and $Br(B_{d,s}\to \mu^+\mu^-)$  
considering only the HI solution for $\Delta C_7^\mathrm{eff}$. Dark (light) areas correspond to the
$68\%$ ($95\%$) probability region.}
\label{fig:raredechC7}
\end{center}
\end{figure}

Since we have two separate ranges for $\Delta
C_7^\mathrm{eff}$, in the following we will also present
separately the results corresponding to the ``LOW" or ``HI" solution
for $\Delta C_7^\mathrm{eff}$ (see Figures~\ref{fig:raredeclC7}
and \ref{fig:raredechC7}).

As can be seen in Figure~\ref{fig:Cs} we have two solutions for $\Delta
C$, one close to the SM and the other corresponding to 
reversing the sign of $C$. We recall that $C \approx 0.81$ in the SM.
The ranges obtained are
\begin{eqnarray}
  \Delta C &=& (-0.16 \pm 0.53) \cup (-2.15 \pm 0.08) \mathrm{~at~68\%~probability,} 
  \nonumber \\
  \Delta C &=& [-1.25,0.44] \cup  [-2.39,-1.45]\mathrm{~at~95\%~probability.}
\label{eq:Crange}
\end{eqnarray} 
From the plot of $\Delta C$ vs $\Delta C_7^\mathrm{eff}$ in
Figure~\ref{fig:Cs}, it is evident that the situation is different for
the HI and LOW solutions for $\Delta C_7^\mathrm{eff}$. Indeed,
the two solutions correspond to the following ranges for $\Delta C$:
\begin{eqnarray}
  \mathrm{LOW:~}\Delta C &=& (-0.03 \pm 0.41) \cup (-2.18 \pm 0.02) 
  \mathrm{~at~68\%~probability,} 
  \nonumber \\
  \mathrm{LOW:~}\Delta C &=& [-0.75,0.50] \cup [-2.49,-1.60] \mathrm{~at~95\%~probability,}
  \nonumber \\
  \mathrm{HI:~}\Delta C &=& (-0.68 \pm 0.58)
  \mathrm{~at~68\%~probability,} 
  \nonumber \\
  \mathrm{HI:~}\Delta C &=& [-1.98,0.04] \mathrm{~at~95\%~probability.}
\label{eq:CrangeHILOW}
\end{eqnarray}

These results are easy to understand. For the LOW solution the
solutions with $\Delta C$ being positive and negative are consistent
with the data on $B\to X_s l^+l^-$. On the other hand for the HI
solution, $\Delta C < 0$ is favoured as with $\Delta C>0$ the
difficulty with a too high $Br(B\to X_s l^+l^-)$ becomes more acute.

\begin{figure}[htb!]
\begin{center}
\includegraphics*[width=0.48\textwidth]{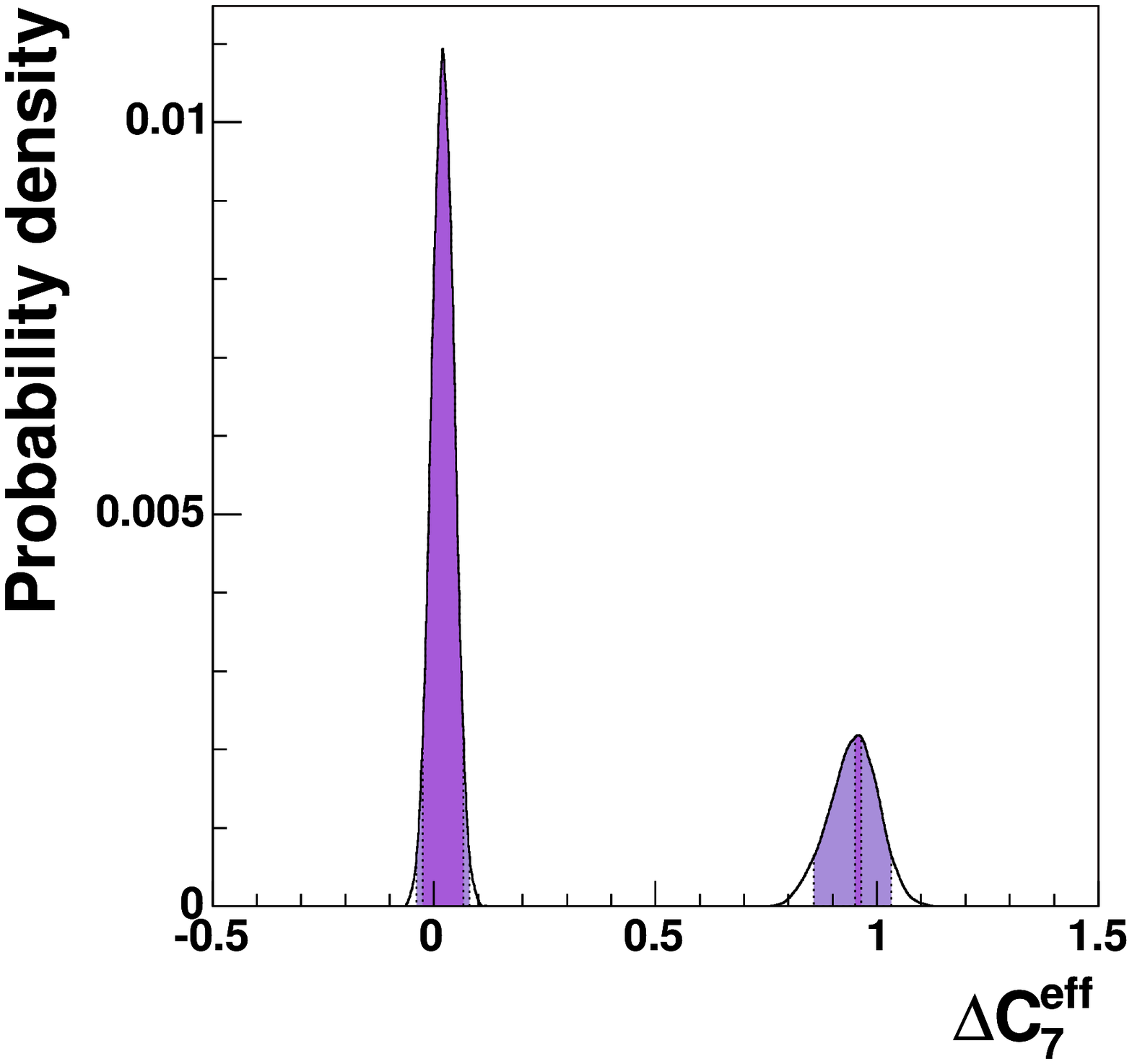}
\includegraphics*[width=0.48\textwidth]{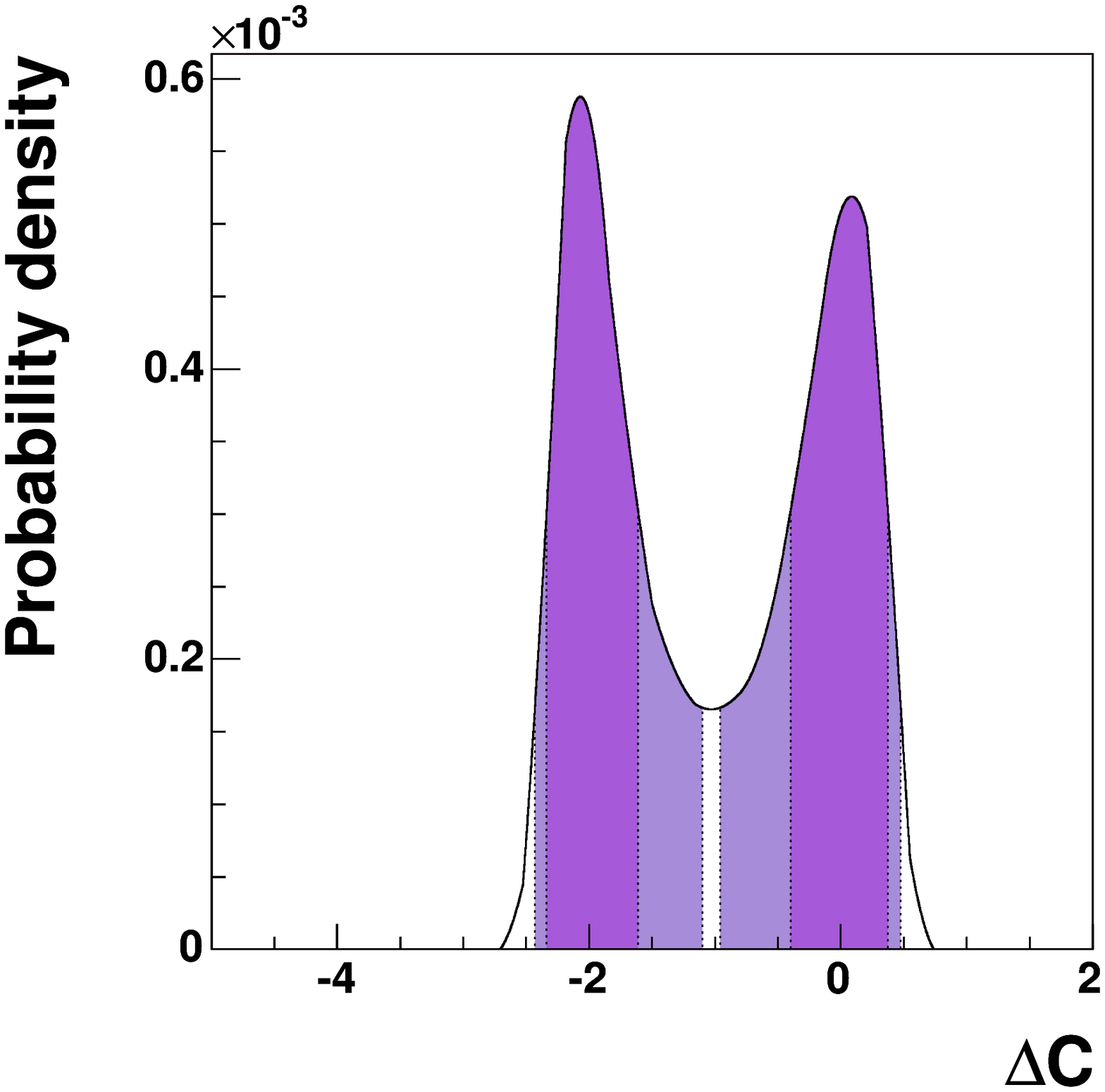}
\includegraphics*[width=0.48\textwidth]{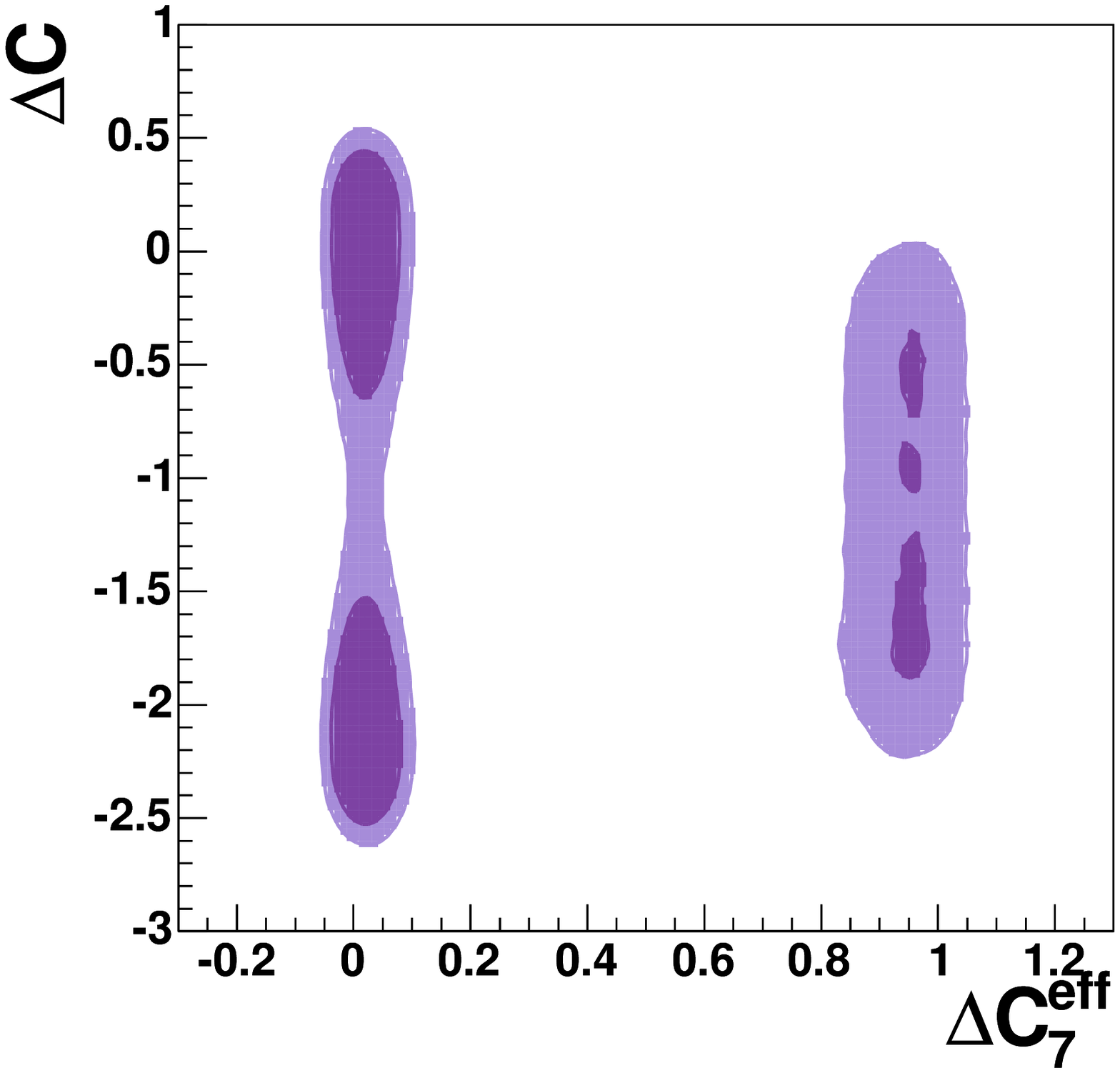}
\includegraphics*[width=0.48\textwidth]{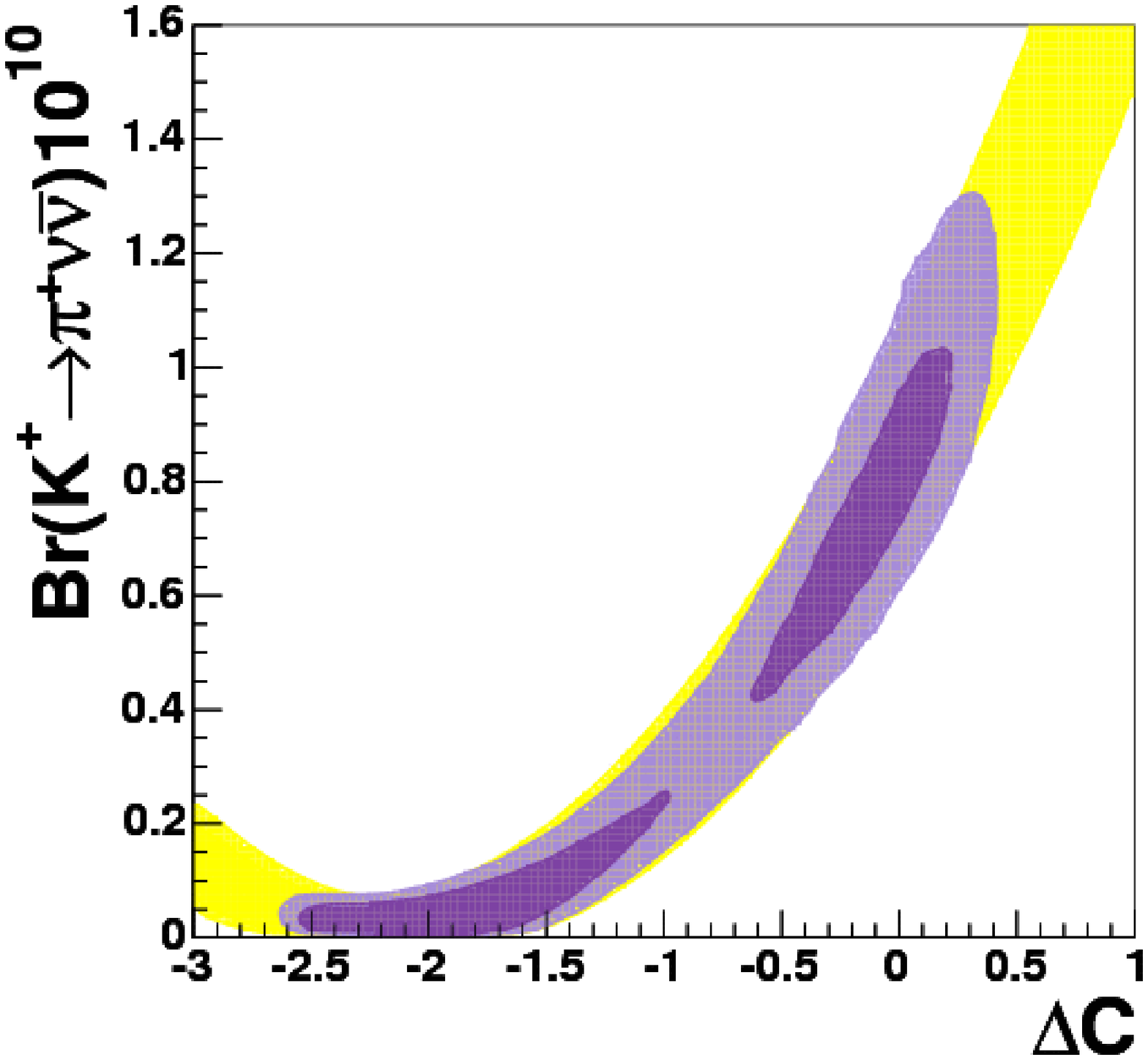}
\caption{%
  \it P.d.f.'s for $\Delta C_7^\mathrm{eff}$ (top-left), $\Delta C$
  (top-right), $\Delta C$ vs.  $\Delta C_7^\mathrm{eff}$ (bottom-left)
  and $Br(\kpn)$ vs $\Delta C$ (bottom-right) obtained without using
  $Br(\kpn)$ as a constraint. Dark (light) areas correspond to the
  $68\%$ ($95\%$) probability region.}
\label{fig:Csrad}
\end{center}
\end{figure}

For the reader's convenience, we report in Table~\ref{tab:XYZ}
  the values of the $X$, $Y$ and $Z$ functions obtained by summing SM
  and NP contributions and by applying all the available experimental
  constraints.

\begin{table*}[t]
\small{
\begin{center}
\begin{tabular}{|c|c|c|}
\hline
{Function} &  MFV (68\%) &  MFV (95\%)
 \\ \hline
$X$ & $[-0.71,-0.55] \cup [0.86,1.90]$ & $[-0.86,0.10] \cup [0.30,1.95]$ 
\\ \hline
$Y$ & $[-1.23,-1.06] \cup [0.33,1.37]$ & $[-1.38,-0.44] \cup [-0.24,1.43]$ 
\\ \hline
$Z$ & $[-1.51,-1.40] \cup [-0.25,1.31]$ & $[-1.74,-1.05] \cup [-0.92,1.46]$ 
\\ \hline
\end{tabular}
\caption[]{\it Values at $68\%$ and $95\%$ probability for the
  functions $X$, $Y$ and $Z$.  See the text for details.  }
\label{tab:XYZ}
\end{center}
}
\end{table*}

The impact of $Br(\kpn)$ on the bounds on NP contributions can be seen
by comparing Figure~\ref{fig:Cs} with Figure~\ref{fig:Csrad}, where
$Br(\kpn)$ was not used as a constraint.\footnote{In order to fully
  exploit the experimental information on $Br(\kpn)$, we use directly
  the likelihood function obtained by deriving the experimental CL
  \cite{kpnexpl}.}  As can be seen from Figure~\ref{fig:Csrad}, the
role of $Br(\kpn)$ is to suppress the solution with $\Delta C \sim
-2$, which corresponds to destructive interference with the SM in
$Br(\kpn)$ and in the other rare decays.  In this respect, a further
improvement of the experimental error on $Br(\kpn)$ will be extremely
useful in further reducing the importance of this
negative-interference solution for $\Delta C$, which is responsible
for the peaks around zero for all the rare decays in
Figure~\ref{fig:BRs}.

We also note that eliminating $\Delta C <0$ by means of
$K^+\to\pi^+\nu\bar\nu$ would basically also eliminate the HI solution
for $\Delta C_7^{\rm eff}$.  We therefore conclude that finding
$Br(K^+\to\pi^+\nu\bar\nu)$ larger than the SM value would help in
eliminating the positive sign of $C_7^{\rm eff}$. To our knowledge
this triple correlation between $K^+\to\pi^+\nu\bar\nu$, $B\to
X_s\gamma$ and $B\to X_s l^+l^-$ has not been discussed in the
literature so far. It is very peculiar to MFV and is generally not
present in models with new flavour violating contributions.

The upper bound on $Br(\kpn)$ in Table~\ref{brMFV} has been
  obtained using the experimental information on this decay. It
  corresponds to the following $95\%$ probabilty ranges:
\begin{eqnarray}
Br(\kpn) &=& [0,0.17] \cup [0.24,1.19]\nonumber \\
&~&(\mathrm{LOW:~} [0,0.12]\cup [0.39,1.26],
~\mathrm{HI:~} [0,0.81])\times 10^{-10}\,.
\label{eq:kpn}
\end{eqnarray} 
If we do not use the experimental result on $Br(\kpn)$, we obtain instead:
\begin{eqnarray}
Br(\kpn) &=& [0,0.15] \cup [0.28,1.12] \times 10^{-10}\,,
\label{eq:kpnrad}
\end{eqnarray} 
corresponding to an upper bound of $11.2 \times 10^{-11}$ at $95\%$
probability.  

We have also analyzed the decays $K_L \to \pi^0 e^+ e^-$ and $K_L \to
\pi^0 \mu^+ \mu^-$ using the formulae of~\cite{BI03,Isidori:2004rb}. In
the models with MFV these decays are dominated by the contribution
from the indirect CP violation that is basically fixed by the measured
values of $\epsilon_K$ and $K_S \to \pi^0 l^+ l^- $. The dependence on
$C(v)$ enters only in the subdominant direct CP-violating component
and the interference of indirect and direct CP-violating
contributions. We find that $Br(K_L \to \pi^0 e^+ e^-)$ and $Br(K_L
\to \pi^0 \mu^+ \mu^-)$ can be enhanced with respect to the SM value
by at most $8 \%$ and $10 \%$, respectively. In view of theoretical
uncertainties in these decays that are larger than these enhancements,
a clear signal of new physics within the MFV scenario is rather
unlikely from the present perspective. Therefore we do not show the
corresponding p.d.f.s.
 
\begin{figure}[htb!]
\begin{center}
\includegraphics*[width=0.70\textwidth]{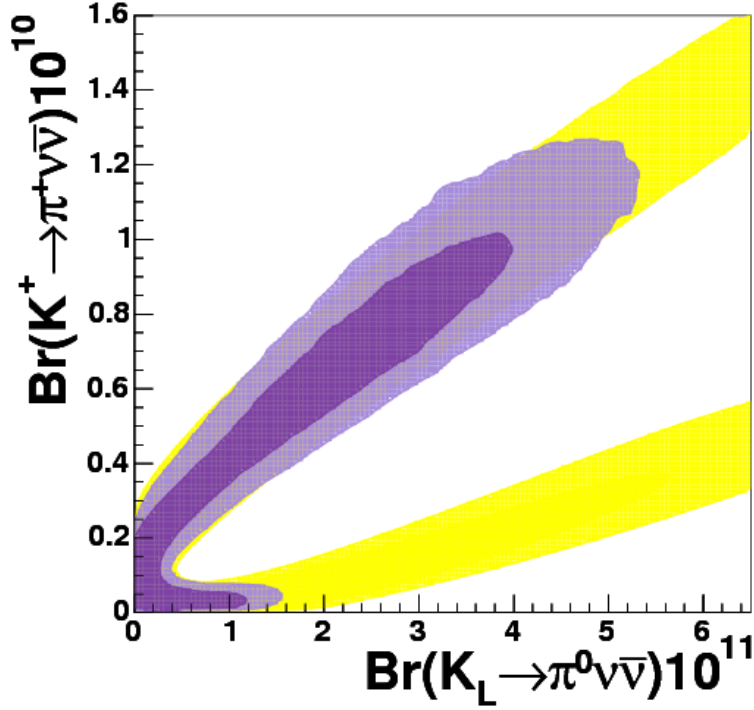}
\caption{%
  \it P.d.f. for the branching ratios of the rare decays $Br(\klpn)$
  vs $Br(\kpn)$.  Dark (light) areas correspond to the $68\%$ ($95\%$)
  probability region. Very light areas correspond to the range
  obtained without using the experimental information.}
\label{fig:kpvskl}
\end{center}
\end{figure}

Concerning $Br(\klpn)$, its $95\%$ probability ranges are given by
\begin{eqnarray}
Br(\klpn) &=& [0,4.59] ~(\mathrm{LOW:~} [0,4.83],~\mathrm{HI:~} [0,2.84])\times 10^{-11}\,
\label{eq:klpn}
\end{eqnarray} 
(see Figures~\ref{fig:BRs}, \ref{fig:raredeclC7} and
\ref{fig:raredechC7}).  In Figure~\ref{fig:kpvskl} we see explicitly
the correlation between the charged and neutral Kaon decay modes. We
observe a very strong correlation, a peculiarity of models with
MFV~\cite{BF01}. In particular, a large enhancement of $Br(\klpn)$
characteristic of models with new complex phases is not
possible~\cite{newphase}. An observation of $Br(\klpn)$ larger than
$6\cdot 10^{-11}$ would be a clear signal of new complex phases or new
flavour changing contributions that violate the correlations between
$B$ and $K$ decays.

The $95\%$ probability ranges for $Br(\kmm)_{\rm SD}$ are
\begin{eqnarray}
Br(\kmm)_{\rm SD} &=& [0,1.36]~(\mathrm{LOW:~} [0,1.44],~\mathrm{HI:~} [0,0.74])
\times 10^{-9}\,.
\label{eq:kmm}
\end{eqnarray} 
As in the previous cases, the HI solution corresponds to a much lower
upper bound.

Let us now consider $B$ decays:
\begin{eqnarray}
Br(B\to X_s\nu\bar\nu) &=& [0,5.17] ~(\mathrm{LOW:~}[0,1.56] \cup [1.59,5.4] ,~\mathrm{HI:~} 
[0,3.22])\times 10^{-5}\,, \nonumber \\
Br(B\to X_d\nu\bar\nu) &=& [0,2.17] ~(\mathrm{LOW:~} [0,2.26] ,~\mathrm{HI:~}[0,1.34]) 
\times 10^{-6}\,, \nonumber \\
Br(B_s\to \mu\bar \mu) &=& [0,7.42] ~(\mathrm{LOW:~} [0,7.91] ,~\mathrm{HI:~} [0,3.94])
\times 10^{-9}\,, \nonumber \\
Br(B_d\to \mu\bar \mu) &=& [0,2.20]~(\mathrm{LOW:~} [0,2.37],~\mathrm{HI:~} [0,1.15])
\times 10^{-10}\,.
\label{eq:bdec}
\end{eqnarray} 

\begin{figure}[htb!]
\begin{center}
\includegraphics*[width=0.48\textwidth]{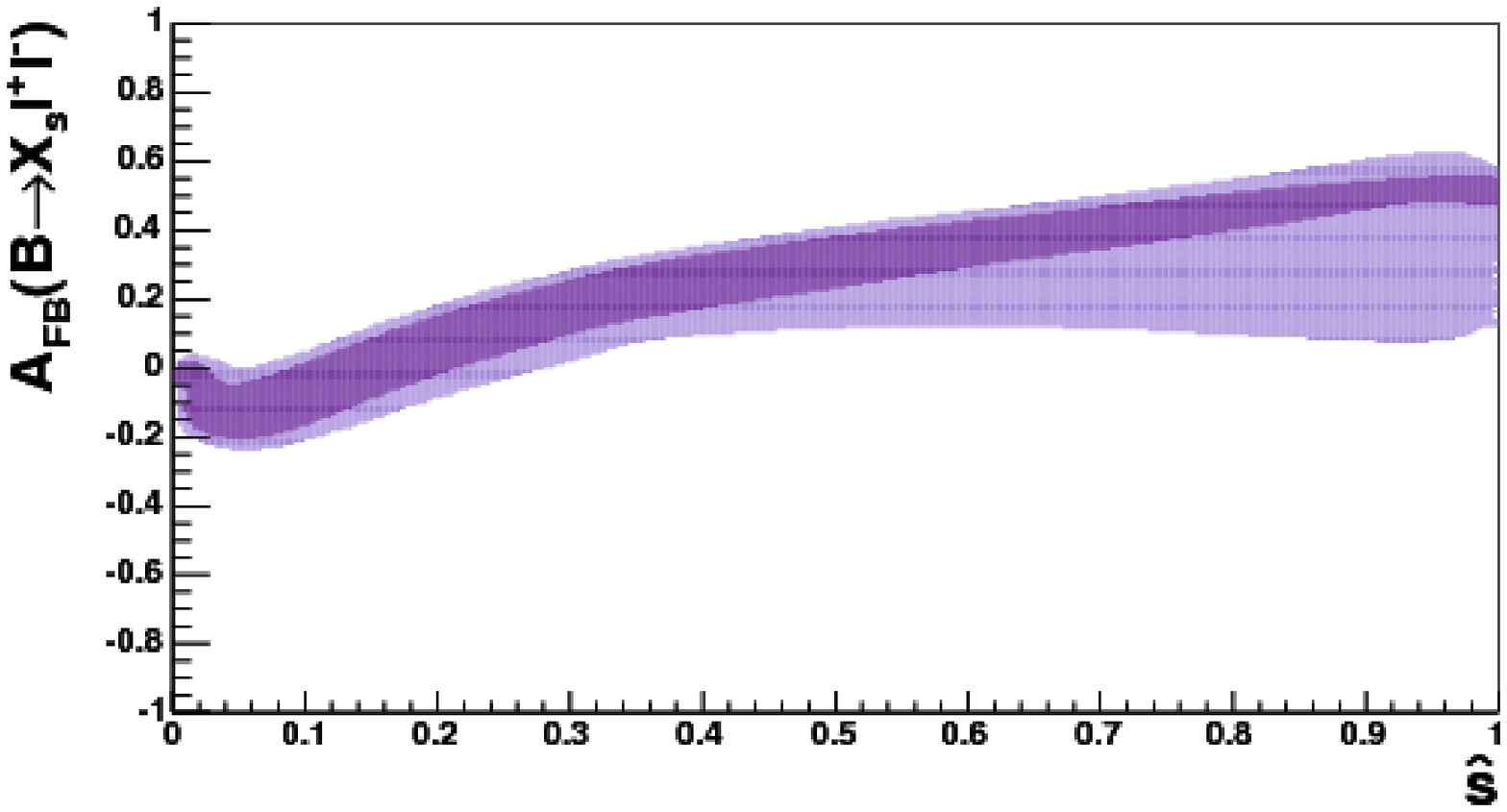}
\includegraphics*[width=0.48\textwidth]{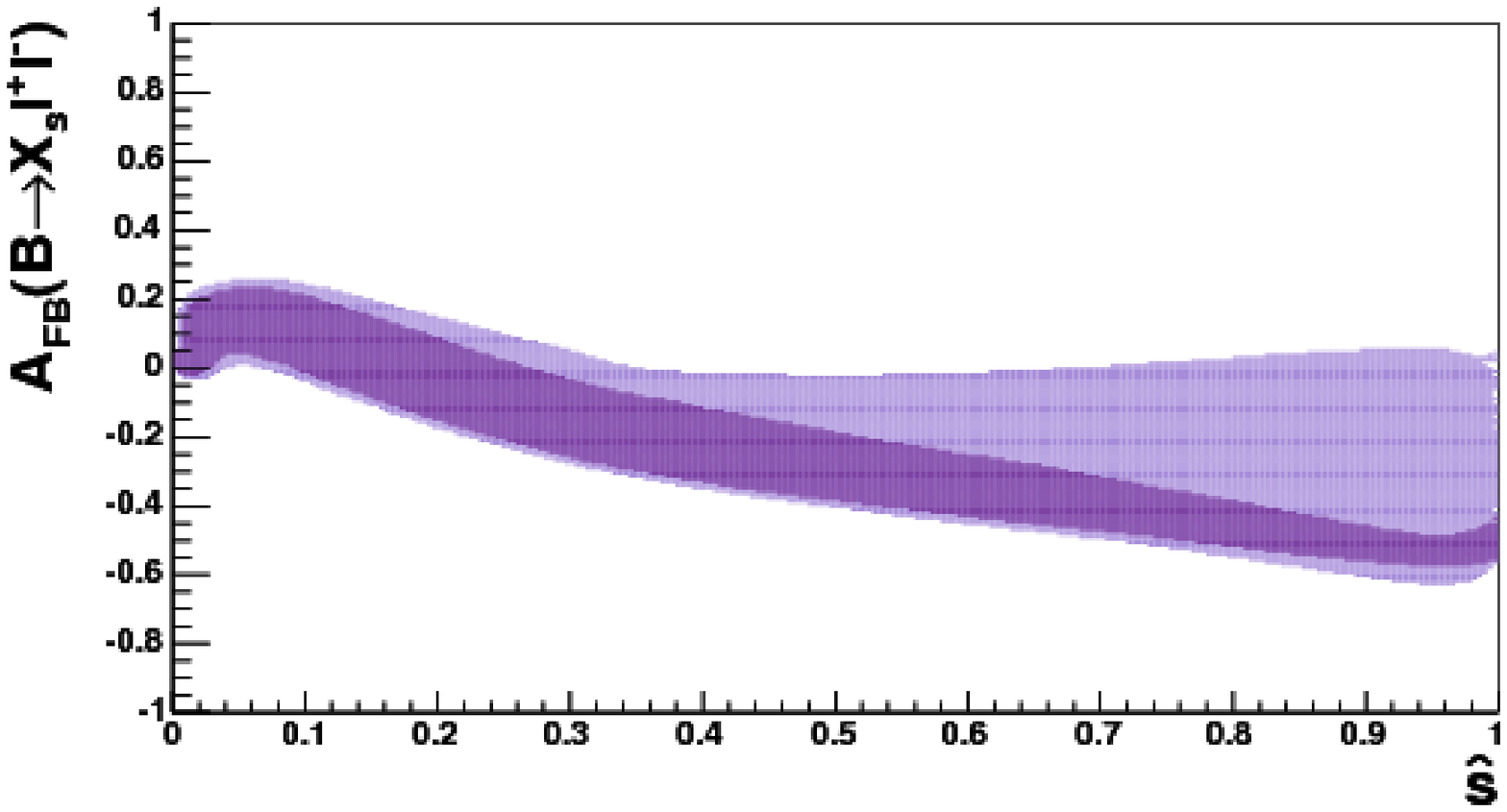}
\includegraphics*[width=0.48\textwidth]{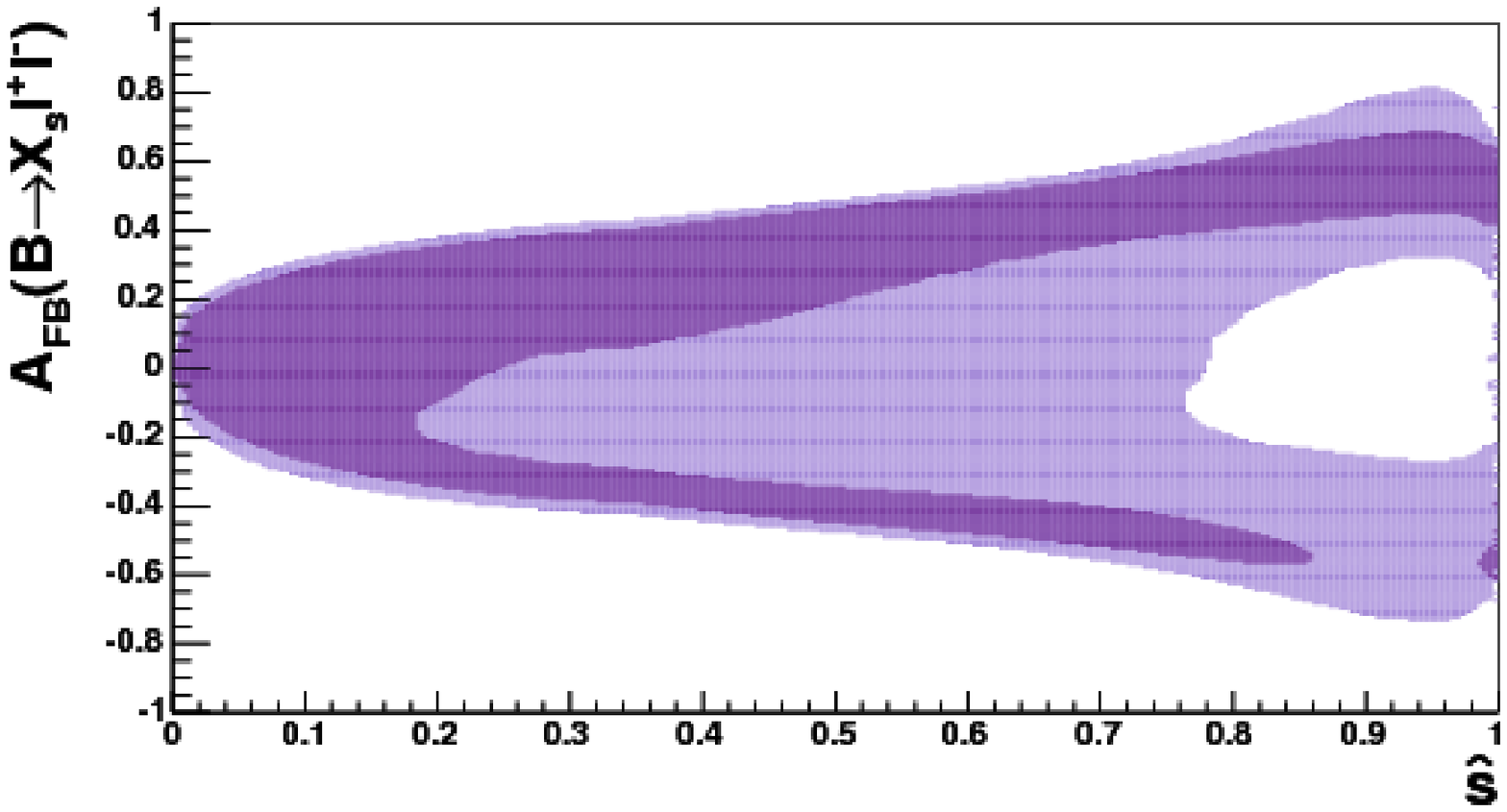}
\caption{%
  \it P.d.f.'s for the normalized forward-backward asymmetry in
  $\BXsll$ for the LOW solution for $\Delta C_7^\mathrm{eff}$ with
  $\Delta C> -1$ (left), for the LOW solution with $\Delta C < -1$
  (center) and for the HI solution for $\Delta C_7^\mathrm{eff}$
  (right).  Dark (light) areas correspond to the $68\%$ ($95\%$)
  probability region.}
\label{fig:AFB}
\end{center}
\end{figure}

The reader may wonder whether other observables could help improving
the constraints on $\Delta C$ and testing MFV models. In particular,
the Forward-Backward asymmetry in $\BXsll$ is known to be a very
sensitive probe of $C_7^{\mathrm{eff}}$ and of $C$ \cite{AFB}.
Indeed, the HI and LOW solutions for $\Delta C_7^{\mathrm{eff}}$ and
corresponding possible values of $\Delta C$ give rise to different
profiles of the normalized $\bar A_{\rm FB}$, defined as
\begin{equation}
  \bar A_{\rm FB}(\hat s)=\frac{\int_{-1}^1 d \cos \theta_l \frac{ d^2
      \Gamma (b \to s \mu^+\mu^-)} { d\hat s d\cos \theta_l} {\rm sgn}
    (\cos \theta_l)}{\int_{-1}^1 d \cos \theta_l \frac{ d^2
      \Gamma (b \to s \mu^+\mu^-)} { d\hat s d\cos \theta_l}}\,.
\end{equation} 
This can be seen explicitly in Figure~\ref{fig:AFB}.
Therefore, a measurement of $\bar A_{\rm FB}(\hat s)$ at a
  Super B factory will be extremely helpful in distinguishing the
  various scenarios discussed above~\cite{SUPERB}. On the other
hand, concerning the CP asymmetry in $\BXsgamma$ decays \cite{KN}, it
turns out that in MFV models its value is reduced with respect to the
SM, once the constraint on the branching ratio is taken into account,
so that it is not expected to play a significant role in present and
future analyses \cite{hep-ph/0312260}.

In Figure~\ref{fig:raredeclC7} we show the
p.d.f.'s for the branching ratios of rare decays for the LOW solution.
The corresponding result for the HI solution is given in
Figure~\ref{fig:raredechC7}. Clearly the branching ratios of various
decays are larger in the case of the LOW solution.

Before concluding this section, let us make a few steps towards the
future and consider a realistic scenario for the projected integrated
luminosities of 
 Belle and BaBar, plus a $10\%$ measurement of $Br(\kpn$).
For concreteness, let us assume the following 2010 experimental data:
\begin{eqnarray}
Br(\BXsll)_{0.04 < q^2 \mathrm{(GeV)} < 1}&=&(1.13 \pm 0.25) \times 10^{-6}\,,\nonumber \\
Br(\BXsll)_{1 < q^2 \mathrm{(GeV)} < 6}&=&
\left\lbrace 
\begin{array}{l}
(1.49 \pm 0.21)  \times 10^{-6} \mathrm{(Belle)} \\
(1.80 \pm 0.18)  \times 10^{-6} \mathrm{(BaBar)}
\end{array} \right.  \nonumber \\
Br(\BXsll)_{14.4 < q^2 \mathrm{(GeV)} < 25}&=&
\left\lbrace 
\begin{array}{l}
(4.18 \pm 0.48)  \times 10^{-7} \mathrm{(Belle)} \\
(5.00 \pm 0.93)  \times 10^{-7} \mathrm{(BaBar)}
\end{array} \right.  \nonumber \\
Br(\BXsgamma) &=&
\left\lbrace 
\begin{array}{l}
(3.51 \pm 0.16)  \times 10^{-4} \mathrm{(Belle)} \\
(3.67 \pm 0.16)  \times 10^{-4} \mathrm{(BaBar~incl.)}\\
(3.29 \pm 0.16)  \times 10^{-4} \mathrm{(BaBar~semincl.)}
\end{array} \right.  
\label{eq:futureexp}
\end{eqnarray} 
corresponding to an integrated luminosity of $1 ~{\rm ab}^{-1}$ and
$700 ~{\rm fb}^{-1}$ for Belle and BaBar, respectively. Additionally a
reduction to $5 \%$ of the theoretical uncertainty in $Br(\BXsgamma)$
thanks to the ongoing NNLO computation is
assumed~\cite{BSGNNLO}.~\footnote{The future results for
  $Br(\BXsgamma)$ are referred to the same kinematic ranges as the
  present results.}

\begin{figure}[htb!]
\begin{center}
\includegraphics*[width=0.48\textwidth]{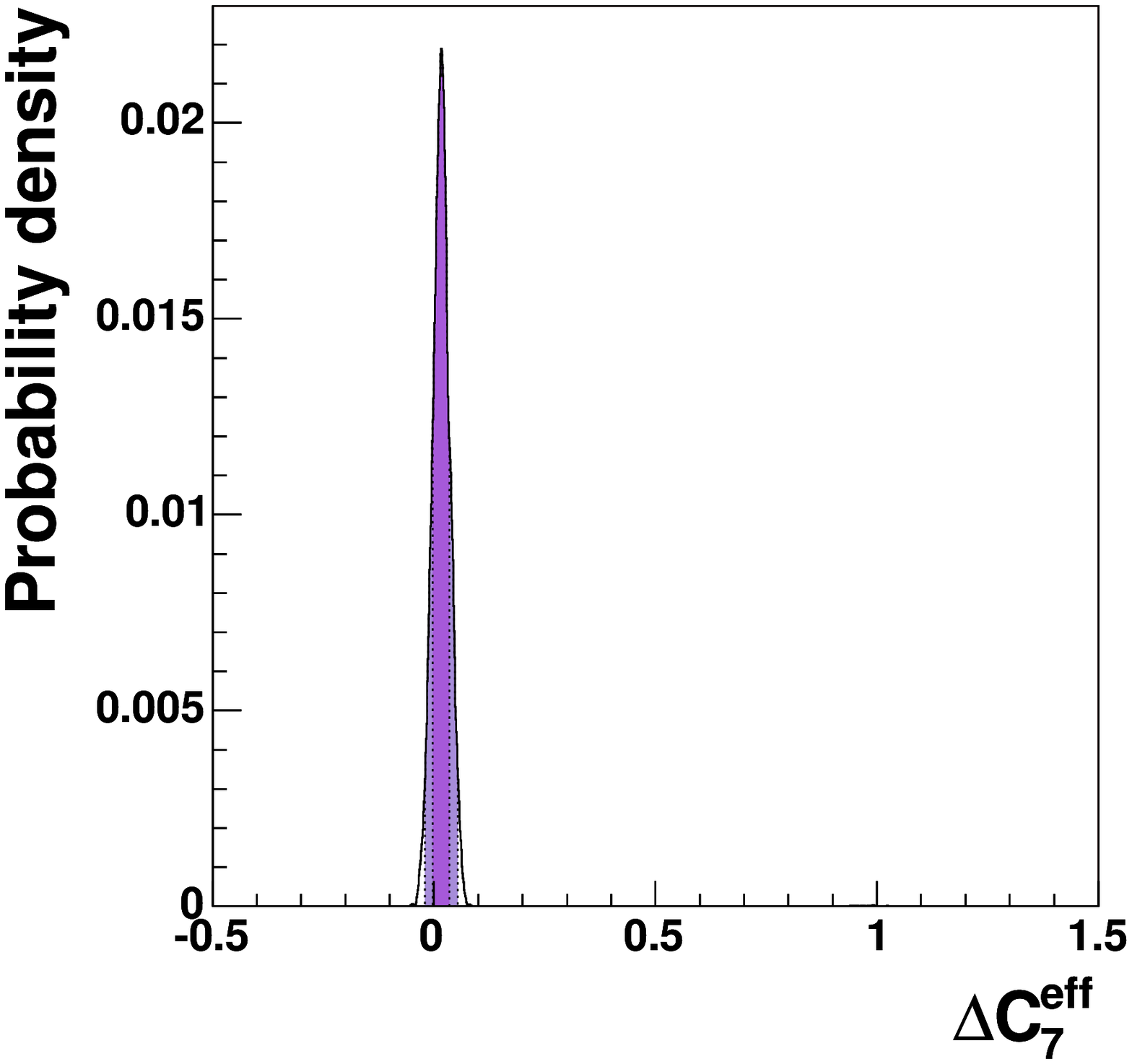}
\includegraphics*[width=0.48\textwidth]{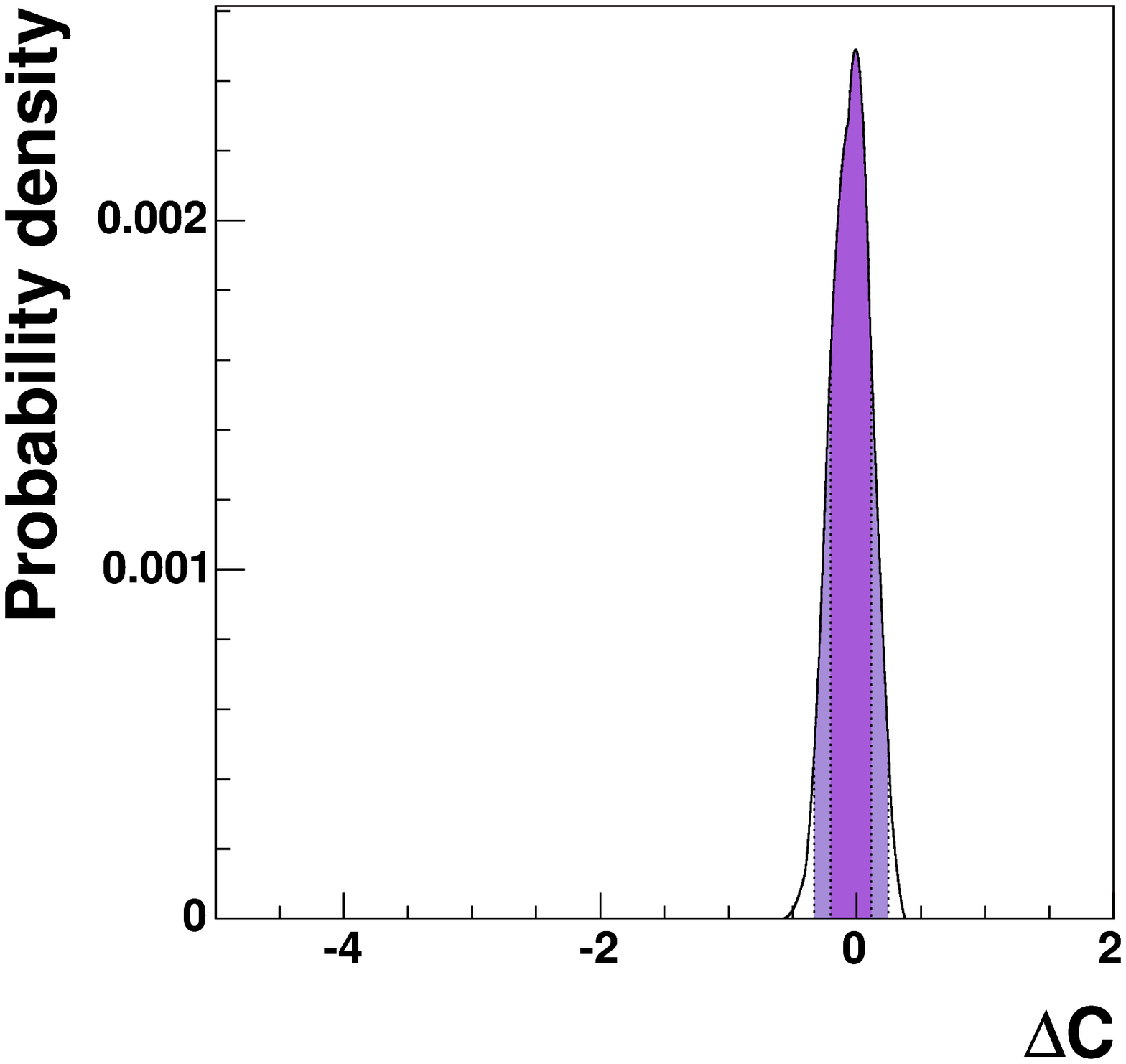}
\caption{%
  \it P.d.f.'s for $\Delta C_7^\mathrm{eff}$ (left) and $\Delta C$
  (right) in the future scenario specified by
  eq.~(\ref{eq:futureexp}). Dark (light) areas correspond to the
  $68\%$ ($95\%$) probability region.}
\label{fig:Csfuture}
\end{center}
\end{figure}

\begin{figure}[htb!]
\begin{center}
\includegraphics*[width=0.32\textwidth]{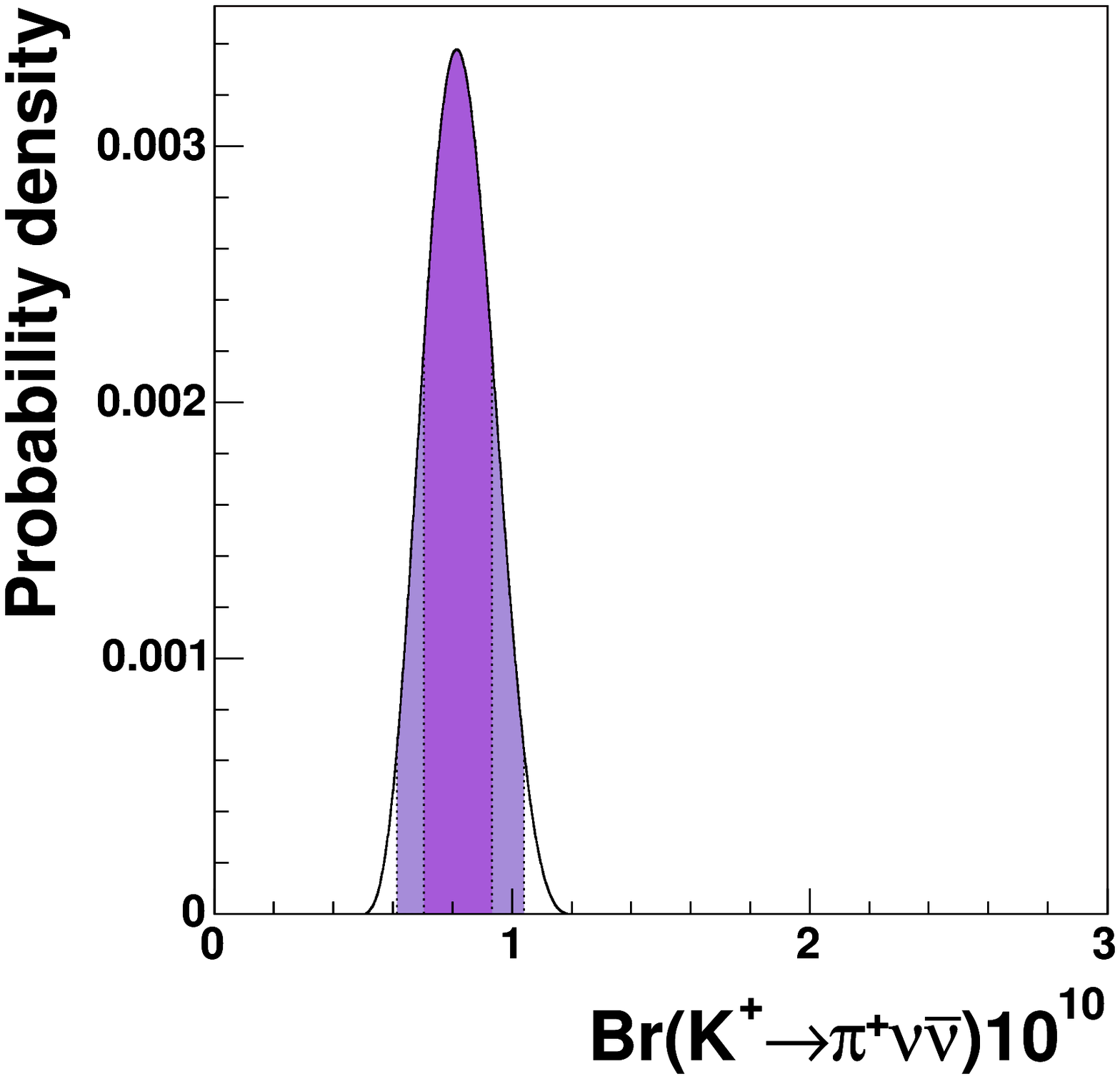}
\includegraphics*[width=0.32\textwidth]{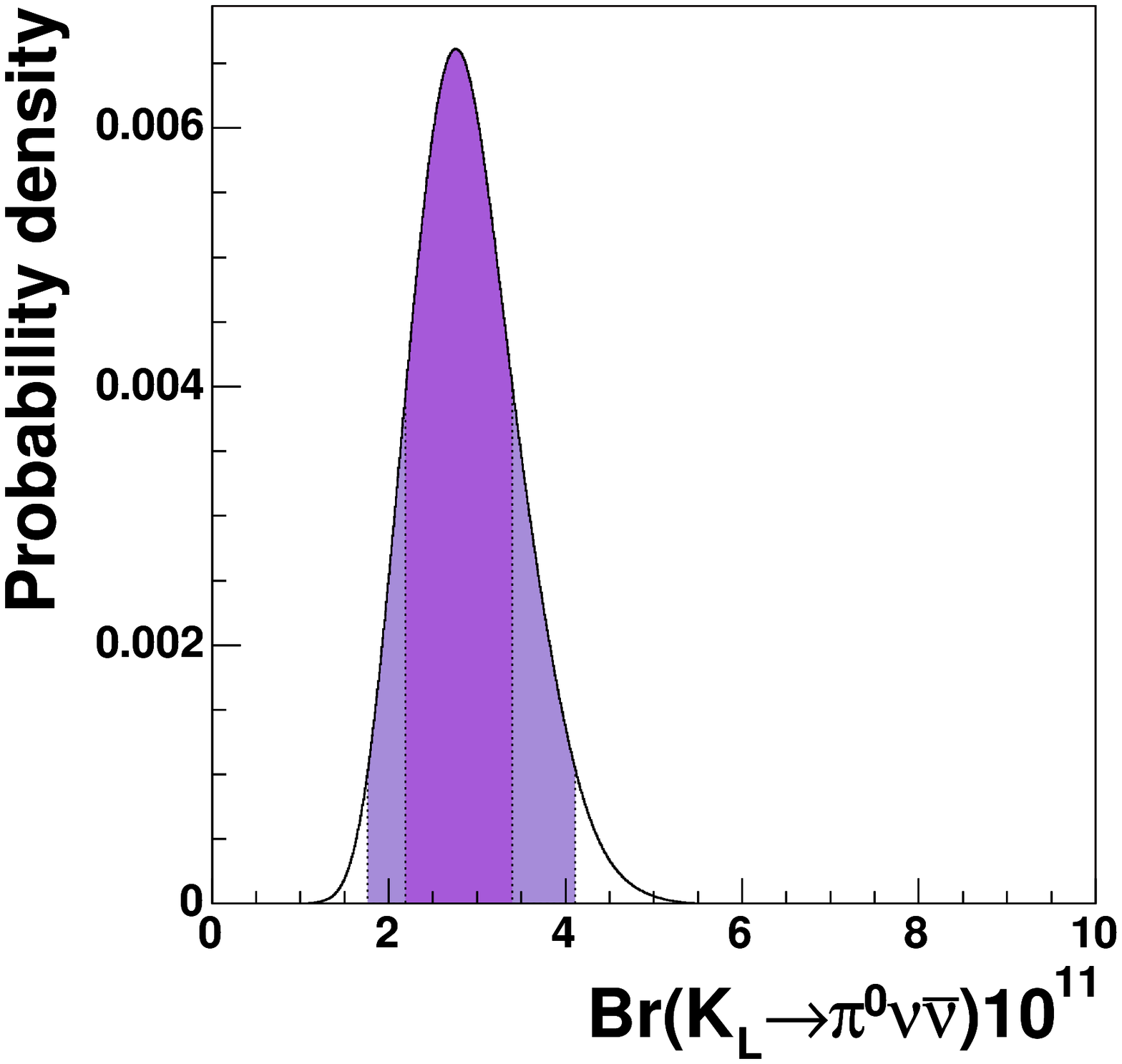}
\includegraphics*[width=0.32\textwidth]{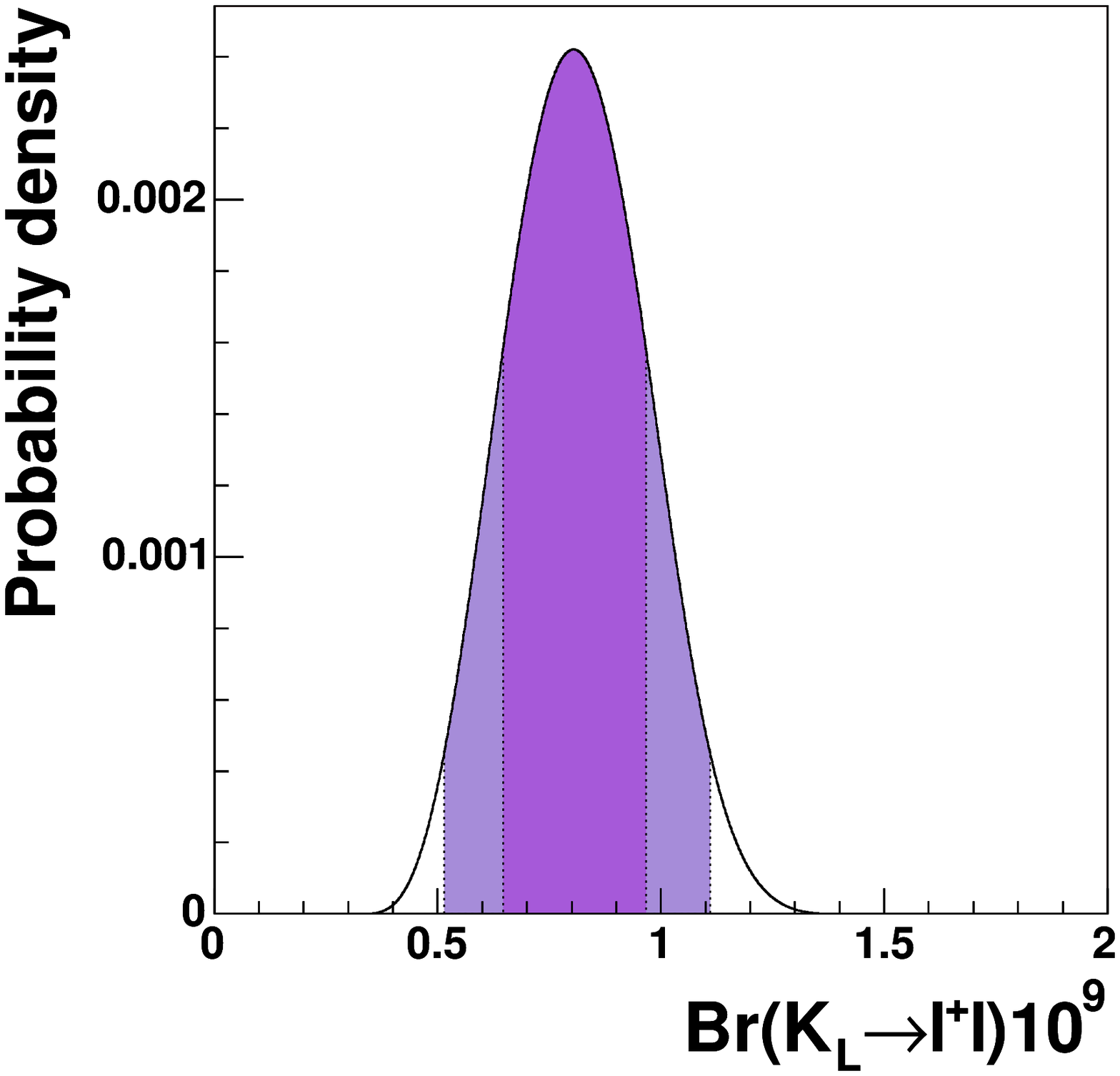}
\includegraphics*[width=0.32\textwidth]{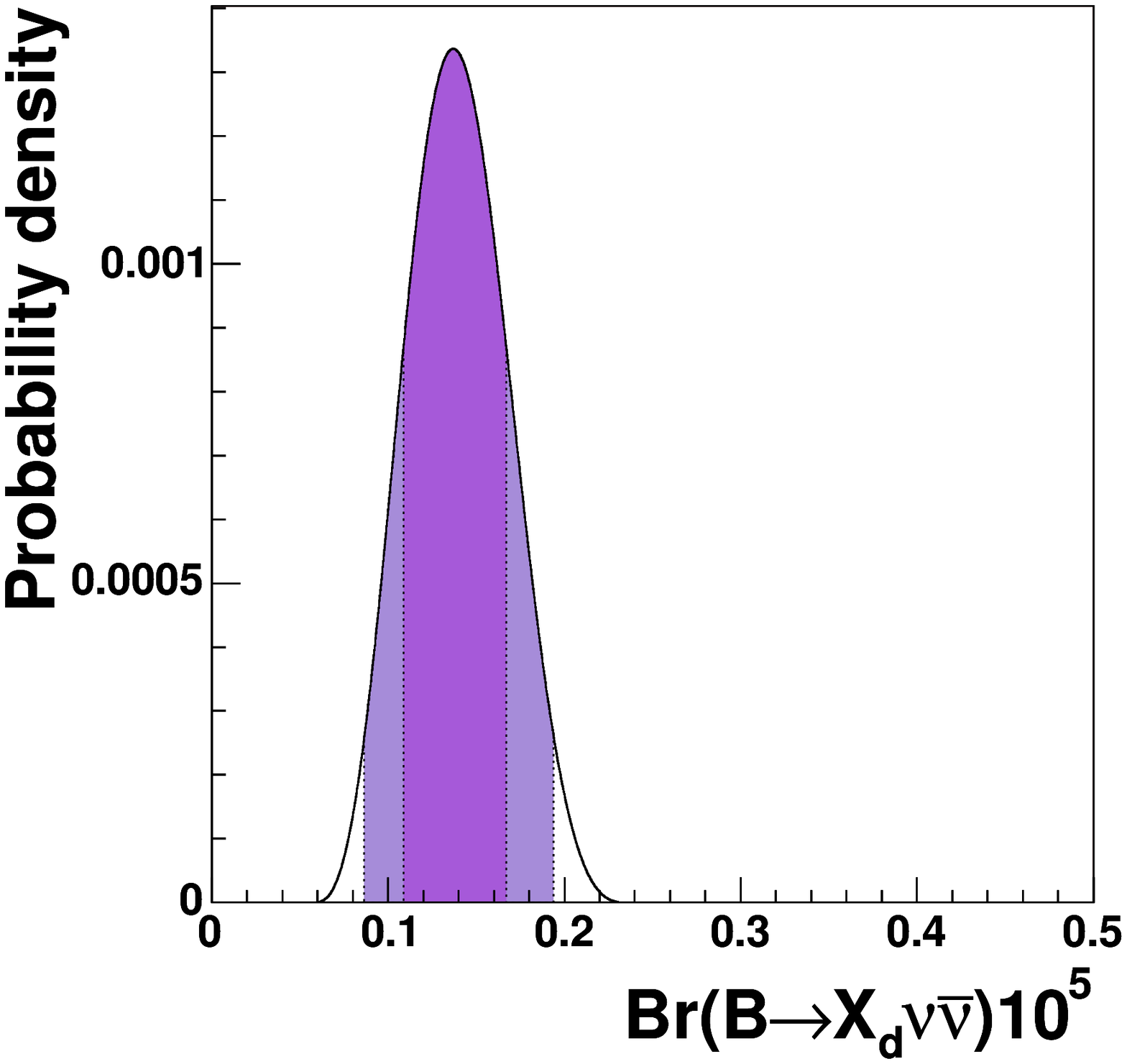}
\includegraphics*[width=0.32\textwidth]{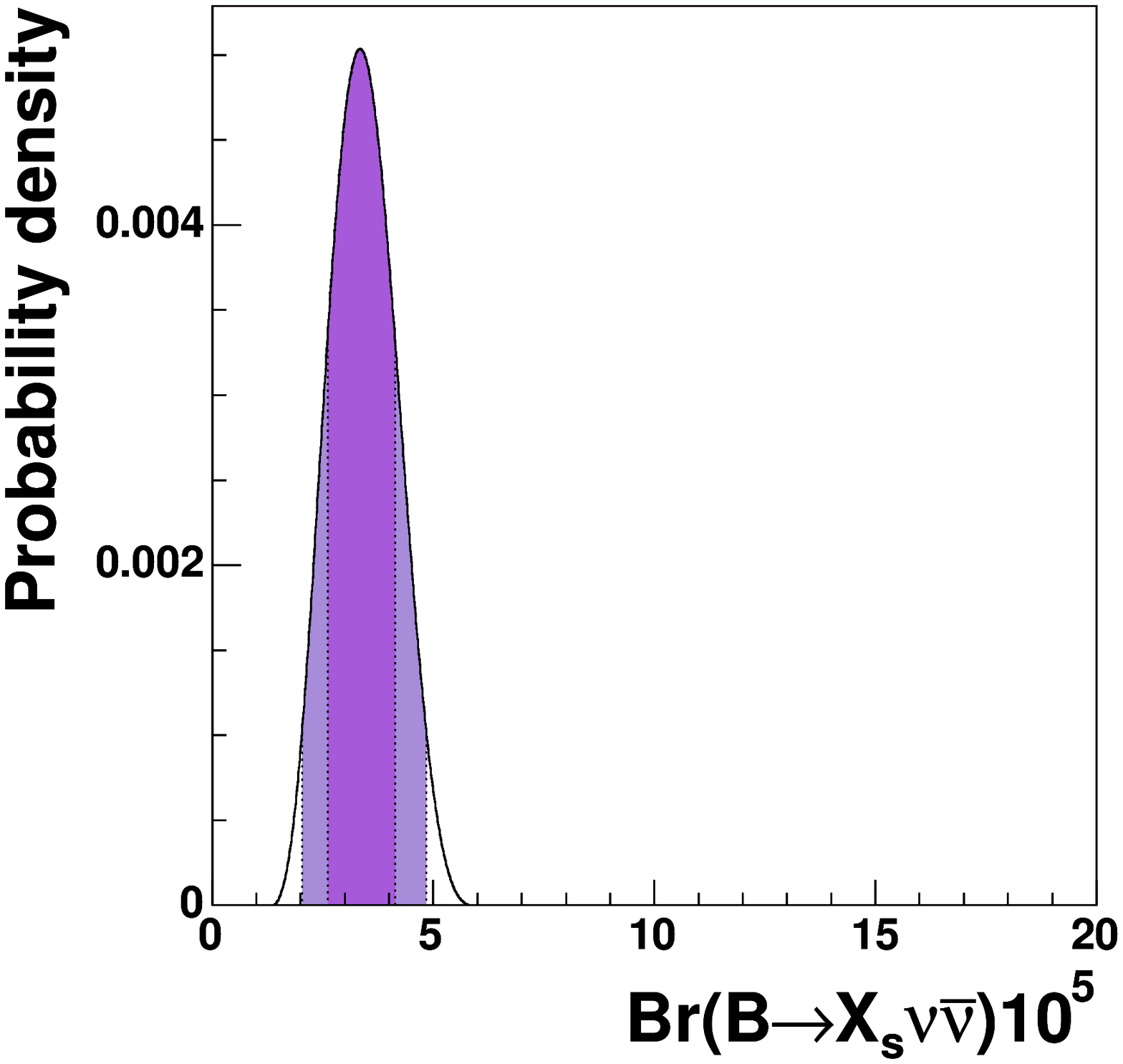}
\includegraphics*[width=0.32\textwidth]{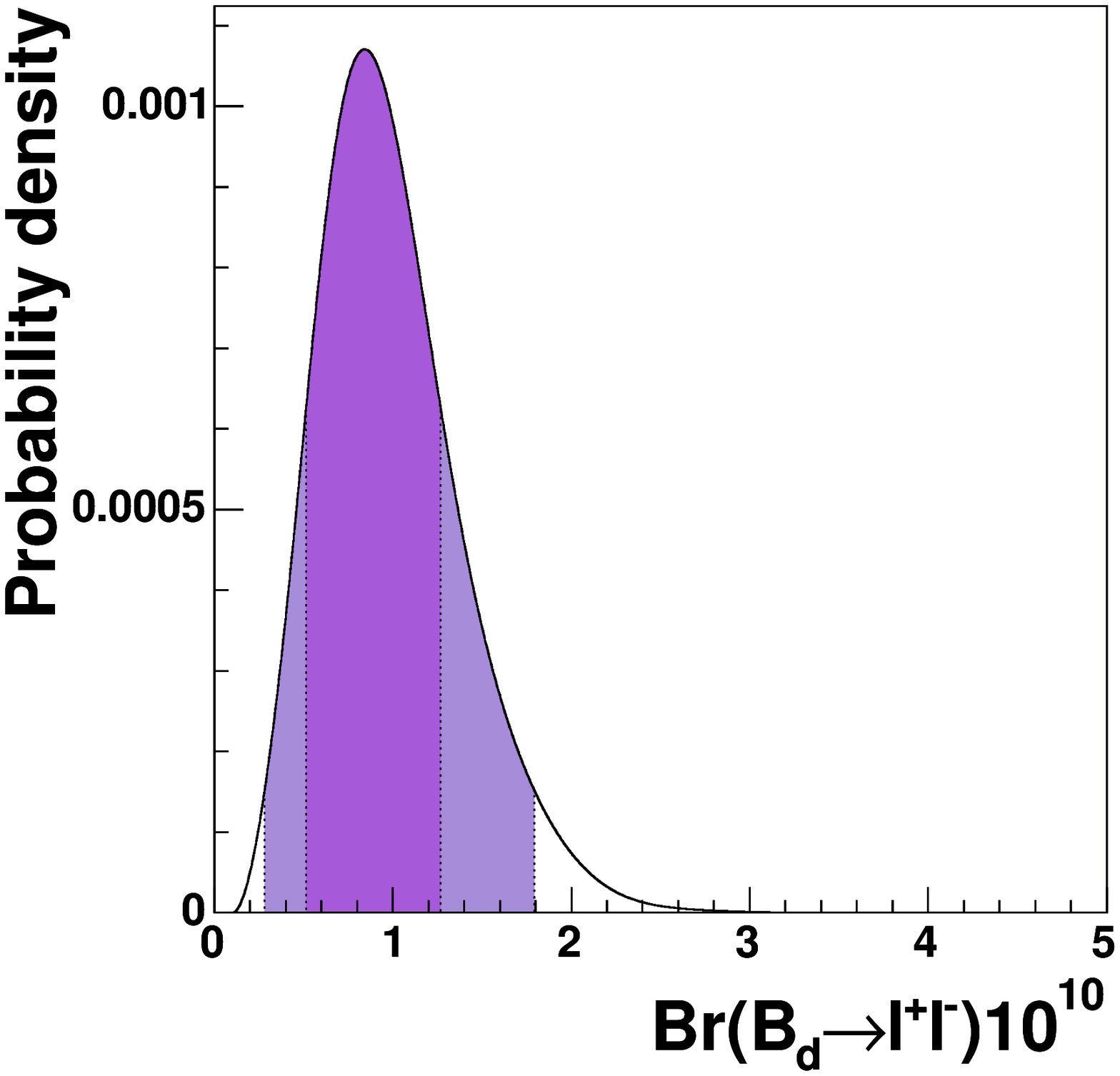}
\includegraphics*[width=0.32\textwidth]{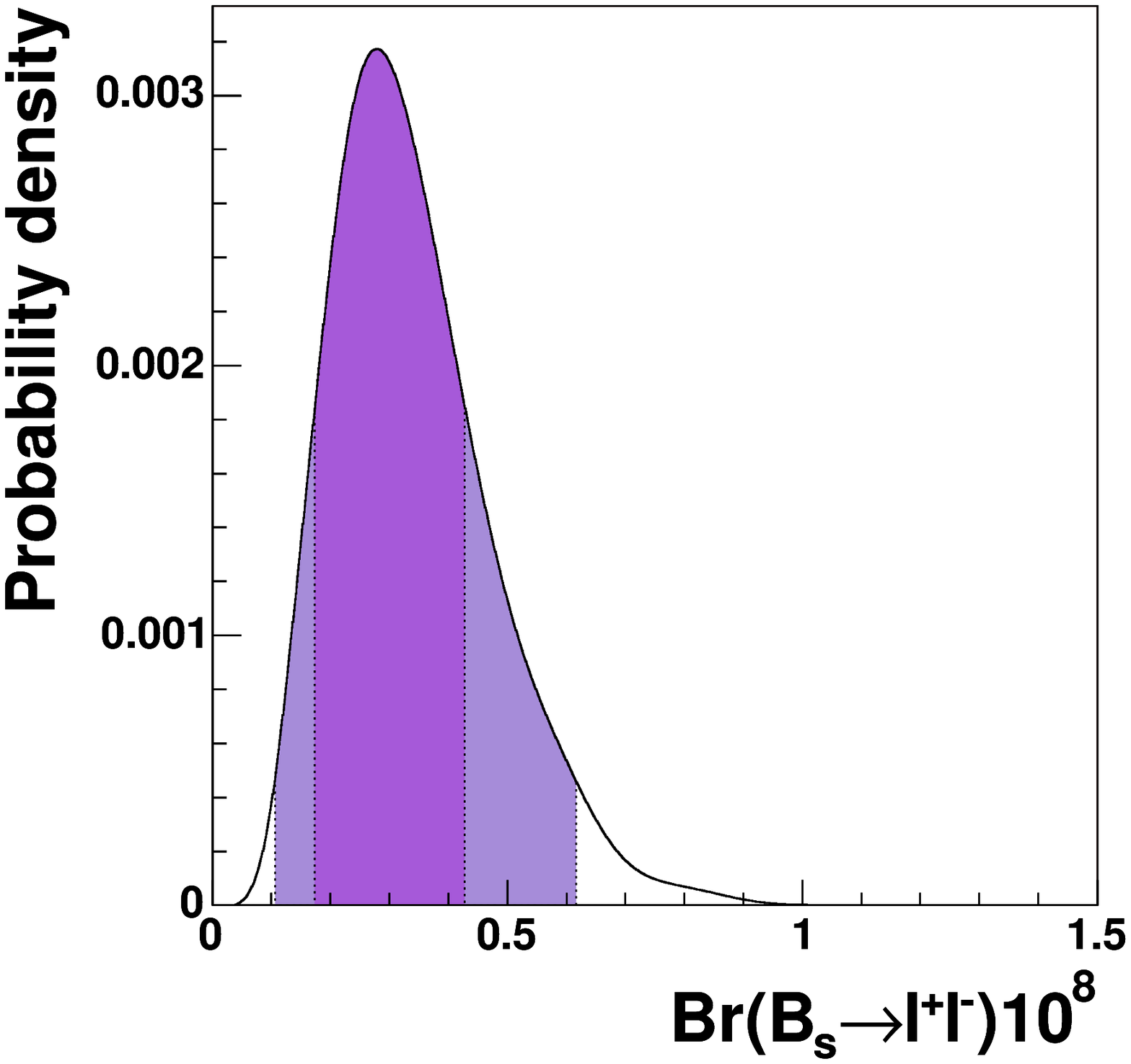}
\caption{%
  \it P.d.f.'s for the branching ratios of the rare decays $Br(\kpn)$,
  $Br(\klpn)$, $Br(\kmm)_{\rm SD}$, $Br(B\to X_{d,s}\nu\bar\nu)$, and
  $Br(B_{d,s}\to \mu^+\mu^-)$ in the future scenario specified by
  eq.~(\ref{eq:futureexp}). Dark (light) areas correspond to the
  $68\%$ ($95\%$) probability region.}
\label{fig:BRsfuture}
\end{center}
\end{figure}

\begin{figure}[htb!]
\begin{center}
\includegraphics*[width=0.48\textwidth]{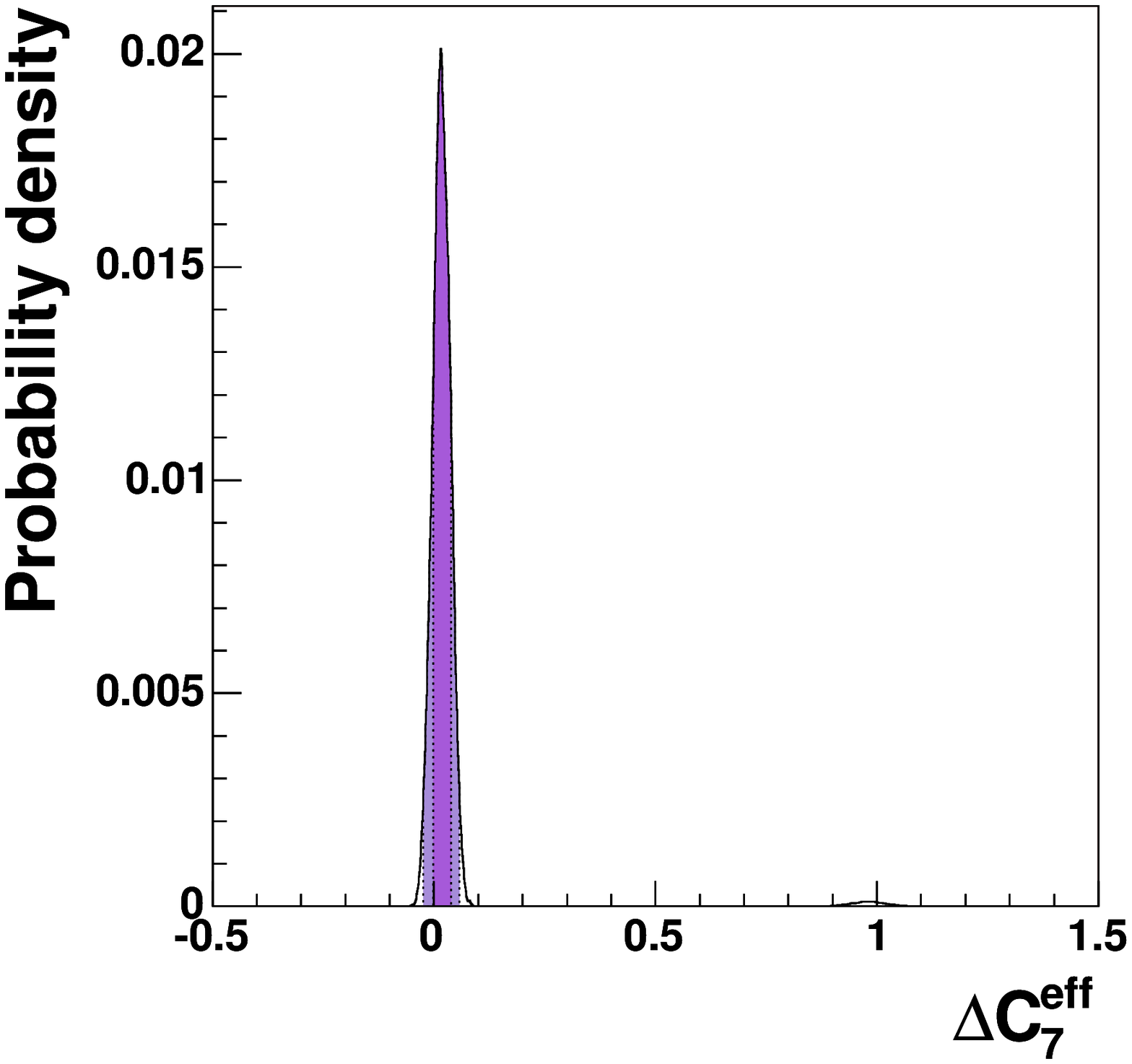}
\includegraphics*[width=0.48\textwidth]{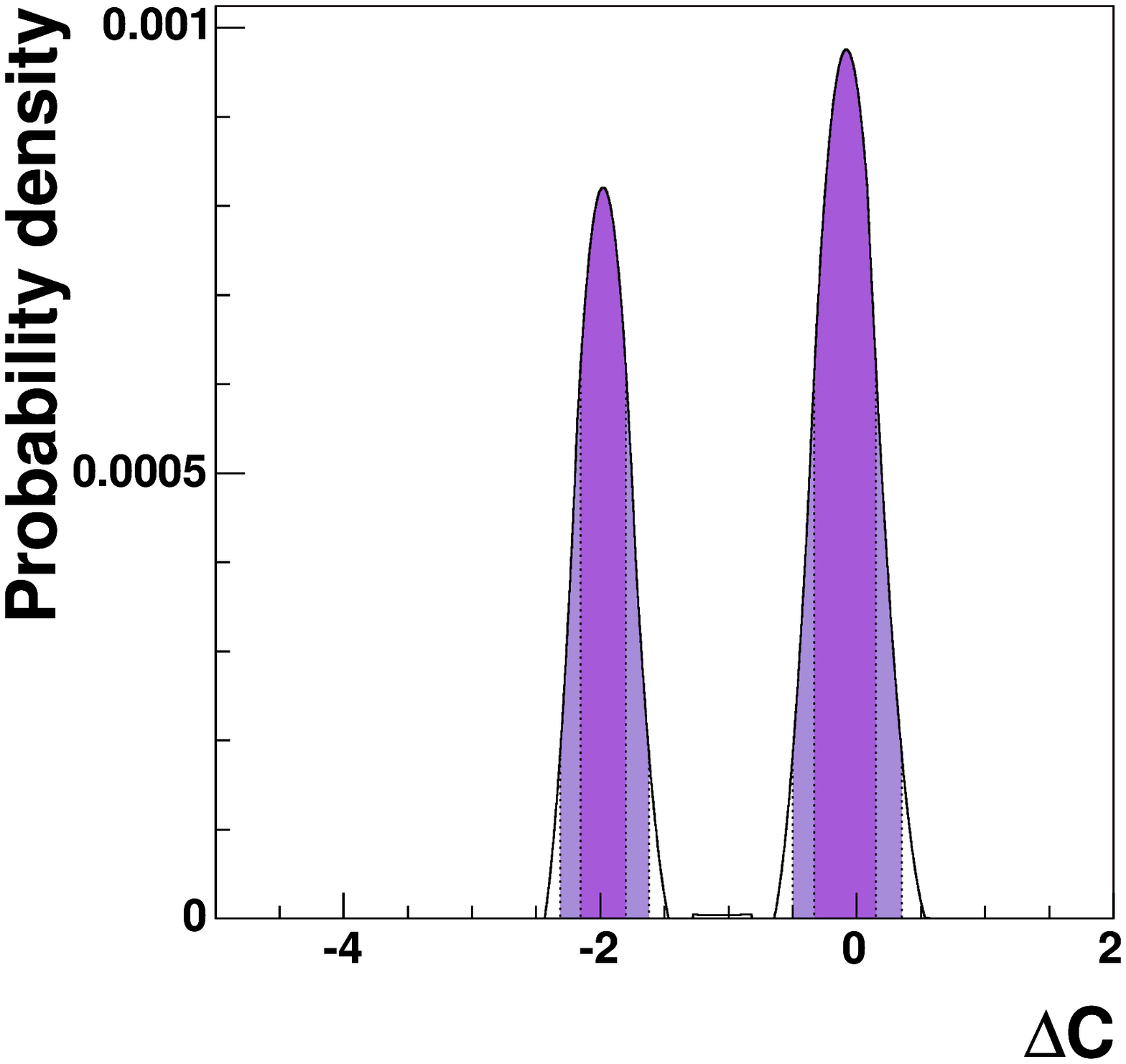}
\includegraphics*[width=0.48\textwidth]{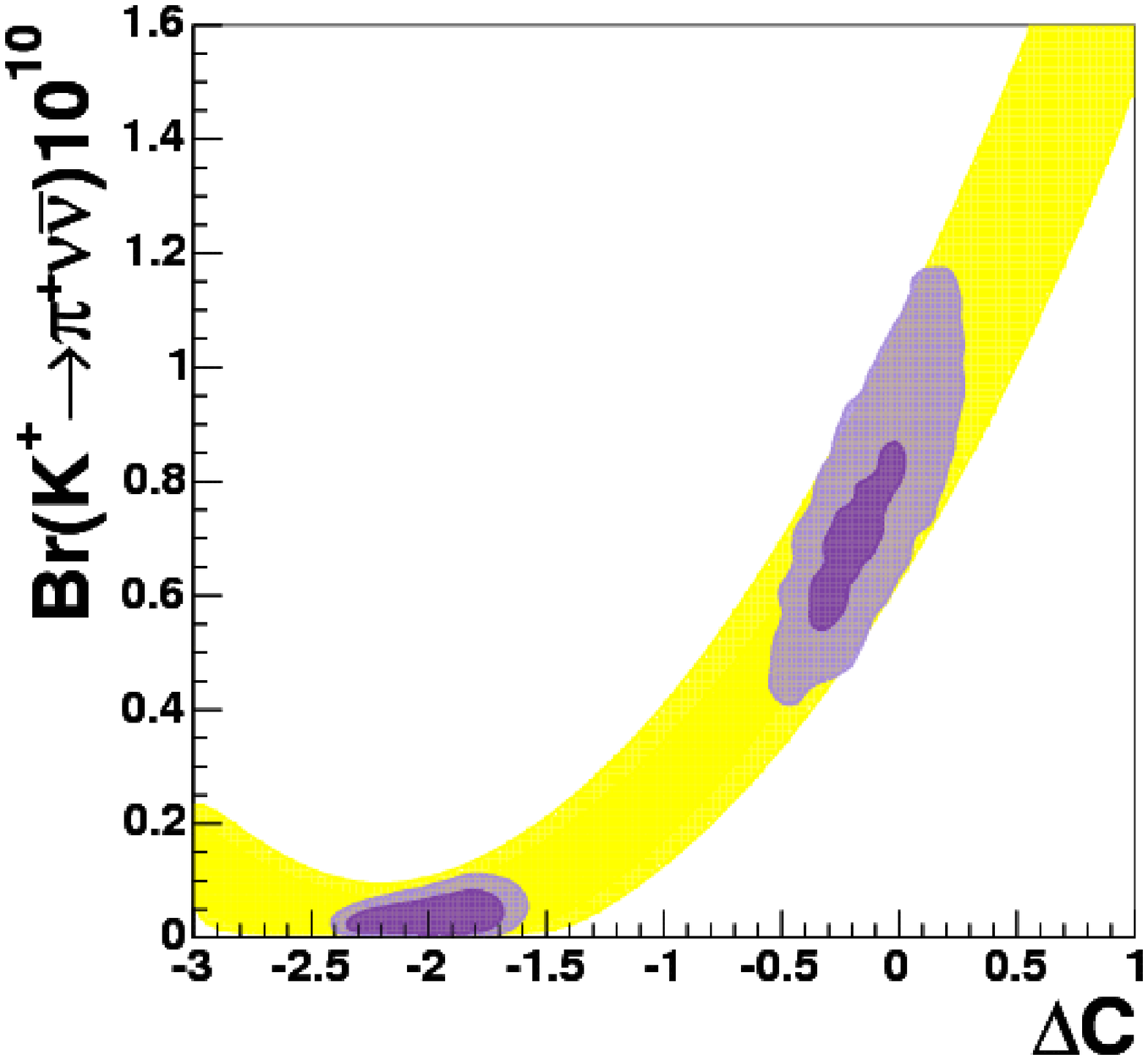}
\caption{%
  \it P.d.f.'s for $\Delta C_7^\mathrm{eff}$ (top-left), $\Delta C$
  (top-right) and $Br(\kpn)$ vs $\Delta C$ (bottom) obtained without
  using $Br(\kpn)$ as a constraint, in the future scenario specified
  by eq.~(\ref{eq:futureexp}). Dark (light) areas correspond to the
  $68\%$ ($95\%$) probability region.}
\label{fig:Csradfuture}
\end{center}
\end{figure}

We can see the dramatic effect of these improvements in
Figures~\ref{fig:Csfuture}-\ref{fig:Csradfuture}.  $B$-factory data
will completely eliminate the non-standard solution for $\Delta
C_7^\mathrm{eff}$, while they cannot distinguish the two solutions for
$\Delta C$ (considering only branching ratio measurements), see
Figure~\ref{fig:Csradfuture}. However, this ambiguity is perfectly
resolved by $Br(\kpn)$, leading to the impressive results in
Figures~\ref{fig:Csfuture} and \ref{fig:BRsfuture}.

With so powerful experimental data, one can even think of generalizing
our analysis by allowing for substantial deviations from the SM in box
diagrams. If the size of new physics contributions to box diagrams is
comparable to the SM ones, the results of our ``future" analysis would
not change sizably. On the other hand, a dramatic modification could
occur for contributions to box diagrams much larger than the SM ones;
however, it is very difficult to conceive new-physics models in which
this possibility can be realized.

%
%
\section{Messages}
\setcounter{equation}{0}

The main message of our paper is the following one:

The existing constraints coming from $K^+\to\pi^+\nu\bar\nu$, $B\to
X_s\gamma$ and $B\to X_s l^+l^-$ do not allow within the MFV scenario
of \cite{UUT} for substantial departures of the branching ratios for
all rare $K$ and $B$ decays from the SM estimates. This is evident
{}from Table~\ref{brMFV}.

There are other messages signalled by our analysis. These are:
\begin{itemize}
\item The decays $B\to X_{s,d} l^+l^-$ will not offer a precise value
  for the function $C$ even in the presence of precise measurements of
  their branching ratios, unless the theoretical errors in these
  decays and $B\to X_s\gamma$ and the experimental error on the
  branching ratio of the latter decay are reduced substantially. This
  is clearly seen in Figure~\ref{fig:constraints}.
\item The situation is considerably better in the case of $B_{d,s}\to
  \mu^+\mu^-$ but as seen in Figure~\ref{fig:raredec}, for a given
  value of $Br(B_{d,s}\to \mu^+\mu^-)$ there are generally two
  solutions for $\Delta C$ and $C$, that cannot be disentangled on the
  basis of these decays alone.
\item The great potential of the decays $K^+\to\pi^+\nu\bar\nu$ and
  $K_L\to\pi^0\nu\bar\nu$ in measuring the function $C$ is clearly
  visible in Figures~\ref{fig:constraints} and~\ref{fig:raredec}, with
  the unique value obtained in the case of $K^+\to\pi^+\nu\bar\nu$ in
  the full allowed range of $C$. In the case of
  $K_L\to\pi^0\nu\bar\nu$ the two solutions are only present for
  $Br(K_L\to\pi^0\nu\bar\nu)$ significant smaller that the SM value.
  Similar comment applies to $B\to X_{s,d}\nu\bar\nu$.
\item Assuming that future more precise measurements of the
  $K\to\pi\nu\bar\nu$ branching ratios will be consistent with the MFV
  upper bounds presented here, the determination of $C$ through these
  decays will imply much sharper predictions for various branching
  ratios that could confirm or rule out the MFV scenario. In this
  context the correlations between various branching ratios discussed
  in \cite{Zakopane} will play the crucial role.
\item One of such correlations predicts that the measurement of $\sin
  2\beta$ and of $Br(K^+\to\pi^+\nu\bar\nu)$ implies only two values
  of $Br(K_L\to\pi^0\nu\bar\nu)$ in the full class of MFV models that
  correspond to two signs of the function $X$ \cite{BF01}.
  Figure~\ref{fig:kpvskl} demonstrates that the solution with $X<0$,
   corresponding to the values in the left lower corner, is
  practically ruled out so that a unique prediction for
  $Br(K_L\to\pi^0\nu\bar\nu)$ can in the future be obtained.
\item A strong violation of any of the $95\%$ probability upper bounds
  on the branching ratios considered here by future measurements will
  imply a failure of MFV as defined in \cite{UUT}, unless an
  explicit MFV scenario can be found in which the contributions of box
  diagrams are significantly larger than assumed here. Dimensional
  arguments \cite{newphase} and explicit calculations indicate that
  such a possibility is rather remote.
\item If the only violation of the upper bounds in
    Table~\ref{brMFV} occurs in $\Bsmumu$ and $\Bdmumu$, it will be
    most likely due to new operators beyond the SM ones. For example,
    the scalar operators which arise in MFV SUSY models at large $\tan
    \beta$ can enhance $Br(\Bsmumu)$ up to the present experimental
    upper bound \cite{AMGIISST,hep-ph/0310219,largetb}.
\item Conversely, a violation of the upper bounds for the
    other channels in Table~\ref{brMFV} would signal the presence of
    new sources of flavour and in particular of CP violation. This can
    be confirmed observing a violation of the correlations between $K$
    and $B$ decays discussed above.
\item In particular, recalling that in most extensions of the SM the
  decays $K\to\pi\nu\bar\nu$ are governed by the single $(V-A)\otimes
  (V-A)$ operator, the violation of the upper bounds on at least one
  of the $K\to\pi\nu\bar\nu$ branching ratios, will either signal the
  presence of new complex weak phases at work or new contributions
  that violate the correlations between the $B$ decays and $K$ decays.
\end{itemize}

Assuming that the MFV scenario will survive future tests, the next step
will be to identify the correct model in this class. Clearly, direct searches
at high energy colliders can rule out or identify specific extensions of the
SM. But also FCNC processes can play an important role in this context,
provided the theoretical and experimental uncertainties in some of them will
be sufficiently decreased. In this case, by studying simultaneously several 
branching ratios it should be in principle possible to select the correct MFV 
models by just identifying the pattern of enhancements and suppressions
relative to the SM that is specific to a given model. If this pattern 
is independent of the values of the parameters defining the model, no 
detailed quantitative analysis of the enhancements and suppressions 
is required in order to rule it out. As an example the
distinction between the MSSM with MFV and the models with one universal 
extra dimension should be straightforward:
\begin{itemize}
\item In the MSSM with MFV the branching ratios for $\kpn$, $\klpn$,
  $B\to X_d\nu\bar\nu$ and $B_d\to\mu^+\mu^-$ are generally suppressed
  relative to the SM expectations, while those governed by $V_{ts}$
  like $B\to X_s\nu\bar\nu$, $B_s\to\mu^+\mu^-$ and $B\to X_s\gamma$
  can be enhanced or suppressed depending on the values of parameters
  involved~\cite{hep-ph/0007313}.
\item In the model with one universal extra dimension analyzed in~\cite{BSW02,BPSW}, branching 
ratios for essentially all rare decays are enhanced, the enhancement
being stronger for the decays governed by $V_{ts}$ than for those where $V_{td}$ 
is involved. A prominent exception is the suppression of $B\to X_{s,d}\gamma$~\cite{BPSW,hep-ph/0105084}.
\end{itemize} 
Finally, if MFV will be confirmed, and some new particles
  will be observed, the rare processes discussed in this work will
  constitute a most powerful tool to probe the spectrum of the NP
  model, which might not be entirely accessible via direct studies at
  the LHC.

%
%

\section*{Acknowledgements}
We would like to thank Christoph Greub, Paolo Gambino and David E.
Jaffe for very informative discussions.  We are grateful to the {\utfit}
Collaboration for letting us use the UUT results prior to publication.
T.E. has been supported by the Swiss National Foundation; RTN,
BBW-Contract No.01.0357 and EC-Contract HPRN-CT-2002-00311
(EURIDICE). C.B. has been supported by the DOE under Grant DE-FG03-97ER40546.
This work has been supported in part by Bundesministerium f\"ur
Bildung und Forschung under the contract 05HT4WOA/3, by the
German-Israeli Foundation under the contract G-698-22.7/2002 and by
the EU network "The quest for unification" under the contract
MRTN-CT-2004-503369.

\renewcommand{\baselinestretch}{0.95}

\vfill\eject


\begin{thebibliography}{99}


\bibitem{UTfitSM}
M.~Bona {\it et al.}  [UTfit Collaboration],
arXiv:hep-ph/0501199.

\bibitem{CKMfitter}
J.~Charles {\it et al.}  [CKMfitter Group],
arXiv:hep-ph/0406184.

\bibitem{hep-ph/0307195}
M.~Ciuchini, E.~Franco, F.~Parodi, V.~Lubicz, L.~Silvestrini and A.~Stocchi,
eConf {\bf C0304052}, WG306 (2003)
[arXiv:hep-ph/0307195].

\bibitem{UTfitNP}
M.~Bona {\it et al.}  [UTfit Collaboration], in preparation; Talk given by M. Pierini at the CKM 2005
Workshop, 
http://ckm2005.ucsd.edu/speakers/WG6/Pierini-WG6-fri2.pdf.

\bibitem{UUT}
A.~J.~Buras, P.~Gambino, M.~Gorbahn, S.~Jager and L.~Silvestrini,
Phys.\ Lett.\ B {\bf 500}, 161 (2001)
[arXiv:hep-ph/0007085].

\bibitem{CAB}
N.~Cabibbo,
Phys.\ Rev.\ Lett.\  {\bf 10}, 531 (1963).

\bibitem{KM}
M.~Kobayashi and T.~Maskawa,
Prog.\ Theor.\ Phys.\  {\bf 49}, 652 (1973).

\bibitem{Zakopane}
A.~J.~Buras,
Acta Phys.\ Polon.\ B {\bf 34}, 5615 (2003)
[arXiv:hep-ph/0310208].

\bibitem{AMGIISST}
G.~D'Ambrosio, G.~F.~Giudice, G.~Isidori and A.~Strumia,
Nucl.\ Phys.\ B {\bf 645}, 155 (2002)
[arXiv:hep-ph/0207036].

\bibitem{BOEWKRUR}
C.~Bobeth, T.~Ewerth, F.~Kr\"{u}ger and J.~Urban,
Phys.\ Rev.\ D {\bf 66}, 074021 (2002)
[arXiv:hep-ph/0204225].

\bibitem{hep-ph/0303060}
A.~J.~Buras,
Phys.\ Lett.\ B {\bf 566}, 115 (2003)
[arXiv:hep-ph/0303060].

\bibitem{hep-ph/0112300}
A.~Ali, E.~Lunghi, C.~Greub and G.~Hiller,
Phys.\ Rev.\ D {\bf 66}, 034002 (2002)
[arXiv:hep-ph/0112300].

\bibitem{hep-ph/0310219}
G.~Hiller and F.~Kr\"{u}ger,
Phys.\ Rev.\ D {\bf 69}, 074020 (2004)
[arXiv:hep-ph/0310219].

\bibitem{hep-ph/0410155}
P.~Gambino, U.~Haisch and M.~Misiak,
Phys.\ Rev.\ Lett.\ {\bf 94}, 061803 (2005)
[arXiv:hep-ph/0410155].

\bibitem{BSU}
A.~J.~Buras, F.~Schwab and S.~Uhlig,
arXiv:hep-ph/0405132.

\bibitem{gino}
G.~Isidori,
Annales Henri Poincare {\bf 4}, S97 (2003)
[arXiv:hep-ph/0301159];
G.~Isidori,
eConf {\bf C0304052}, WG304 (2003)
[arXiv:hep-ph/0307014].

\bibitem{Adler970}
S.~C.~Adler {\it et al.}  [E787 Collaboration],
Phys.\ Rev.\ Lett.\  {\bf 79}, 2204 (1997)
[arXiv:hep-ex/9708031];
S.~C.~Adler {\it et al.}  [E787 Collaboration],
Phys.\ Rev.\ Lett.\  {\bf 84}, 3768 (2000)
[arXiv:hep-ex/0002015].

\bibitem{Adler02}
S.~Adler {\it et al.}  [E787 Collaboration],
Phys.\ Rev.\ Lett.\  {\bf 88}, 041803 (2002)
[arXiv:hep-ex/0111091];
S.~Adler {\it et al.}  [E787 Collaboration],
Phys.\ Rev.\ D {\bf 70}, 037102 (2004)
[arXiv:hep-ex/0403034].

\bibitem{E949}
V.~V.~Anisimovsky {\it et al.}  [E949 Collaboration],
Phys.\ Rev.\ Lett.\  {\bf 93}, 031801 (2004)
[arXiv:hep-ex/0403036].

\bibitem{hep-ex/0408119}
K.~Abe {\it et al.}  [Belle Collaboration],
arXiv:hep-ex/0408119.

\bibitem{hep-ex/0404006}
B.~Aubert {\it et al.}  [BABAR Collaboration],
Phys.\ Rev.\ Lett.\  {\bf 93}, 081802 (2004)
[arXiv:hep-ex/0404006].

\bibitem{BSW02}
A.~J.~Buras, M.~Spranger and A.~Weiler,
Nucl.\ Phys.\ B {\bf 660}, 225 (2003)
[arXiv:hep-ph/0212143].

\bibitem{BPSW}
A.~J.~Buras, A.~Poschenrieder, M.~Spranger and A.~Weiler,
Nucl.\ Phys.\ B {\bf 678}, 455 (2004)
[arXiv:hep-ph/0306158].

\bibitem{LittleHiggs}
S.~R.~Choudhury, N.~Gaur, A.~Goyal and N.~Mahajan,
Phys.\ Lett.\ B {\bf 601}, 164 (2004)
[arXiv:hep-ph/0407050];
A.~J.~Buras, A.~Poschenrieder and S.~Uhlig,
arXiv:hep-ph/0410309;
A.~J.~Buras, A.~Poschenrieder and S.~Uhlig,
arXiv:hep-ph/0501230.

\bibitem{PBE}
G.~Buchalla, A.~J.~Buras and M.~K.~Harlander,
Nucl.\ Phys.\ B {\bf 349}, 1 (1991).

\bibitem{BH92}
A.~J.~Buras and M.~K.~Harlander,
Adv.\ Ser.\ Direct.\ High Energy Phys.\  {\bf 10}, 58 (1992).

\bibitem{BBL}
G.~Buchalla, A.~J.~Buras and M.~E.~Lautenbacher,
Rev.\ Mod.\ Phys.\  {\bf 68}, 1125 (1996)
[arXiv:hep-ph/9512380].

\bibitem{Isidori:2005xm}
G.~Isidori, F.~Mescia and C.~Smith,
[arXiv:hep-ph/0308008].

\bibitem{BI03}
G.~Buchalla, G.~D'Ambrosio and G.~Isidori,
Nucl.\ Phys.\ B {\bf 672}, 387 (2003)
[arXiv:hep-ph/0308008].

\bibitem{Isidori:2004rb}
G.~Isidori, C.~Smith and R.~Unterdorfer,
Eur.\ Phys.\ J.\ C {\bf 36}, 57 (2004)
[arXiv:hep-ph/0404127].

\bibitem{b2sgammanlo}
A.~Ali and C.~Greub,
Z.\ Phys.\ C {\bf 49}, 431 (1991);
A.~Ali and C.~Greub,
Phys.\ Lett.\ B {\bf 259}, 182 (1991);
A.~Ali and C.~Greub,
Phys.\ Lett.\ B {\bf 361}, 146 (1995)
[arXiv:hep-ph/9506374];
A.~J.~Buras, A.~Czarnecki, M.~Misiak and J.~Urban,
Nucl.\ Phys.\ B {\bf 631}, 219 (2002)
[arXiv:hep-ph/0203135];
K.~Adel and Y.~P.~Yao,
Phys.\ Rev.\ D {\bf 49}, 4945 (1994)
[arXiv:hep-ph/9308349];
C.~Greub and T.~Hurth,
Phys.\ Rev.\ D {\bf 56}, 2934 (1997)
[arXiv:hep-ph/9703349];
A.~J.~Buras, A.~Kwiatkowski and N.~Pott,
Nucl.\ Phys.\ B {\bf 517}, 353 (1998)
[arXiv:hep-ph/9710336];
M.~Ciuchini, G.~Degrassi, P.~Gambino and G.~F.~Giudice,
Nucl.\ Phys.\ B {\bf 527}, 21 (1998)
[arXiv:hep-ph/9710335];
C.~Bobeth, M.~Misiak and J.~Urban,
Nucl.\ Phys.\ B {\bf 567}, 153 (2000)
[arXiv:hep-ph/9904413];
K.~G.~Chetyrkin, M.~Misiak and M.~Munz,
Nucl.\ Phys.\ B {\bf 520}, 279 (1998)
[arXiv:hep-ph/9711280];
M.~Misiak and M.~Munz,
Phys.\ Lett.\ B {\bf 344}, 308 (1995)
[arXiv:hep-ph/9409454];
N.~Pott,
Phys.\ Rev.\ D {\bf 54}, 938 (1996)
[arXiv:hep-ph/9512252];
Z.~Ligeti, M.~E.~Luke, A.~V.~Manohar and M.~B.~Wise,
Phys.\ Rev.\ D {\bf 60}, 034019 (1999)
[arXiv:hep-ph/9903305];
C.~Greub, T.~Hurth and D.~Wyler,
Phys.\ Lett.\ B {\bf 380}, 385 (1996)
[arXiv:hep-ph/9602281];
C.~Greub, T.~Hurth and D.~Wyler,
Phys.\ Rev.\ D {\bf 54}, 3350 (1996)
[arXiv:hep-ph/9603404];
A.~J.~Buras, A.~Czarnecki, M.~Misiak and J.~Urban,
Nucl.\ Phys.\ B {\bf 611}, 488 (2001)
[arXiv:hep-ph/0105160].

\bibitem{b2sllnnlo}
A.~Ghinculov, T.~Hurth, G.~Isidori and Y.~P.~Yao,
Nucl.\ Phys.\ B {\bf 685}, 351 (2004)
[arXiv:hep-ph/0312128];
C.~Bobeth, P.~Gambino, M.~Gorbahn and U.~Haisch,
JHEP {\bf 0404}, 071 (2004)
[arXiv:hep-ph/0312090];
C.~Bobeth, M.~Misiak and J.~Urban,
Nucl.\ Phys.\ B {\bf 574}, 291 (2000)
[arXiv:hep-ph/9910220];
H.~H.~Asatryan, H.~M.~Asatrian, C.~Greub and M.~Walker,
Phys.\ Rev.\ D {\bf 65}, 074004 (2002)
[arXiv:hep-ph/0109140];
H.~H.~Asatryan, H.~M.~Asatrian, C.~Greub and M.~Walker,
Phys.\ Rev.\ D {\bf 66}, 034009 (2002)
[arXiv:hep-ph/0204341];
A.~Ghinculov, T.~Hurth, G.~Isidori and Y.~P.~Yao,
Nucl.\ Phys.\ B {\bf 648}, 254 (2003)
[arXiv:hep-ph/0208088];
H.~M.~Asatrian, K.~Bieri, C.~Greub and A.~Hovhannisyan,
Phys.\ Rev.\ D {\bf 66}, 094013 (2002)
[arXiv:hep-ph/0209006];
H.~M.~Asatrian, H.~H.~Asatryan, A.~Hovhannisyan and V.~Poghosyan,
Mod.\ Phys.\ Lett.\ A {\bf 19}, 603 (2004)
[arXiv:hep-ph/0311187];
A.~Ghinculov, T.~Hurth, G.~Isidori and Y.~P.~Yao,
Eur.\ Phys.\ J.\ C {\bf 33}, S288 (2004)
[arXiv:hep-ph/0310187].

\bibitem{hep-ex/0108032}
S.~Chen {\it et al.}  [CLEO Collaboration],
Phys.\ Rev.\ Lett.\  {\bf 87}, 251807 (2001)
[arXiv:hep-ex/0108032].

\bibitem{hep-ex/0403004}
P.~Koppenburg {\it et al.}  [Belle Collaboration],
Phys.\ Rev.\ Lett.\  {\bf 93}, 061803 (2004)
[arXiv:hep-ex/0403004];
\bibitem{babarbsg}
J. Walsh for the Babar Collaboration, talk presented at Moriond QCD, 2005, 
http://moriond.in2p3.fr/QCD/2005/SundayAfternoon/Walsh.ppt.

\bibitem{b2sgammacut}
M.~Neubert,
Eur.\ Phys.\ J.\ C {\bf 40}, 165 (2005)
[arXiv:hep-ph/0408179];
D.~Benson, I.~I.~Bigi and N.~Uraltsev,
Nucl.\ Phys.\ B {\bf 710}, 371 (2005)
[arXiv:hep-ph/0410080].

\bibitem{Harati:1999hd}
A.~Alavi-Harati {\it et al.}  [The E799-II/KTeV Collaboration],
Phys.\ Rev.\ D {\bf 61}, 072006 (2000)
[arXiv:hep-ex/9907014].

\bibitem{Barate:2000rc}
R.~Barate {\it et al.}  [ALEPH Collaboration],
Eur.\ Phys.\ J.\ C {\bf 19}, 213 (2001)
[arXiv:hep-ex/0010022].

\bibitem{Herndon:2004tk}
M.~Herndon[CDF and D0 Collaborations],
FERMILAB-CONF-04-391-E
SPIRES entry
{\it To appear in the proceedings of 32nd International Conference on High-Energy Physics (ICHEP 04), Beijing, China, 16-22 Aug 2004}

\bibitem{utfitmethod}
M.~Ciuchini {\it et al.},
JHEP {\bf 0107}, 013 (2001)
[arXiv:hep-ph/0012308].



\bibitem{kpnexpl}
The experimental CL can be found at 
http://www.phy.bnl.gov/e949/E949Archive/br\_cls.dat,
and the likelihood we use can be found at
http://www.utfit.org/kpinunubar/ckm-kpinunubar.html.


\bibitem{BF01}
A.~J.~Buras and R.~Fleischer,
Phys.\ Rev.\ D {\bf 64} (2001) 115010
[arXiv:hep-ph/0104238].

\bibitem{newphase}
A.~J.~Buras, A.~Romanino and L.~Silvestrini,
Nucl.\ Phys.\ B {\bf 520}, 3 (1998)
[arXiv:hep-ph/9712398];
G.~Colangelo and G.~Isidori,
JHEP {\bf 9809}, 009 (1998)
[arXiv:hep-ph/9808487];
A.~J.~Buras, G.~Colangelo, G.~Isidori, A.~Romanino and L.~Silvestrini,
Nucl.\ Phys.\ B {\bf 566}, 3 (2000)
[arXiv:hep-ph/9908371];
G.~Buchalla, G.~Hiller and G.~Isidori,
Phys.\ Rev.\ D {\bf 63}, 014015 (2001)
[arXiv:hep-ph/0006136];
D.~Atwood and G.~Hiller,
arXiv:hep-ph/0307251;
A.~J.~Buras, R.~Fleischer, S.~Recksiegel and F.~Schwab,
Nucl.\ Phys.\ B {\bf 697}, 133 (2004)
[arXiv:hep-ph/0402112];
A.~J.~Buras, T.~Ewerth, S.~Jager and J.~Rosiek,
Nucl.\ Phys.\ B {\bf 714}, 103 (2005)
[arXiv:hep-ph/0408142].

\bibitem{AFB}
G.~Burdman,
Phys.\ Rev.\ D {\bf 57}, 4254 (1998)
[arXiv:hep-ph/9710550].

\bibitem{SUPERB}
A.~G.~Akeroyd {\it et al.}[SuperKEKB Physics Working Group],
arXiv:hep-ex/0406071;
J.~L.~Hewett {\it et al.},
arXiv:hep-ph/0503261.

\bibitem{KN}
A.~L.~Kagan and M.~Neubert,
Phys.\ Rev.\ D {\bf 58}, 094012 (1998)
[arXiv:hep-ph/9803368].

\bibitem{hep-ph/0312260}
T.~Hurth, E.~Lunghi and W.~Porod,
Nucl.\ Phys.\ B {\bf 704}, 56 (2005)
[arXiv:hep-ph/0312260].

\bibitem{BSGNNLO}
 K.~Bieri, C.~Greub and M.~Steinhauser,
  Phys.\ Rev.\ D {\bf 67} (2003) 114019
  [[arXiv:hep-ph/0302051];
M.~Misiak and M.~Steinhauser,
Nucl.\ Phys.\ B {\bf 683}, 277 (2004)
[arXiv:hep-ph/0401041];
M.~Gorbahn and U.~Haisch,
  Nucl.\ Phys.\ B {\bf 713} (2005) 291
  [arXiv:hep-ph/0411071];
M.~Gorbahn, U.~Haisch and M.~Misiak,
arXiv:hep-ph/0504194;
H.~M.~Asatrian, C.~Greub, A.~Hovhannisyan, T.~Hurth and V.~Poghosyan,
  arXiv:hep-ph/0505068.

\bibitem{largetb}
K.~S.~Babu and C.~F.~Kolda,
Phys.\ Rev.\ Lett.\ {\bf 84}, 228 (2000)
[arXiv:hep-ph/9909476];
C.~S.~Huang, W.~Liao and Q.~S.~Yan,
Phys.\ Rev.\ D {\bf 59}, 011701 (1999)
[arXiv:hep-ph/9803460];
S.~R.~Choudhury and N.~Gaur,
Phys.\ Lett.\ B {\bf 451}, 86 (1999)
[arXiv:hep-ph/9810307];
C.~S.~Huang, W.~Liao, Q.~S.~Yan and S.~H.~Zhu,
Phys.\ Rev.\ D {\bf 63}, 114021 (2001)
[Erratum-ibid.\ D {\bf 64}, 059902 (2001)]
[arXiv:hep-ph/0006250];
  C.~Bobeth, T.~Ewerth, F.~Kruger and J.~Urban,
  Phys.\ Rev.\ D {\bf 64} (2001) 074014
  [arXiv:hep-ph/0104284];
A.~Dedes,
Mod.\ Phys.\ Lett.\ A {\bf 18}, 2627 (2003)
[arXiv:hep-ph/0309233];
A.~J.~Buras, P.~H.~Chankowski, J.~Rosiek and L.~Slawianowska,
Phys.\ Lett.\ B {\bf 546}, 96 (2002)
[arXiv:hep-ph/0207241];
A.~J.~Buras, P.~H.~Chankowski, J.~Rosiek and L.~Slawianowska,
Nucl.\ Phys.\ B {\bf 659}, 3 (2003)
[arXiv:hep-ph/0210145];
S.~Baek, P.~Ko and W.~Y.~Song,
JHEP {\bf 0303}, 054 (2003)
[arXiv:hep-ph/0208112];
G.~L.~Kane, C.~Kolda and J.~E.~Lennon,
arXiv:hep-ph/0310042;
A.~Dedes and B.~T.~Huffman,
Phys.\ Lett.\ B {\bf 600}, 261 (2004)
[arXiv:hep-ph/0407285].
\bibitem{hep-ph/0007313}
  A.~J.~Buras, P.~Gambino, M.~Gorbahn, S.~Jager and L.~Silvestrini,
  Nucl.\ Phys.\ B {\bf 592}, 55 (2001)
  [arXiv:hep-ph/0007313].
\bibitem{hep-ph/0105084}
  K.~Agashe, N.~G.~Deshpande and G.~H.~Wu,
  Phys.\ Lett.\ B {\bf 514}, 309 (2001)
  [arXiv:hep-ph/0105084].
\end{thebibliography}
\end{document}